# From Glucose Patterns to Health Outcomes: A Generalizable Foundation Model for Continuous Glucose Monitor Data Analysis

Guy Lutsker[1,2,3], Gal Sapir[1,4], Smadar Shilo[1,2,5,6], Jordi Merino[7,8,9], Anastasia Godneva[1,2], Jerry R Greenfield[10,11,12], Dorit Samocha-Bonet[10,11], Raja Dhir[13], Francisco Gude[14,15], Shie Mannor[3], Eli Meirom[3], Gal Chechik[3], Hagai Rossman[4,*] & Eran Segal[1,16,*]

## Author affiliations

[1] Department of Computer Science and Applied Mathematics, Weizmann Institute of Science, Rehovot, Israel.
[2] Department of Molecular Cell Biology, Weizmann Institute of Science, Rehovot, Israel.
[3] NVIDIA, Tel Aviv, Israel.
[4] Pheno.AI, Tel-Aviv, Israel.
[5] Faculty of Medical and Health Sciences, Tel Aviv University, Tel-Aviv, Israel
[6] The Jesse Z and Sara Lea Shafer Institute for Endocrinology and Diabetes, National Center for Childhood Diabetes, Schneider Children's Medical Center of Israel, Petah Tikva, Israel
[7] Novo Nordisk Foundation Center for Basic Metabolic Research, University of Copenhagen, Copenhagen, Denmark.
[8] Diabetes Unit, Endocrine Division, Massachusetts General Hospital, Boston, MA, USA.
[9] Center for Genomic Medicine, Massachusetts General Hospital, Boston, MA, USA.
[10] Clinical Diabetes, Appetite and Metabolism Lab, Garvan Institute of Medical Research, Sydney, Australia
[11] St Vincent's Clinical School, University of NSW, Sydney, Australia
[12] Department of Endocrinology and Diabetes, St Vincent's Hospital, Sydney, Australia
[13] Swiss Institute of Allergy and Asthma Research (SIAF), University of Zurich, Davos, Switzerland.
[14] Department of Medicine, University of Santiago de Compostela, Spain.
[15] Concepción Arenal Primary Care Center, Santiago de Compostela, Spain.
[16] Mohamed bin Zayed University of Artificial Intelligence, Abu Dhabi, UAE.

[*] Corresponding authors:

Prof. Eran Segal, Department of Molecular Cell Biology and Department of Computer Science and Applied Mathematics, Weizmann Institute of Science, Rehovot, Israel, Tel: 972-8-934-3540, Fax: 972-8-934-4122, Email: eran.segal@weizmann.ac.il, ORCID: 0000-0002-6859-1164.

Dr. Hagai Rossman, Tel-Aviv, Israel, Email: hagai@pheno.ai, ORCID: 0000-0002-5613-8004

## Abstract

Recent advances in self-supervised learning enabled novel medical AI models, known as foundation models (FMs), offer great potential for better characterizing health from diverse biomedical data. Continuous glucose monitoring (CGM) provides rich, temporal data on glycemic patterns, but its full potential for predicting broader health outcomes remains underutilized. Here, we introduce GluFormer, a generative foundation model for CGM data designed to learn glycemic patterns and produce representations that reflect aspects of metabolic health. Trained on over 10 million CGM measurements from 10,812 adults, primarily without diabetes, GluFormer uses autoregressive token prediction to capture longitudinal glucose dynamics. We show that GluFormer generalizes to 19 external cohorts (n=6,044) spanning different ethnicities and ages, 5 different countries, 8 different CGM devices, and diverse pathophysiological states (e.g., prediabetes, type 1 & type 2 diabetes, gestational diabetes, and obesity). Our results suggest that GluFormer's representations can improve upon current CGM metrics, such as the Glucose Management Indicator (GMI), for forecasting clinical measures. In individuals with prediabetes, GluFormer stratifies those likely to experience clinically significant increases in HbA1C% over a two-year period - outperforming baseline HbA1C%. In a longitudinal study of 580 adults with CGM data and 12-year follow-up, GluFormer identifies individuals at elevated risk of developing diabetes more effectively than blood HbA1C%, capturing 66% of all new-onset diabetes diagnoses in the top quartile versus 7% in the bottom quartile. Similarly, 69% of cardiovascular-death events occurred in the top quartile and none in the bottom quartile, indicating GluFormer-based stratification could capture additional risk signals not reflected by traditional glycemic metrics. We also show that CGM representations from pre-intervention periods in Randomized Clinical Trials (RCTs) outperform well established CGM analysis methods in predicting primary and secondary outcomes. Finally, when integrating dietary data into GluFormer, we show that the multi-modal version of the model can generate CGM data based on glucose and dietary intake data, and predict individual responses to specific foods.

## Introduction

The emergence of self-supervised learning (SSL) in healthcare marks a shift in medical AI, enabling the creation of foundation models (FMs) capable of processing and analyzing vast amounts of unlabelled data[1,2] and performing diverse downstream tasks. Examples include FMs for retinal images[3,4], which improved detection rates of ophthalmic diseases in a community setting[5]; FM from wearable data has demonstrated the potential of SSL in analyzing continuous physiological signals from wearable devices[6], and FMs for sleep analysis have improved detection of sleep disorders[7]. Furthermore, FM in pathology have shown remarkable accuracy in diagnosing complex diseases from histopathological images [8]. SSL facilitates efficient training of AI systems without extensive annotated datasets, accelerating their integration into clinical practice[9]. This convergence of AI advancements and the availability of high-quality datasets produces opportunities for major improvements in healthcare delivery and outcomes, particularly in managing chronic conditions like diabetes, which remains one of the leading causes of death and disability worldwide[10].

Diabetes is a chronic disease characterized by progressive deterioration of glycemic control that affects individuals across age groups and geographical regions. The prevalence of diabetes is increasing dramatically - currently affecting one in ten individuals globally, with >500 million people affected in 2021 and global expenditure estimated at over $900 billion per annum[10]. If current trends persist, projections indicate the age-standardized prevalence will increase by 60%, resulting in 1.31 billion people living with diabetes by 2050. Type 2 diabetes (T2D), the most common form of diabetes, is largely driven by preventable risk factors, such as poor diet and lack of appropriate physical activity[10]. Diabetes doubles the risk of cardiovascular complications and is a major risk factor for numerous other comorbidities, including liver disease, lung diseases, cancer conditions, chronic kidney disease, and mental health issues, many of which are leading causes of global morbidity and mortality[11,12,13,14–18]. As the impact of diabetes continues to grow, Continuous Glucose Monitoring (CGM) has emerged as a crucial tool in both managing the disease and enhancing overall patient care.

CGM has shown several advantages over traditional self-monitoring of blood glucose (SMBG) in individuals with diabetes. These include improving glycemic control in adults[19] and in children[20], reducing hypoglycemic events[19], and improving overall quality of life[19]. Recently, in an important consensus statement endorsed by the American Diabetes Association (ADA) and the European Association for the Study of Diabetes (EASD), clinicians and researchers recommended that CGM-derived metrics be incorporated in clinical trials for diabetes[21], paving the way for wider adoption. The usage of CGM devices has also been studied in the non-diabetic population. It may assist in detecting early signs of glucose dysregulation[22], sports related performance enhancement[22], and receiving personalized advice on diet[23]. Additionally, recent research using CGM in non-diabetic adults has revealed substantial day-to-day variability in fasting glucose levels, suggesting the potential for CGM to refine glycemic status assessment [24]. Importantly, the Food and Drug Administration (FDA) recently approved the first over-the-counter CGM device[25], representing a significant shift towards widespread accessibility and glucose monitoring for the non-diabetic population.

Here, we present GluFormer, a generative model for CGM data based on the transformer architecture[26], which efficiently learns over long sequences. It is trained in a self-supervised way using CGM data from the Human Phenotype Project (https://humanphenotypeproject.org/home) dataset, a large-scale, prospective, longitudinal study[27]. During the HPP study, CGM was recorded for 10,812 participants without prior diagnosis of diabetes for two weeks, with simultaneous meal logging. Participants underwent extensive clinical testing as part of the HPP research protocol[27]. To validate GluFormer's generalizability, we evaluated its performance on 19 external cohorts encompassing diverse populations, clinical conditions, and CGM devices (Table 1 in Methods). Importantly, these cohorts differ significantly from the training cohort in terms of glycemic status, comorbidities, study designs, and ancestry.

Despite differences in study populations, GluFormer generalizes to out-of-distribution (OOD) data, accurately predicting clinical outcomes across diverse geographical areas and metabolic conditions. The model generates representations capable of predicting clinical measures both concurrent with CGM recording and up to 12 years into the future. Moreover, we show that the embeddings GluFormer derives from pre-intervention CGM data in clinical trial participants exhibit significant associations with observed post-intervention outcomes, suggesting GluFormer’s potential utility in treatment response prediction. This approach may inform future analyses of glycemic and cardiovascular health parameters and suggests possible applications in diabetes care, including risk stratification and treatment planning..

## Results

Figure 1 provides a schematic overview of the GluFormer model developed in this study. Initially, the model was trained on data from 10,812 mostly non-diabetic participants in the HPP cohort (See population statistics in the supplementary information, Figures S11-S16), with >10,000,000 glucose measurements in total. To facilitate efficient training, CGM data was tokenized such that each measurement within each sample was a discrete token (See Methods). GluFormer was trained using next token prediction, and thus is capable of generating, or having been prompted by some CGM, continuing CGM time series. In addition the model is capable of computing model CGM representations, also known as *embeddings* (see Methods), which can be used in various downstream tasks. Our approach includes a series of evaluations to demonstrate GluFormer's capabilities and versatility: (1) Generative modeling (Figure 1A); (2) Generalizability on 19 external validation cohorts, diverse in geographical location, CGM devices used, and underlying disorders; (3) Multiple downstream tasks: disease prediction, and benchmarking against conventional CGM parameters (Figure 1B); (4) Forecasting clinical trial results in primary and secondary outcomes based on baseline CGM measurements (Figure 1C); and (5) Dietary response generation: Integrating CGM with dietary data, a multimodal version of the model stimulates glucose responses to various foods based on their nutritional content (Figure 1D).

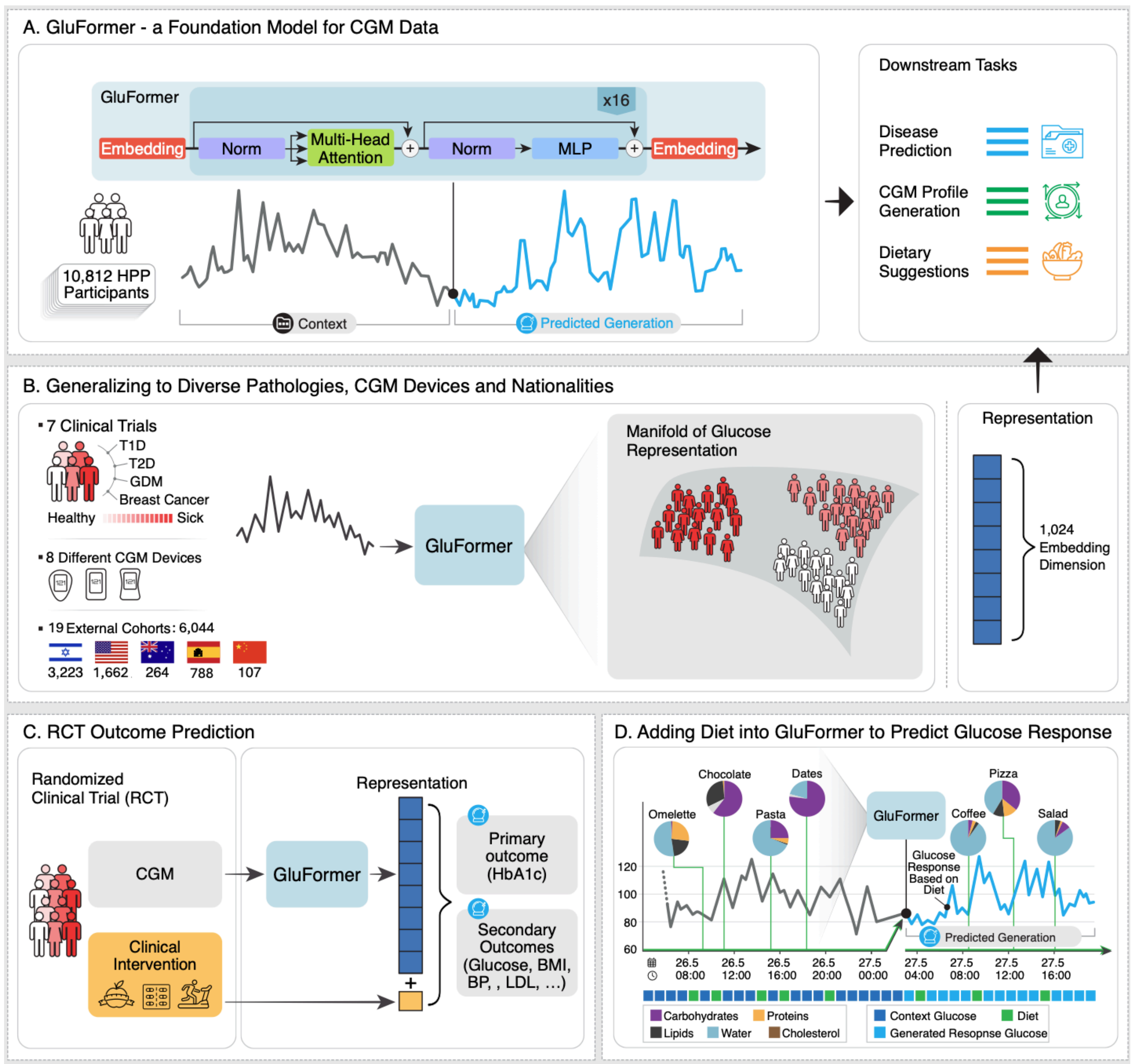

**Figure 1: Overview of GluFormer Architecture, Training Pipeline and Downstream Tasks.**

**A.** Pretraining of the GluFormer model on CGM data from 10,812 individuals in the HPP cohort, with the objective of predicting subsequent glucose measurements (next token prediction). We evaluate GluFormers utility on a variety of downstream tasks, including generating CGM time-series data, fine-tuning for predicting clinical measures, and creating representations (embeddings) that can be used by simple linear models to predict medical outcomes. **B.** We test GluFormer's clinical measure prediction generalization ability on 19 different external datasets including 7 clinical trials across various pathologies (T1DM, T2DM, GDM, etc.) using 8 different CGM devices from 5 different countries. **C.** GluFormer's CGM representations are utilized to forecast treatment outcomes in clinical trial participants - using pre-intervention CGM data to predict clinical outcomes of interventions. **D.** An extended, multimodal, version of GluFormer with an added input of date and time information, and can accept tokens from both glucose modality, as well as dietary tokens.

**GluFormers Representation Space Encodes Physiological Parameters and Generates Reliable CGM Signals**

To evaluate the ability of GluFormer to capture clinically relevant information from CGM data, we analyzed the model's output representations using uniform manifold approximation and projection (UMAP)[28]. We processed CGM time series through GluFormer to generate representation vectors, and these representation vectors were then aggregated using max-pooling to extract features[29] (see Methods). We applied UMAP to visualize these representations in a two-dimensional space, with each point corresponding to a CGM sample from the HPP. When we colored the UMAP by fasting plasma glucose (FPG), we found patterns related to glucose homeostasis. Similarly, when we colored the UMAP by postprandial glucose response (PPGR), we observed distinct gradients indicating varying levels of glucose tolerance. These results imply that GluFormer encodes clinically relevant information, capturing glycemic profiles with related yet distinct characteristics (Figure S2A and S2B).

To further evaluate GluFormer's representations within and between participants, we compared the cosine distances[30] between representations. We measured intra-participant distances using representations from the same participant on different days, and inter-participant distances using representations from different participants. Our results show that inter-participant distances are significantly higher than intra-participant distances (Mann-Whitney, p-value $< 0.001$), indicating that GluFormer captures individual-specific glycemic patterns (Figure S2C).

To assess GluFormer's ability to predict HbA1C% - the most relevant glycemic marker in diabetes control and management with important implications for cardiometabolic outcomes [31,32] - we examined its performance across different model architectures. We compared GluFormer's predictions to those of other model types, including convolutional neural networks[33,34] (CNNs) and multilayer perceptrons (MLPs), as well as Glucose Management Indicator (GMI), which serves as a clinical standard for estimating HbA1C% from CGM data[35]. Our analysis reveals that pre-training through self-supervised learning[36] (SSL) enhances GluFormer's predictive accuracy for HbA1C%. Notably, transformer architectures outperform CNNs and MLPs in capturing complex CGM patterns. Furthermore, frozen representations from the pre-trained model yield better predictions than GMI - the clinical standard, as well as the CGM-based composite scores (i.e., commonly used CGM-derived metrics like mean glucose, time in range, and glycemic variability indices)[37,38], suggesting more effective feature capture (Figure. S2D).

Overall, while HbA1C% is a risk factor rather than a direct clinical outcome, these findings demonstrate the potential utility of pre-trained models in predicting such glycemic markers from CGM data. The importance of relevant marker predictions and the relationship to long-term diabetes outcomes is further explored in the following section, where we examine its role in identifying individuals at risk of worsening glycemic control.

**GluFormer‑Derived HbA1C% Identifies Individuals at Risk of Worsening Glycemic Control and Predicts Long‑Term Diabetes Outcomes**

To assess whether GluFormer‑predicted HbA1C% conveys clinically relevant risk information beyond standard HbA1C% measurements, we first analyzed a subcohort of 337 prediabetic individuals (baseline HbA1C% of 5.7–6.4%) from the HPP. We stratified participants into quartiles according to either their GluFormer‑predicted HbA1C% or their measured (blood) HbA1C% at baseline and examined changes in actual HbA1C% at two‑year follow‑up (Figure. 2A). Notably, stratification by GluFormer‑predicted HbA1C% distinguished individuals whose glycemic control deteriorated over two years from those who improved or remained stable. Participants in the top GluFormer quartile showed a mean increase of 0.18% in HbA1C%, whereas those in the bottom quartile decreased by 0.13% ($p < 0.001$ by Mann–Whitney). In contrast, when participants were stratified according to baseline blood HbA1C% alone, no significant difference in future HbA1C% trajectories emerged among quartiles. These findings suggest that, despite all individuals being in the prediabetic range at baseline, GluFormer captures subtle glycemic‑risk signals that enhance prediction of short‑term progression beyond what is discernible from a baseline measurement of HbA1C%. The observed changes in HbA1C in the top and bottom GluFormer quartiles (0.18% increase and 0.13% decrease, respectively) may carry substantial clinical relevance. For example, the Diabetes Prevention Program (DPP) study demonstrated that a lifestyle intervention achieving an HbA1C reduction of approximately 0.1% within the first year of follow-up led to a 58% reduction in diabetes incidence among individuals with elevated fasting and post-load plasma glucose levels compared to the placebo group over an average follow-up period of 2.8 years[39,40].

Next, we investigated whether GluFormer also outperforms standard HbA1C% measurements in predicting long‑term diabetes incidence and cardiovascular mortality (Figure. 2B). For this purpose, we analyzed the AEGIS[41] cohort, a longitudinal study of 580 adults followed for 12 years. Each participant was ranked by either GluFormer‑predicted HbA1C% or their baseline blood HbA1C%, and we compared the top vs. bottom quartiles in their risk of developing diabetes or experiencing cardiovascular death. Individuals in the top 25% of GluFormer predictions exhibited a significantly elevated risk of diabetes (log‑rank $p = 2.3 \times 10^{-6}$), capturing 65.8% of all new‑onset cases over 12 years; by contrast, 7.3% of diagnoses were observed in the bottom 25%. Stratification by measured HbA1C% yielded overlapping incidence curves (log‑rank $p = 0.71$), indicating no meaningful difference in diabetes‑free incidence among quartiles.

A similar pattern was observed for cardiovascular death: 69.2% of events occurred in the GluFormer top quartile, whereas none were detected in the bottom quartile (log‑rank $p = 0.001$). In contrast, stratification by blood HbA1C% was not significantly associated with death risk ($p = 0.25$). Thus, although AEGIS[41] lies fully outside GluFormer's training distribution, GluFormer‑predicted HbA1C% nonetheless captures critical risk signals for both long‑term diabetes and cardiovascular mortality that standard HbA1C% measurements fail to detect. These results demonstrate GluFormer's potential to improve personalized risk stratification and identify individuals who stand to benefit most from early intervention.

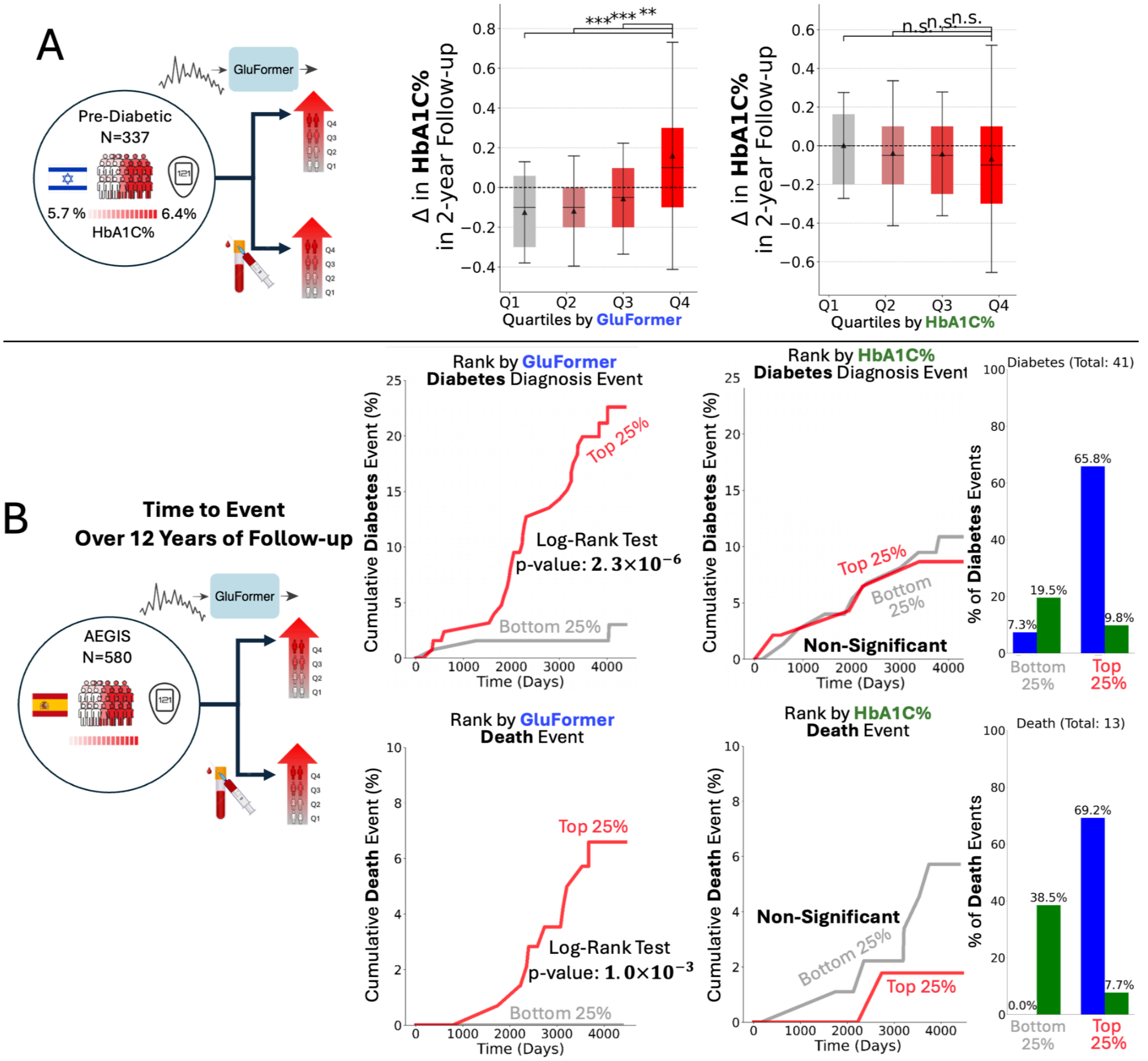

**Figure 2. GluFormer-derived HbA1C% Outperforms Measured (blood) HbA1C% in Identifying Individuals At Risk for Future Increases in HbA1C% And Long-term Diabetes Incidence.**

**(A)** Analysis of 337 prediabetic individuals (baseline HbA1C% of 5.7–6.4%) from the HPP cohort, comparing 2-year HbA1C% changes when stratified into quartiles by GluFormer-predicted HbA1C% versus baseline (blood) HbA1C%. The quartile cutoffs for GluFormer are Q1: 5.17–5.37, Q2: 5.37–5.46, Q3: 5.46–5.60, Q4: 5.60–6.56, while for baseline HbA1C% they are Q1: 5.70–5.73, Q2: 5.73–5.85, Q3: 5.85–6.00, Q4: 6.00–6.40. GluFormer-based quartiles (top right) exhibit significant differences in 2-year HbA1C% trajectories ($p < 0.001$ by Mann–Whitney): participants in the top quartile (red) show a mean increase of +0.18% whereas those in the bottom quartile (grey) decrease by −0.13%.

By contrast, quartiles defined by baseline HbA1C% alone (bottom right) do not show statistically significant separation over the same period.

**(B)** Kaplan–Meier analysis of 580 adults from the AEGIS[41] study, followed for 12 years.

**(Top)** *Diabetes Incidence*. Individuals in the top 25% of GluFormer‑predicted HbA1C% (red) have significantly shorter diabetes‑free survival than those in the bottom 25% (grey), with $p = 2.3 \times 10^{-6}$ by log‑rank test. In contrast, stratification into quartiles of measured HbA1C% produces survival curves that are not statistically different ($p = 0.71$). The bar plot (right) shows that 65.8% of all diabetes diagnoses occur in the top quartile identified by GluFormer, versus only 7.3% in the bottom quartile.

**(Bottom)** *Cardiovascular Death.* Similarly, individuals in the top 25% by GluFormer‑predicted HbA1C% have a higher risk of cardiovascular death ($p = 0.001$ by log‑rank test), whereas measured HbA1C% quartiles are not significantly associated with death risk ($p = 0.25$). The bar plot (right) indicates that 69.2% of cardiovascular‑death events are captured in the GluFormer top quartile, compared to 0% in the bottom quartile.

The quartile cutoffs for GluFormer are Q1: 4.86 - 5.22, Q4: 5.62 - 8.99; while for baseline HbA1C% they are Q1: 3.9 - 5.2, Q4: 5.8 - 9.9.

### GluFormer Generates CGM Signals and Captures Innate Characteristics of Glycemic Dynamics, and Generalizes to Diverse Populations

Generative models have shown great promise in healthcare by producing high-quality synthetic data, helping to mitigate issues of data scarcity and imbalance in medical research. For instance, MINIM[42], a recently introduced foundation model for multimodal medical imaging, demonstrated how large-scale synthetic image generation can bolster diagnostic performance and improve clinical applications. Here we evaluate GluFormer's ability to generate realistic CGM time series, we compared observed and generated CGM data using both visual inspection and quantitative metrics. We generated CGM signals for individuals and compared them to their actual recordings using the CGM-based composite scores[37,38]. These results show that the model captures glucose dynamics and generates signals that align with individuals' glycemic characteristics. The radar plots demonstrate the model's ability to reproduce clinically important variables, though some discrepancies occur due to unmodeled personal actions such as changes in daily dietary habits (Figure. 3A).

To assess the accuracy of generated signals across multiple glycemic metrics, we compared key CGM-based composite scores from real CGM signals to the average of three generated CGM signals. Our analysis reveals significant Pearson correlations between the CGM-based composite scores of generated and original CGM curves. For instance, we observed correlations of $r=0.98$ ($p<0.001$) for mean glucose, $r=0.98$ ($p<0.001$) for GMI, and $r=0.89$ ($p<0.001$) for % time glucose below 70 mg/dL, indicating the model's ability to reproduce essential glycemic features (Figure. 3B). We also noted that while variability measures (e.g., CV and SD) correlated strongly for the HPP and Israeli T2D cohorts, they yielded lower correlations in several other cohorts, suggesting that capturing short-term glucose fluctuations can be more challenging for GluFormer in diverse populations.

To test the model's generalizability, we evaluated its performance on external data from cohorts in new geographical areas not included in the training set. We generated time-series for participants with diverse glycemic characteristics, including those with type 2 diabetes mellitus (T2DM) or gestational diabetes mellitus (GDM), across different continents (Asia, Australia and North America). The results show that the model generates time-series with correlated glycemic metrics for these diverse populations. For the T2D cohort, all correlations were above 0.8 ($p<0.001$), demonstrating the model's adaptability to different populations (Figure. 3C).

To examine the impact of input data length on model performance, we varied the duration of CGM data provided to the model. We found that extending the input time series from 0 to 10 days increased the agreement between the model’s generated signals and the actual CGM data, with the average correlation (based on Pearson correlation between their CGM-derived metrics) rising from 0.46 to 0.90 ($p < 0.001$ for both). This highlights that providing longer sequences of past glucose readings markedly enhances GluFormer’s generative accuracy (see Figure. S3, S4).

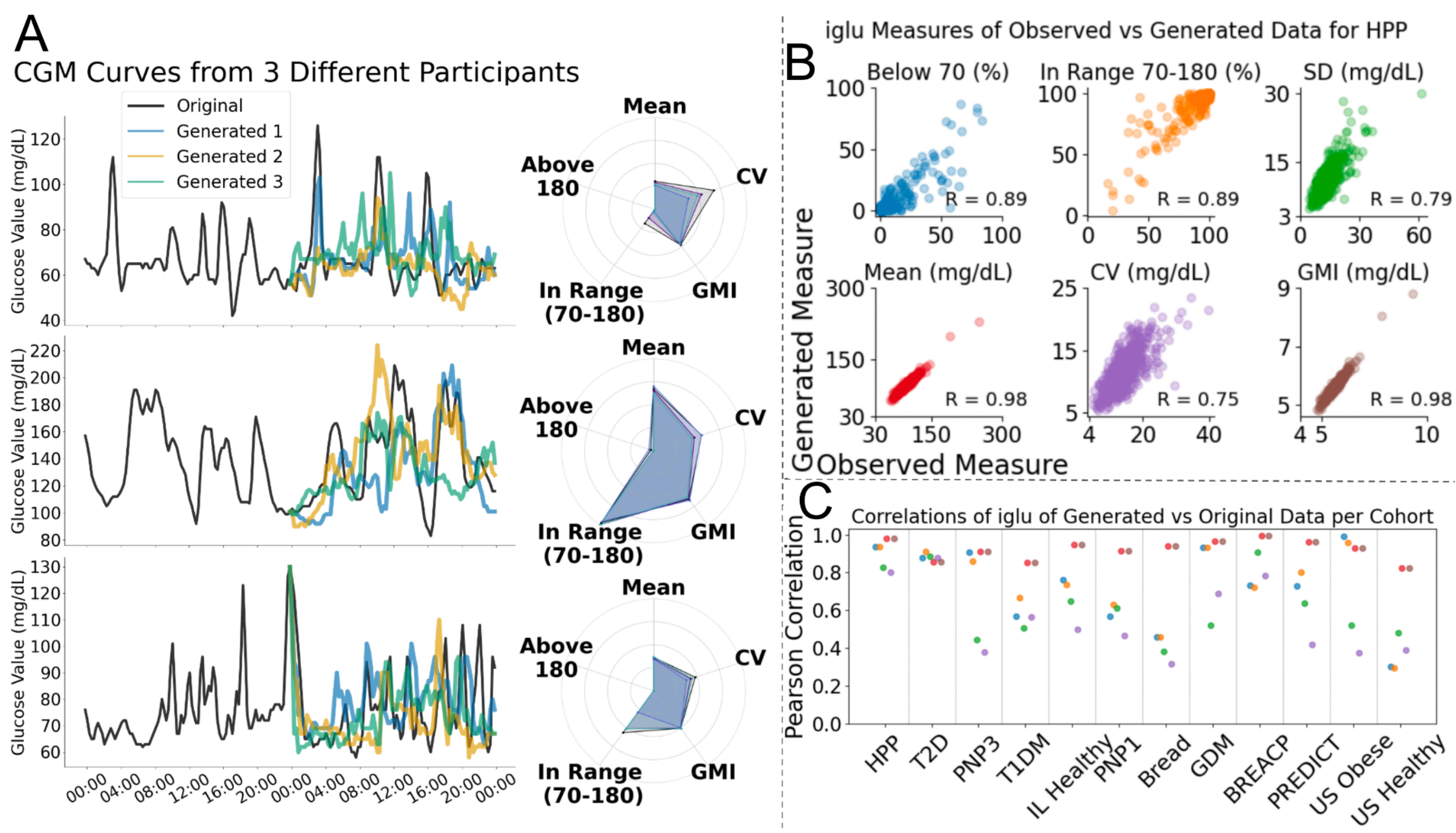

**Figure 3: Evaluation of GluFormer's Capabilities in Simulating and Analyzing CGM Data**

**A.** Day-by-day analysis for three participants (one in each row) from the HPP test set. Each panel on the right column compares the observed CGM readings (black curve) with three predicted time-series generated by GluFormer using different seeds (colored in blue, orange, and green). Each time series spans two full days, with the first day's CGM data providing context for the model prediction of the second day, which is compared with the true CGM profiles. Radar charts on the left evaluate CGM-based composite scores (Mean, Coefficient of Variation (CV), and Glucose Management Indicator (GMI), In Range 70-180, Above 180) and show similarity between observed CGM-based composite scores, and generated CGM-based composite scores.

**B.** CGM-based composite scores of observed versus generated time series across the test participants from the HPP cohort. For each test participant from the HPP, we generated 3 CGM days and calculated their CGM-based composite scores. To get a single CGM-based score calculation per person, we averaged the 3 calculated composite scores on the generated data. Each scatter plot depicts a different CGM-based score, where each point corresponds to a sample from the test set from a particular day - observed and generated. CGM-based composite scores shown here are: Below 70, In Range 70-180, Standard Deviation (SD), Mean, Coefficient of Variation (CV), and Glucose Management Indicator (GMI). We observe a strong agreement (all Pearson correlations are above 0.75, p <

0.001) between observed and predicted metrics, validating GluFormer's precision in adhering to clinically significant metrics, affirming its potential for large-scale diagnostic and predictive healthcare applications.

**C.** Plot illustrating the correlation of key CGM-based composite scores between observed and generated CGM signals across cohorts. Metrics are the same from panel B. External cohorts include (from left to right):

HPP: n=10,812, Israel, Healthy, FSLP, FSL IQ
PNP3: n=225, Israel, Pre-diabetes, FSL
T2D IL: n=23, Israel, T2DM, FSL
T1DM PNP: n=121, Israel, T1DM, Dexcom, Freestyle Navigator, FSL
IL Healthy: n=1,159, Israel, Healthy, iPro2; Medtronic
PNP1: n=926, Israel, Healthy, FSL
Bread Study: n=20, Israel, Healthy, iPro2
GDM: n=549, Israel, GDM, FSL
BREACP: n=200, Israel, Breast Cancer Survivors, FSL
PREDICT: n=264, Australia, T2DM and prediabetes, FSL
US Obese: n=156, US, Obese / pre-diabetes, FSL
US Healthy: n=327, US, Healthy, FSL

FSL: Freestyle Libre Pro.

Pearson correlation is significantly above random for all metrics and all cohorts, and is particularly high for T2D, PNP3 and BREACP.

**GluFormer Encodes Clinical Information with Long-Term Associations**

In addition to evaluating diabetes onset and cardiovascular mortality, we next examined GluFormer's ability to predict a wide range of clinical parameters both at baseline and longitudinally, we compared GluFormer against GMI using ridge regression models. We assessed predictions at the time of CGM recording and for prospective time points up to 4 years in the future to evaluate the model's capability for both immediate and longer-term health assessments. Even among adults without a diabetes diagnosis, CGM data has been shown to correlate with various metabolic and clinical measures[43]. In our cohort, GluFormer achieves higher albeit still modest correlation coefficients compared with GMI across several clinical parameters, indicating significant predictive improvements (Figure. 4). At baseline (Figure. 4A), GluFormer shows higher predictive performance for visceral adipose tissue (VAT) ($r = 0.41$ vs. 0.20 for GMI, $p < 0.001$), liver attenuation ($r = 0.19$ vs. 0.04, $p < 0.001$).

The model also effectively predicts sleep-related apnea-hypopnea index (AHI) ($r = 0.22$ vs. $0.11$, $p < 0.001$) and systolic blood pressure (SBP) ($r = 0.26$ vs. $0.05$, $p < 0.001$). Importantly, GluFormer's computational representations from CGM at baseline extend to clinical parameters 2 and 4 years following the baseline measurements. At a 2-year horizon (Figure. 4B), the model maintains its higher performance, particularly for VAT ($r = 0.41$ vs. $0.27$, $p < 0.001$) and fasting glucose levels ($r = 0.48$ vs. $0.33$, $p < 0.001$). Even at a 4-year horizon (Figure. 4C), GluFormer continues to outperform GMI in predicting fasting glucose levels ($r = 0.47$ vs. $0.29$, $p < 0.001$) and maintains significant predictive power for VAT ($r = 0.21$ vs. $0.14$, $p < 0.001$), although the overall correlation strength decreases over time. These findings demonstrate GluFormer's ability to extract meaningful health insights from CGM data, surpassing traditional metrics in both immediate and long-term health assessments. The model's capacity to predict a wide range of clinical measures up to four years in advance underscores its potential for long-term health monitoring and risk assessment.

Beyond encoding continuous clinical measures, we also evaluated GluFormer's ability to discriminate which individuals would develop diabetes over these same time horizons (Figure. 4, left and middle panels). At baseline (Figure. 4A, left panel), GluFormer achieved a higher ROC AUC (0.75) for classifying individuals at risk of diabetes compared to GMI (0.66). The decision curve analysis (Figure. 4A, middle panel) demonstrates that GluFormer provides superior clinical utility across a wide range of threshold probabilities (0.2-0.8), where the threshold probabilities represent the clinician's chosen trade-off between the benefits of identifying true positives and the harms of false positives [44] [45]. This indicates the potential for GluFormer's classifications to be more clinically beneficial when compared to GMI across intervention strategies. This pattern of enhanced discriminative performance and clinical utility persisted at both 2-year and 4-year horizons (Figure. 4B,C, left and middle panels), underscoring GluFormer's value for both immediate and long-term risk assessment.

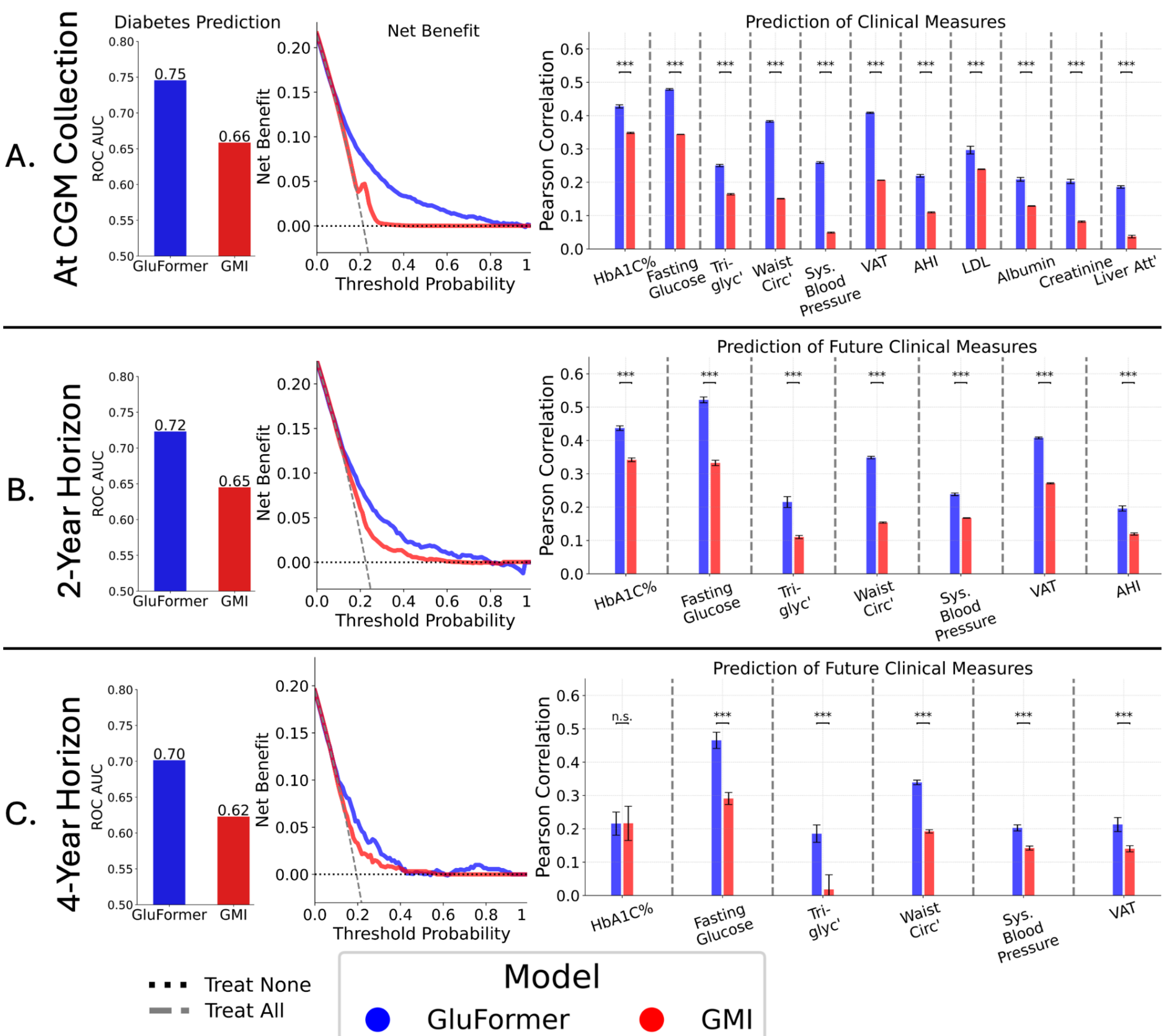

**Figure 4: Predictive Performance of Clinical Measures using GluFormer Representations and GMI**

**A.** Prediction results of clinical measures from GluFormer representations / GMI at the time of CGM recording.

**B.** Prediction results of clinical measures from GluFormer representations / GMI 2 years past CGM recording.

**C.** Prediction results of clinical measures from GluFormer representations / GMI 4 years past CGM recording.

**(Left panels)** Bar plots comparing ROC AUC for diabetes prediction at three time points: time of CGM collection, 2-year horizon, and 4-year horizon. GluFormer (blue) consistently achieves higher AUC values than GMI (red), indicating better discrimination in identifying individuals at risk for diabetes.

**(Middle panels)** Decision curve analysis depicting net benefit as a function of threshold probability for the same three time points. Net benefit offers a clinically interpretable measure balancing the true positive benefit against the false positive cost. The dotted black line indicates a “treat none” strategy, while the dashed gray line indicates “treat all.” Across all horizons, GluFormer (blue) provides almost always a higher net benefit than GMI (red).

**(Right Panels)** Pearson correlations for various clinical measures at different time points using representations from GluFormer (blue) and GMI (red), processed through a ridge regression model. Each data point represents the average of 10 prediction iterations with different random seeds, and error bars show standard deviations. Clinical measures include HbA1C%, glucose, triglycerides (Tri-glic’), waist circumference (Waist Circ’), systolic blood pressure (Sys. Blood Pressure), visceral adipose tissue (VAT), apnea-hypopnea index (AHI), LDL cholesterol, albumin, creatinine, and liver attenuation (Liver Att). Statistical significance between the two groups (GluFormer and GMI) was determined using the Mann-Whitney U test, marked by asterisks (*$p < 0.05$, **$p < 0.01$, ***$p < 0.001$, n.s. not significant). GluFormer outperforms GMI measures, showing significant improvements in predictive performance in all comparisons across panels A and B, while in panel C, GluFormer outperforms GMI in all measures except for HbA1C%.

**GluFormer representations predict glycemic and other clinically relevant outcomes for external data, across geographical areas**

To test whether GluFormer generalizes across diverse cohorts, geographies, devices, and diseases, we applied its representations to predict outcomes using CGM data from external research cohorts. We evaluated the model's performance on varied populations.

Our results show that GluFormer successfully created representations reflecting aspects of personal clinical pictures and predicted several clinical outcomes more accurately than GMI, consistently across these diverse populations (Figure 5). The representations demonstrated higher Pearson correlations with several disease-specific features. In T2DM patients, GluFormer predicted creatinine levels ($r = 0.27$, $p < 0.001$), a key indicator of kidney function. For breast cancer survivors, the model estimated albumin ($r=0.19$ vs GMI $r=0.05$, $p<0.001$) and creatinine levels ($r=0.12$ vs GMI $r=0.02$, $p<0.001$), which reflect the impact of treatments on renal function and overall health. In pregnant women with gestational diabetes, GluFormer accurately predicted hemoglobin ($r=0.42$ vs GMI $r=0.15$, $p<0.001$) and platelet counts ($r=0.35$ vs GMI $r=0.12$, $p<0.001$), important markers for maternal and fetal health. These improved correlations compared to GMI across both populations suggest that GluFormer captures underlying physiological patterns in glucose dynamics that are more informative of treatment-related effects on organ function and maternal health status.

These findings may indicate that GluFormer's latent space encompasses a broad spectrum of health indicators beyond glycemic metrics, reflecting overall health status and disease-specific characteristics.

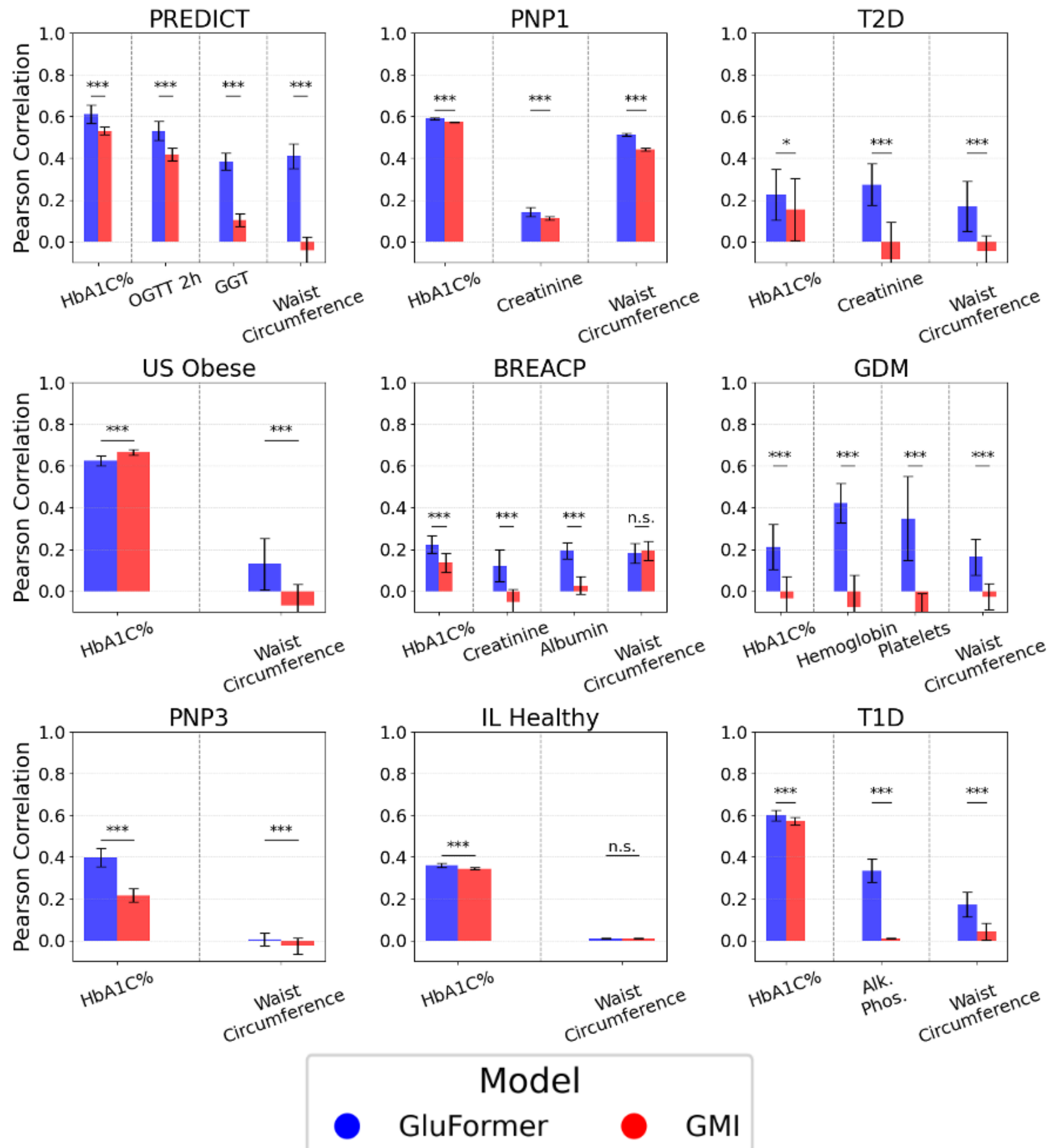

**Figure 5: Predictive Performance of Clinical Measures Across Out-Of-Distribution Cohorts using GluFormer Representations and GMI.**

Pearson correlations for various clinical measures across diverse patient populations not seen during model training. These populations include the cohorts: PREDICT (Australia), PNP1 (Israel), T2D (Israel), US Obese (United States), BREACP (Israel), GDM (Israel), PNP3 (Israel), IL Healthy (Israel), and T1D (Israel). Clinical measures include HbA1C%, oral glucose tolerance test (OGTT 2h), gamma-glutamyl transferase (GGT), waist circumference, creatinine, albumin, hemoglobin, and platelets, Predictions were generated using a ridge regression model using GluFormer representations (blue) or GMI measures (red). Each data point represents the average of 10 prediction iterations with different random seeds, and error bars show standard deviations.

Statistical significance between the two groups (representations and GMI) was determined using the Mann-Whitney U test, marked by asterisks (*$p < 0.05$, **$p < 0.01$, ***$p < 0.001$, n.s. not significant). The results demonstrate that GluFormer generally outperforms GMI measures, showing significant improvements in predictive performance in 12 out of the examined comparisons, while GMI only outperformed the representations significantly in two instances.

**GluFormer CGM representations predict outcomes of clinical trials**

A critical challenge in clinical trials is predicting which participants will respond best to specific interventions. To address this, we investigated whether we could predict clinical trial outcomes from baseline data, potentially enabling personalized treatment decisions and improved patient care. We used pre-intervention CGM data from completed clinical trials and used GluFormer to compute CGM representations, and to predict the studies' clinical outcomes, comparing our results to the traditional CGM derived GMI.

Our results show that GluFormer representations consistently outperform GMI in forecasting clinical outcomes across diverse studies (Figure 6A). In the PREDICT cohort, GluFormer showed higher predictive power relative to GMI for HbA1C%, creatinine, and waist circumference ($p < 0.001$). In the BREACP study, the model significantly improved predictions for HbA1C%, body fat percentage ($p < 0.001$), lymphocyte levels and creatinine. For the PNP3 diet intervention study, GluFormer achieved improvements in predicting changes in HbA1C%, LDL, and glucose levels ($p < 0.001$). These predictions used only pre-intervention CGM data and a binary variable indicating the intervention arm. GluFormer also outperformed GMI in predicting primary outcomes for open access clinical trials with CGM data (Figure 6B). These results demonstrate the model's ability to predict medical intervention effects across various measures and studies. This suggests that CGM-derived GluFormer representations could potentially benefit precision health and clinical trial design by providing insights into patient-specific treatment responses based on pre-intervention metabolic states.

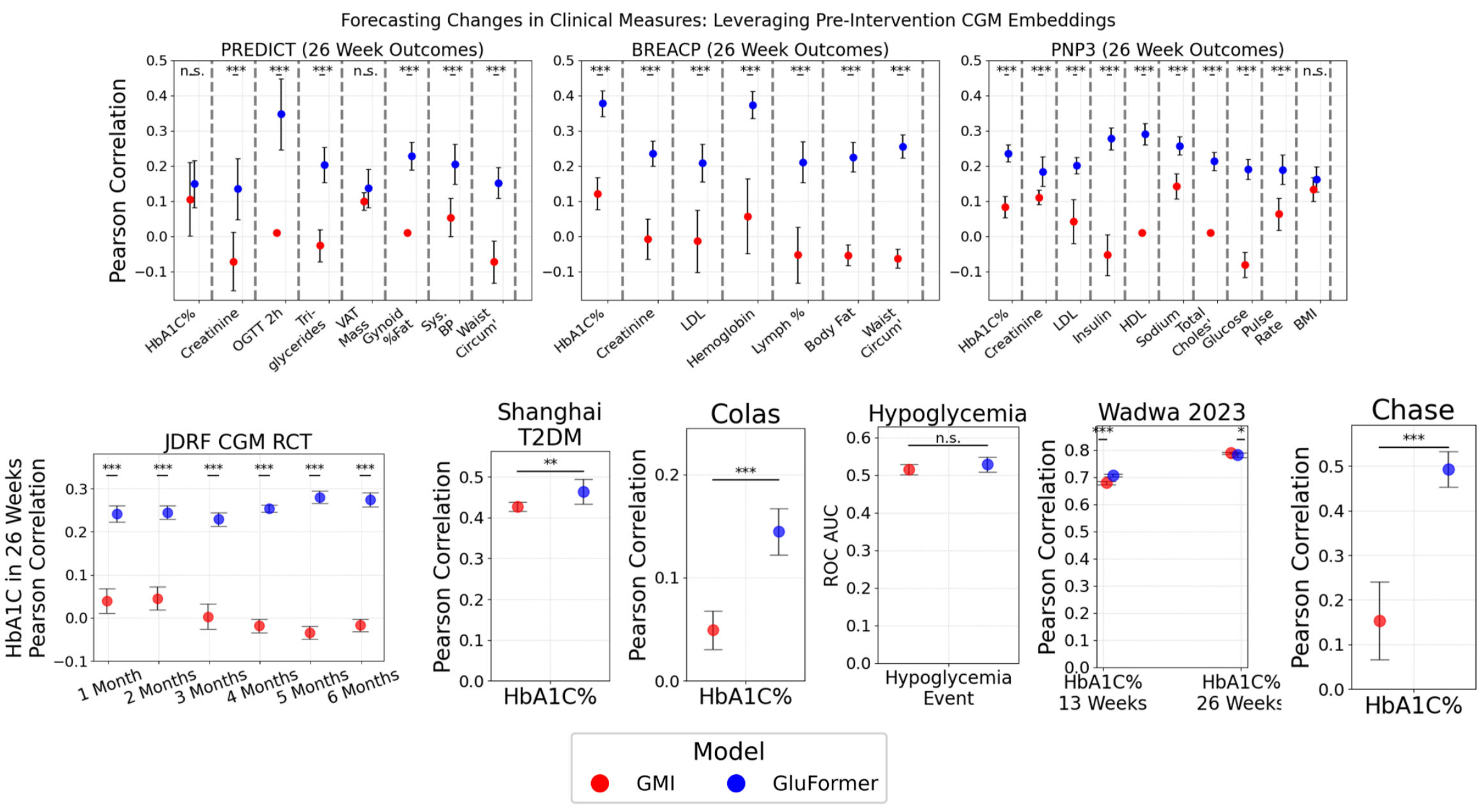

**Figure 6: Forecasting Clinical Outcomes Following Interventional Studies Using Pre-Intervention CGM Representations**

**(Top)** Pearson correlations for clinical measures post-intervention across three different clinical studies, each featuring two intervention arms. Each panel represents data from one of the studies, specifically focusing on populations from Australia (metformin plus diet intervention), Breast Cancer (metformin intervention‎), and PNP3 (diet intervention). The measures evaluated include Hemoglobin A1c (HbA1C%), creatinine, LDL cholesterol, hemoglobin, lymphocyte count, visceral adipose tissue (VAT) mass, body fat percentage, waist circumference (Waist Circum'), and more, depending on the specific conditions and interventions of each study.

Correlations shown here measure the efficacy of using CGM-derived representations (blue points) versus GMI (red points) to forecast changes in clinical measures 26 weeks following the intervention. The representations are derived from CGM data recorded pre-intervention, while the clinical outcomes are analyzed post-intervention, incorporating the binary intervention variable (0 / 1) representing the absence or presence of a specific clinical intervention.

**(Bottom)** Open-access clinical trial prediction results. In each trial, we chose the primary outcome and predicted it using GluFormer representations versus GMI. Specifically:

**JDRM CGM RCT:** Predicting HbA1C% at 26 weeks, using CGM from 1–6 months of measurements.

**Shanghai T2D:** Predicting HbA1C% at 26 weeks from 2 weeks of CGM measurement.

**Colas:** Predicting baseline HbA1C% from 2 weeks of CGM measurement.

**Hypoglycemia:** Predicting, from a week of CGM, whether there will be a hypoglycemia event in the following week.

**Wadwa 2023:** Predicting HbA1C% at 13 and 26 weeks from baseline CGM measurements.

**Chase:** Predicting baseline HbA1C% from baseline CGM measurements.

Each data point represents the average of 10 predictions, each utilizing a different random seed, to ensure robustness and reliability of the results. Error bars indicate the standard deviation across these predictions. Statistical significance between the prediction capabilities of CGM-derived representations and GMI is assessed using the Mann-Whitney U test, annotated with asterisks indicating levels of significance (*$p < 0.05$, **$p < 0.01$, ***$p < 0.001$, n.s. = not significant).

### Exploring Temporal Encoding in CGM Data Generation

To improve GluFormer's ability to capture temporal information we added date and time into the architecture of GluFormer through learned representations for minute, hour, day of the week, and month (see Methods). We then trained two versions of the model using a pre-processed HPP dataset (see Methods) and evaluated their performance in generating CGM data for test participants (See Figure. S5). The temporal informed model achieved a Pearson correlation of 0.22 ($p < 0.001$) with the observed participant CGM data, outperforming the original GluFormer, which had a correlation of 0.15 ($p < 0.001$).

### Diet Encoding to Create a Multimodal GluFormer Model on HPP

To enhance glucose prediction accuracy, we incorporated dietary information into GluFormer, recognizing that blood glucose dynamics are heavily influenced by daily actions, particularly food intake. Unlike static clinical measures, continuous glucose monitoring captures real-time physiological responses to various inputs, making the integration of dietary data particularly valuable for improving predictive capability. We developed a multimodal version of GluFormer that includes macronutrient content alongside glucose data, tokenizing both glucose measurements and diet macronutrients to create a synchronized multimodal sequence for training. The multimodal model was trained using a next-token prediction strategy, learning to predict subsequent glucose tokens based on the combined sequence of glucose and diet tokens, with diet tokens masked out of the loss function to focus on glucose response prediction (See Methods). We then evaluated the impact of dietary data inclusion by comparing two versions of the model: one learned with diet tokens and one without. These models were tested on a generation task using data from test participants. Our results show that incorporating dietary information significantly enhances prediction accuracy (Figure. 7). The multimodal GluFormer achieved an average Pearson correlation of 0.5 with observed CGM data across test participants, surpassing the original GluFormer's correlation of 0.22 (p-value < 0.001). In addition we show that 91% of test participants showed better correlation (Figure. 7.A.1), and 92% exhibited improved mean absolute error (MAE) when diet information was included (Figure. 7.A.2). These findings are further supported by a qualitative inspection of generated CGM time series plots, which demonstrate that glucose predictions incorporating dietary data more closely track the original glucose curves, particularly around meal times (Figure. 7.C.1, Figure. 7.C.2). To ensure robustness, we conducted tests across different random seeds, with consistent outcomes (Figure. S8). This enhancement suggests a significant advancement in our ability to model and predict glucose responses, potentially leading to improved management strategies for diverse populations.

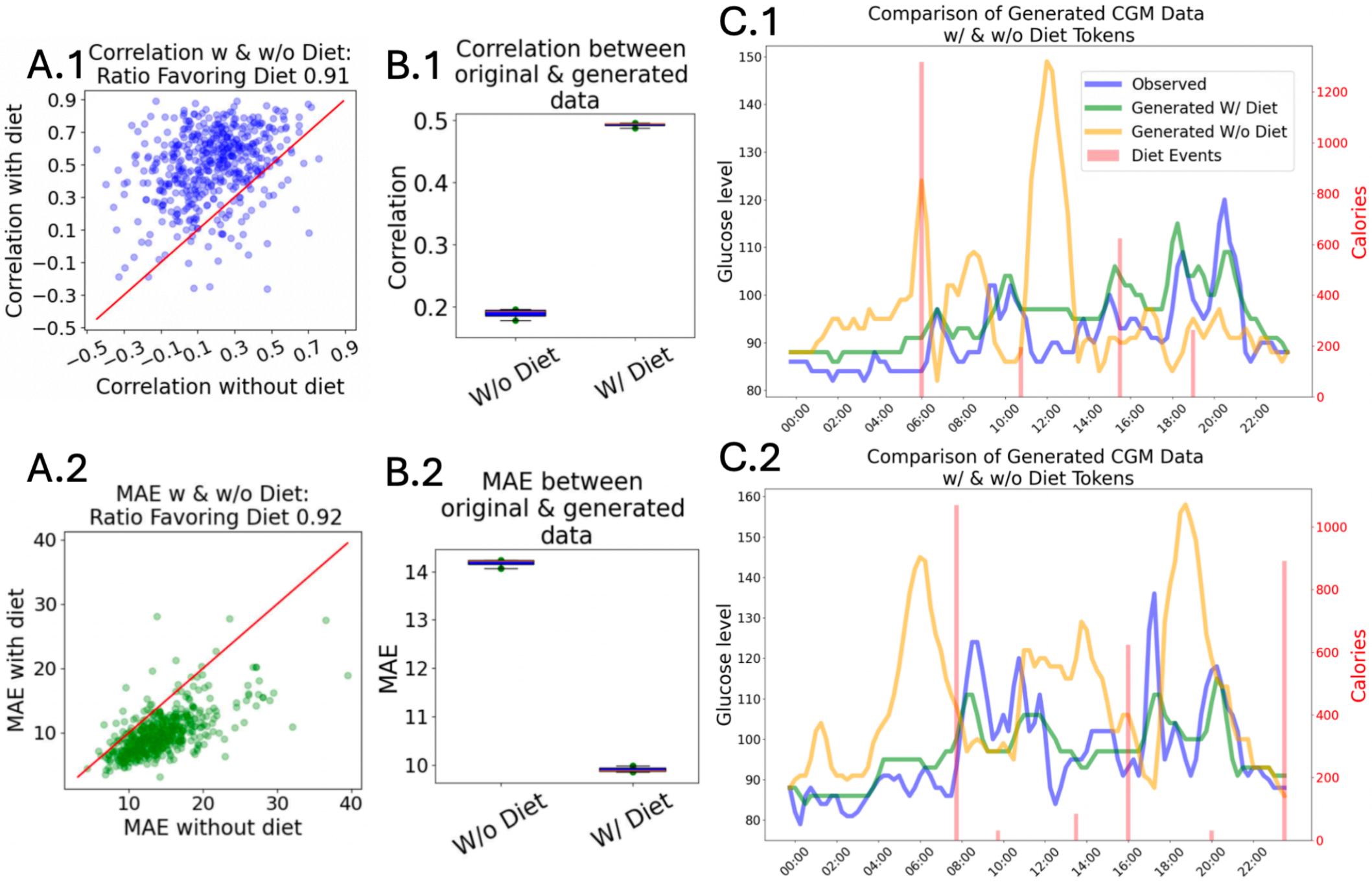

**Figure 7: Impact of Dietary Data on GluFormer Model Performance.**

**A.** Comparison of Pearson correlation **A.1** and mean absolute error (MAE) **A.2** between the original and generated CGM data, with and without the inclusion of dietary data. Scatter plots show the improvements in correlation and MAE when dietary data is included, indicated by the majority of points falling above the diagonal line on correlation, and below on MAE metrics. **B.** Box plots summarizing the overall performance, showing the average correlation **B.1** and MAE **B.2** across all test participants, for 5 different random seeds (used for generation) with lower MAE and higher correlation for models including dietary data. **C.** Time series plots demonstrating glucose level predictions for two example participants **C.1** and **C.2.** The observed CGM data (blue line) is compared to data generated with dietary tokens (green line) and without dietary tokens (orange line). Red bars indicate times of dietary events, highlighting the model's improved performance in capturing glucose spikes when dietary information is included.

## Discussion

This work introduces GluFormer, a foundation model trained on over 10 million CGM measurements from 10,812 adults, primarily without diabetes. The model demonstrates robust generalization across 19 external cohorts spanning different ethnicities, ages, and pathophysiological states. When applied to a 12-year longitudinal study, GluFormer identified individuals at elevated risk of developing diabetes more effectively than blood HbA1C%, capturing 66% of all new-onset diabetes diagnoses in the top quartile versus 7% in the bottom quartile. Similarly, 69% of cardiovascular-death events occurred in the top quartile with none in the bottom quartile, indicating that these GluFormer-based stratifications may capture risk beyond traditional glycemic metrics. The model outperforms conventional CGM-derived metrics in predicting diverse clinical outcomes, including HbA1C%, liver function, and blood lipids, and demonstrated higher net benefit[44,45] across the examined decision thresholds in our decision-curve analysis. Integration of dietary information further enhanced the model's predictive accuracy, enabling simulation of glucose responses to nutritional interventions. These capabilities, validated across multiple geographical regions and metabolic disorders, could potentially establish GluFormer as a tool for analyzing glycemic patterns and predicting long-term health outcomes.

The clinical significance of GluFormer's risk stratification capabilities extends beyond statistical improvements over existing metrics. By identifying individuals at risk of metabolic deterioration before clinical manifestations appear, the model could enable a shift toward preventive intervention. This aligns with findings from landmark studies like the Diabetes Prevention Program[39], where early lifestyle modifications reduced diabetes incidence by 58% in high-risk populations. GluFormer's ability to distinguish future glycemic trajectories in prediabetic individuals - where traditional HbA1C% measurements show no predictive power - offers potential clinical value, as this population represents a critical window for intervention. Furthermore, the model's capacity to simultaneously assess risks for both diabetes and cardiovascular outcomes could help clinicians develop more comprehensive prevention strategies. The performance across different clinical decision thresholds, demonstrated through decision curve analysis, suggests practical utility in diverse healthcare settings where the balance between true positives and false positives must be calibrated according to local resources and intervention capabilities.

Analysis of GluFormer's latent space (represented in Figure S2) reveals how self-supervised learning can capture biologically meaningful patterns in physiological time series data. Just as language model embeddings organize semantic relationships, GluFormer's representations separate distinct aspects of glucose metabolism, distinguishing between postprandial responses (primarily linked to beta cell dysfunction [46]) and fasting glucose levels (primarily associated with hepatic insulin resistance [47]). This unsupervised organization of metabolic features aligns with clinical understanding of diabetes progression, where postprandial hyperglycemia typically precedes fasting hyperglycemia. The model's ability to encode these temporal dynamics without explicit supervision suggests potential applications in monitoring metabolic dysfunction and disease progression.

The choice of next token prediction as our pre-training strategy has proven to be highly effective. This approach, inspired by the success of models like GPT [26,48], compels the model to utilize past knowledge to predict future events. By focusing on the task of predicting subsequent glucose measurements based on previous readings, the model learns intricate patterns and dependencies within the data. This strategy not only provides the model with the ability to generate realistic glucose signals but also prepares it for diverse downstream tasks, as evidenced by its performance across various clinical predictions. The results demonstrate the versatility of this pre-training method, making it a useful approach for self-supervised learning in the biomedical domain.

Our findings demonstrate that pretraining significantly enhances the predictive capabilities of our model, as illustrated in Figure 2D. By utilizing representations generated from the pretrained model without any finetuning, we achieved superior predictions of clinical measures compared to traditional CGM-based metrics. Interestingly, in some scenarios, our model struggled to significantly outperform traditional metrics, which are closely related to diabetes diagnosis and management. This is a well-documented phenomenon in SSL[49], given that traditional CGM-based statistics were specifically designed to proxy measures like HbA1C%.

However, GluFormer demonstrated performance in predicting outcomes not directly tied to these conventional metrics. Notably, it outperforms traditional CGM-based metrics on measures less directly related to glycemic health, such as creatinine and hemoglobin levels. These metrics are critical for assessing kidney function in T2DM, the impact of chemotherapy in cancer survivors, and maternal health in gestational diabetes. GluFormer's performance across this broader range of clinical parameters reflects its general-purpose design, which enables pattern recognition across diverse medical parameters.
Currently, clinicians typically rely on limited metrics such as fasting glucose and HbA1C% to assess metabolic health status. Even for individuals with CGM, clinical focus is often restricted to time in range, hypo/hyperglycemic events, GMI/eHbA1C%, and coefficient of variation[50]. Our unsupervised analysis approach may provide additional predictive signals that are not fully reflected in standard glycemic metrics used in diabetes care.

The model's generative capabilities further validate its deep representation of glycemic dynamics, successfully producing CGM signals that closely match clinical parameters of original data, even for out-of-distribution populations. We observed significant Pearson correlations with disease-specific features, such as creatinine levels in T2DM patients, albumin and creatinine in cancer survivors, and hemoglobin and platelet counts in pregnancy.

Recent studies have demonstrated that discretizing continuous time series into categorical tokens can enhance neural networks performance[51]. Following this insight, GluFormer employs tokenization of glucose measurements, enabling effective capture of temporal patterns in glycemic data. This approach, inspired by successes in natural language processing, proved particularly valuable when integrating dietary information. By incorporating both glucose and nutrient tokens into a unified sequence, our multimodal version of GluFormer achieved significantly improved prediction accuracy compared to glucose-only models, especially around meal times.

While our model effectively predicts glucose responses to dietary inputs, we acknowledge important limitations - particularly regarding counterfactual prediction modeling that would be necessary for simulating intervention outcomes. Nevertheless, our nutrient-based approach to food representation offers inherent advantages for generalization across dietary patterns and cultural contexts. The success of this multimodal integration suggests promising directions for incorporating additional physiological signals, such as sleep patterns, continuous photoplethysmogram (PPG) signals, and physical activity data, potentially enabling a more comprehensive understanding of health dynamics. This points towards the development of a multimodal health model based on continuous signals and tracking.

The convergence of increasing CGM data availability, in light of recent FDA clearance of over-the-counter devices for non-diabetic use [52], and advancements in AI technology presents an unprecedented opportunity for metabolic health research. GluFormer represents a significant step towards utilizing this wealth of information. In addition, as CGM devices become more affordable and accessible, applications in the wellness realm, such as personalized diet planning for weight loss, are likely to proliferate.

The recent global initiative to deliver precision health in diabetes [53], emphasizes the need for a paradigm shift in understanding diabetes heterogeneity. This initiative calls for redefinition of diabetes subtypes [54], integration of multiple data sources, and development of novel biomarkers. GluFormer aligns closely with these objectives, offering a tool for capturing diabetes heterogeneity across diverse populations. Trained on the HPP dataset, which includes a diverse population encompassing various ethnicities, GluFormer is designed to be applicable to diverse demographics.

Despite the demonstrated performance and broad applicability of GluFormer, several limitations must be acknowledged. First, while our validation methodology ensured participant-level data separation to prevent leakage, the dataset predominantly comprises healthy, non-diabetic individuals from a limited geographical region, which may restrict the model's generalizability to populations with rare metabolic conditions or different ethnic backgrounds [55]. Our benchmarking choices, particularly the use of GMI as baselines, while representing current clinical standards, may not capture all possible approaches to CGM data analysis. The varying lengths of CGM recordings and inconsistent availability of contextual information across participants create additional challenges for standardization. The dietary data integrated into the model relies on self-reported logs, which are prone to inaccuracies and omissions, potentially affecting predictions related to dietary interventions. Additionally, the integration of dietary data required extensive engineering efforts to quantify the nutrient content of each food item, a process that is both expensive and time-consuming, creating barriers to scalability and widespread implementation.

The complexity and interpretability of transformer models, including GluFormer, also pose significant challenges. These models are often regarded as "black boxes", making it difficult to understand the reasoning behind their predictions. The computational requirements for training and deploying such models create additional implementation hurdles in clinical settings. Currently, clinical practice primarily relies on simpler metrics such as HbA1C% and fasting glucose levels, and even metrics like GMI and CGM-based composite scores have not yet been widely adopted. Consequently, while the advanced architecture of GluFormer provides enhanced predictive capabilities for complex tasks, it may still require considerable time before it can be adopted clinically. Additionally, the model may inherit biases present in the training data, which need to be meticulously managed to ensure accurate and fair predictions across different demographic groups.

In conclusion, GluFormer demonstrates versatility in metabolic health analysis, showing improved performance over traditional metrics for specific clinical outcomes across diverse populations and conditions. While the model's representation learning approach reveals patterns in CGM data that extend beyond conventional glycemic metrics, further validation studies are needed to establish their full clinical utility. By introducing an analytical approach that captures longitudinal glucose patterns and their associations with health outcomes, this work contributes to ongoing efforts in precision diabetes care and clinical trial optimization.As CGM technology becomes increasingly accessible, approaches like GluFormer may help advance our understanding of metabolic health and support more targeted intervention strategies, though significant work remains to translate these capabilities into clinical practice.

# Methods

## GluFormer

Data

The training dataset comprises Continuous Glucose Monitoring (CGM) records from 10,812 subjects, each monitored over a two-week period using the Freestyle Libre Pro 2 device (Abbott, US), which captures glucose levels subcutaneously every 15 minutes. These values typically range from 30 to 200 mg/dL. Initial day readings were excluded to mitigate noise from device calibration errors, commonly observed during the first 24 hours. Further, participants with fewer than 100 readings were removed from the dataset. Missing data points, occurring in less than 0.1% of instances, were addressed through linear interpolation to fill gaps due to time skips in measurements.

### Tokenization

We adopt a tokenization approach that transforms continuous glucose measurements into discrete tokens, drawing on compelling evidence that discretization can substantially enhance time-series forecasting performance in neural models[51,56–58]. Notably, large-scale empirical analyses have demonstrated that binning real-valued observations into categories can provide both stability in training and improved predictive accuracy across different architectures, even though it discards both the continuous, as well as the ordinal nature of the raw data in favor of discrete labels[56]. These benefits are believed to arise because tokenization regularizes the input space, effectively reducing noise and forcing the model to learn robust patterns rather than overfit to minor fluctuations. Moreover, the discrete nature of tokens aligns with the protocols used in LLMs, where a fixed vocabulary simplifies the training pipeline. Indeed, recent developments in foundational time-series models have illustrated the potential of direct analogy to natural language processing frameworks: models like Chronos[56] suggest that, by treating time-series signals as “words” within a learned vocabulary, one can leverage the self-supervised learning strategies that have revolutionized NLP[56]. Likewise, Rabanser et al. have shown that discretizing continuous values into tokens can outperform sophisticated continuous-valued methods, often due to better generalization and reduced sensitivity to outliers [56]. These insights also echoes findings in generative modeling of audio and images, where discretizing raw data (e.g., quantized waveforms or pixel intensities) proved crucial for the success of convolutional and transformer-based architectures[57,58]. In this spirit, we quantized glucose readings between 40 and 500 mg/dL into 460 evenly spaced intervals based on their empirical distribution in the training dataset. Each resulting token corresponds to one of these discrete bins, creating a well-defined "vocabulary" of glucose states. Samples with fewer than 1200 measurements were padded with a special <MASK> token to maintain a constant input size of 1200 for training and evaluation.

Data division into training, validation, and testing sets was participant-based rather than sample-based to ensure consistency in model evaluation. The division was made randomly, and was divided into these sizes : train set: 80% of participants, validation set: 10%, test set: 10%.

**Model Architecture**

We employed a transformer-based model[26] structured primarily around an encoder mechanism. The architecture was defined by the following parameters: *embedding dimension* of 1024, 16 attention heads, 16 transformer layers (or blocks), and a feed-forward dimension of 2048. The model was designed to process sequences up to 25,000 tokens in length and was trained with a *context length* of 1200 tokens. These choices were made based on a hyper parameter search.

The *vocabulary size* of the model is 461 - 460 representing glucose values from 40 to 500 md/dL, and one masking token.

The model initially embeds all tokens using the embedding layer (*vocabulary size X embedding dimension*), such that a single sample is embedded into a matrix of *context length X embedding dimension*. These then enter the transformer blocks, and exit having the same dimensions. These are then inputted into the un-embedding layer (same dimension as the embedding matrix), and are transformed to a distribution over possible token - *context length X vocabulary size*.

Each position is trained to predict the next token, and so these distributions are compared with the observed next tokens using cross-entropy to match the observed distribution.

We also added positional encoding [26] to model the global positions on the tokens in the sequence.

To create a generative autoregressive model, we used a causal masking technique [26] during training. This approach ensures that the attention mechanism only considers past tokens as context, effectively preventing access to information from future tokens.

**Pretraining and Optimization**

The objective during pretraining was to forecast subsequent tokens using a causal masking technique. This task was structured to enhance predictive accuracy by learning from unmasked tokens to predict the subsequent masked ones, employing cross-entropy loss calculated across the distribution of the 460 possible tokens.

**Optimization Details**

Model optimization was conducted using the AdamW optimizer [59] with a learning rate of 3e-5, with no weight decay, and dropout at 0.1, with batch size of 32 per GPU (effective batch size of 32X8 = 256). A StepLR scheduler was applied to adjust the learning rate with a decay factor of 0.99 every 10 steps over a total of 100 epochs. Model selection was based on performance metrics from the validation set, accounting for 10% of the data. The final model was evaluated using the test set, also representing 10% of the data. Training was performed on 8 NVIDIA A40.

**Producing an Output Embedding per Sample**

Producing a single embedding for each CGM sample offers a compact, learned summary of an individual's glycemic profile. By compressing long sequences of CGM readings into a single vector, we reduce the dimensionality of the data while retaining meaningful patterns that are critical for downstream tasks such as disease risk classification or outcome prediction. This representation-based approach parallels methods in other domains (e.g., language modeling), where embeddings succinctly capture the essential features of lengthy input sequences in a way that is both interpretable and readily integrable with traditional machine learning algorithms.

To represent each CGM sample, we wanted to use the output of the model, to get the processed version of the sample. Each CGM sample, containing 1200 glucose measurements (tokens), is passed through the transformer, which outputs a 1024-dimensional vector for each token. We keep these vectors in the high-dimensional representation space rather than convert back to token probabilities.

<u>Aggregation of Vectors:</u> We needed to reduce the 1200 vectors from each sample into a single representative vector. We evaluated three pooling methods:

Average Pooling: Calculates the mean of the 1200 vectors, assuming equal contribution from all time points.

Maximum Pooling: Selects the maximum value for each of the 1024 dimensions across the 1200 vectors, highlighting the most prominent features in the sequence.

Minimum Pooling: Selects the minimum value for each dimension, capturing the lowest bounds of the glucose measurements.

Note, we removed the <MASK> token before doing any of the pooling methods, to remove any non-informative data from the representation.

Evaluation and Selection: These methods were evaluated based on their ability to predict clinical outcomes using representations generated from the validation set. Max-pooling emerged as the most effective method, consistently providing better performance in predicting key clinical measures like HbA1C%.

**Initialization of the Embedding Layer.**

Our initial approach for embedding tokens in GluFormer followed standard practice in large language models (LLMs), wherein the embedding matrix is randomly initialized. We experimented with alternative schemes aimed at leveraging prior knowledge of the scalar glucose values, including techniques that enforced uniform Euclidean or cosine distances among tokens and regularization approaches that maintained these distances throughout training. Despite our expectations that such specialized initializations might yield performance gains—given the known meaning of scalar glucose values—none of these methods consistently improved prediction accuracy or other metrics. Consequently, we reverted to random initialization of the embedding layer for the final model. Nevertheless, we believe further research is warranted to explore whether specialized embeddings can more effectively encode the numeric relationships among glucose measurements.

**UMAP**

To visualize the high-dimensional embedding space generated by GluFormer and explore the clustering of glycemic profiles, we employed Uniform Manifold Approximation and Projection (UMAP) using the umap-learn Python package (version 0.5.6). All representations from the Human Phenotype Project (HPP) dataset, encompassing training, validation, and test subsets, were included to ensure a comprehensive representation of the data distribution. Prior to dimensionality reduction, representations were standardized to have zero mean and unit variance to facilitate effective distance calculations during the UMAP process. We utilized UMAP's default parameters, specifically setting n_neighbors to 15, min_dist to 0.1, and the distance metric to 'cosine', aligning with the cosine similarity used in the distance analyses.

The standardized representations were then input into the UMAP algorithm to project the 1,024-dimensional vectors into a two-dimensional space, enabling intuitive visualization of potential clusters and gradients within the data. The choice of 'cosine' as the distance metric ensured consistency with the intra- and inter-participant analyses, preserving the angular relationships between representations. The resulting UMAP projections were subsequently used to generate scatter plots, which were color-coded based on various clinical and demographic attributes to facilitate the examination of underlying patterns and associations within the embedding space. This dimensionality reduction technique provided a meaningful visualization framework to interpret the complex relationships captured by GluFormer's representations, supporting further downstream analyses and validation of the model's performance across diverse clinical parameters.

**Clinical Measures in UMAP:**

HbA1C% (Figure S1.A): When colored by independently measured HbA1C% levels, the UMAP projection exhibits a gradient from lower to higher values, indicating that GluFormer's representations effectively capture long-term glycemic control.

Visceral Fat Mass (Figure S1.B): representations color-coded by DXA-derived visceral fat mass show distinct clustering, suggesting that the model encodes information related to body composition.

Hypoglycemic and Hyperglycemic Events (Figure S1.C): The distribution of representations based on the percentage of time CGM values were above 140 mg/dL and below 70 mg/dL displays clear separations, reflecting the model's sensitivity to glycemic variability.

Demographic Factors (Figure S1.D & E): Age and gender-specific colorations further demonstrate that GluFormer representations encapsulate demographic variations, although to a lesser extent compared to clinical metrics.

**Intra & Inter Distance Analysis**

To evaluate the distinctiveness of GluFormer's representations at both individual and population levels, we conducted an intra- and inter-participant distance analysis using cosine similarity metrics. This analysis aimed to determine whether the model effectively captures unique glycemic patterns specific to each individual while maintaining broader generalizability across the population.

First, representations for each CGM sample in the test set were generated using the pre-trained GluFormer model. Each CGM sample, comprising 1200 glucose measurements, was processed to obtain a 1,024-dimensional embedding vector through the max-pooling aggregation method described earlier. This resulted in an embedding representation for each participant's glycemic profile on a given day.

Distance Calculation: Intra-Participant Distances: For each participant, we computed the cosine distance between all pairs of their representations across different days. Specifically, for a participant with $N$ CGM samples, we calculated $N(N-1)/2$ pairwise cosine distances. This measures the variability of representations within the same individual over time.
Inter-Participant Distances: To assess the diversity between different individuals, we calculated the cosine distance between representations from distinct participants. Given the large number of participants, we employed a stratified sampling approach to ensure computational feasibility while maintaining representativeness. Specifically, for each participant, we randomly selected 100 representations from other participants to compute inter-participant cosine distances.

Statistical Analysis: To statistically compare the intra- and inter-participant distance distributions, we employed the Mann-Whitney U test, a non-parametric test suitable for comparing two independent samples without assuming normal distribution. Given the large sample size, the test provides robust significance levels.

A p-value threshold of $p < 0.001$ was set to determine statistical significance, mitigating the risk of type I errors given the extensive pairwise comparisons.

**Predictive Stratification Analysis**

To evaluate the predictive power of GluFormer-derived HbA1C% compared to baseline HbA1C% in forecasting future glycemic trajectories, we conducted a quartile-based stratification analysis within a cohort of 337 prediabetic individuals. Prediabetes was defined by baseline HbA1C% levels ranging from 5.7% to 6.4%, as per established clinical guidelines. Baseline HbA1C% measurements were obtained from clinical records at the time of Continuous Glucose Monitoring (CGM) data collection. Utilizing the GluFormer model, we generated predicted HbA1C% values based on CGM-derived representations for each participant. Participants were then stratified into quartiles based on either their baseline

HbA1C% or their GluFormer-predicted HbA1C% values. Over a two-year follow-up period, changes in HbA1C% levels were tracked for each quartile. To assess the significance of differences in future HbA1C% changes across quartiles, we employed the Mann-Whitney U test, setting significance thresholds at $^{*}p < 0.05$, $^{**}p < 0.01$, $^{***}p < 0.0001$.

In addition to HbA1C%, similar stratification analyses were performed for fasting glucose levels within the prediabetic cohort and for estimated Glomerular Filtration Rate (eGFR) in individuals classified as eGFR stage 2 (60-100 $mL/min/1.73m^2$). For fasting glucose, participants were stratified into quartiles based on their baseline and GluFormer-predicted fasting glucose values, followed by an evaluation of two-year changes in fasting glucose across quartiles using the Mann-Whitney U test. For eGFR stage 2, individuals were divided into quartiles based on baseline and predicted eGFR values, and subsequent changes in eGFR over two years were similarly analyzed. These analyses were conducted to determine whether GluFormer-derived predictions offer superior stratification capabilities in identifying high- and low-risk subgroups for future clinical parameter changes compared to traditional baseline measurements. All statistical tests were two-tailed, with p-values below 0.05 considered statistically significant.

**Prediction Performance of GluFormer vs. Clinically Validated Parameters and GMI Across Subgroups**

To evaluate the predictive performance of GluFormer in comparison to Clinically Validated Parameters and the Glucose Management Indicator (GMI) across different clinical subgroups, we conducted subgroup-specific regression analyses followed by correlation assessments. The study cohort was stratified into four distinct subgroups based on baseline glycemic status: Healthy, Pre-Diabetic, Diabetic, and a combined Pre-Diabetic & Diabetic group. For each subgroup, we employed ridge regression models to predict clinical outcomes—specifically HbA1C% and Fasting Glucose levels—using three sets of input features: GluFormer-derived representations, Clinically Validated Parameters (as extracted using the iglu package), and GMI values. For the gluformer embeddings we observed on an independent validation set that an alpha of between 50-100 yielded best results, so we chose an alpha value of 80 for all ridge regressions we did with the GluFormer embeddings. For iglu we saw the same effect with an alpha value of between 0.1 and 1, so we chose 0.5. For GMI regularization was not needed as it consists of only one variable - hence we set alpha to be 0. The models were trained and validated using a consistent cross-validation framework to ensure comparability across methods. Following model training, we calculated the Pearson correlation coefficients between the predicted and actual clinical measurements within each subgroup to quantify predictive accuracy.

To determine the statistical significance of GluFormer's performance relative to Clinically Validated Parameters and GMI, we applied the Mann-Whitney U test to compare the distributions of Pearson correlation coefficients across the different prediction methods within each subgroup.

**Survival Analysis (Kaplan–Meier) for Long-Term Outcomes**

We utilized data from the AEGIS[41] study, which tracked 580 participants for 12 years, to determine whether GluFormer representations or standard HbA1C% measurements offer stronger prognostic information for future diabetes onset. Among these participants, 42 individuals developed diabetes during follow-up, and we recorded the time (in days) until diagnosis. To evaluate stratification performance, we split participants into top and bottom quartiles based on either (i) baseline blood HbA1C% or (ii) predicted HbA1C% values derived from GluFormer embeddings. For baseline HbA1C%, this involved simply ranking all participants by measured HbA1C% and taking the highest and lowest 25%. For GluFormer, we used the same ridge regression protocol (with alpha=80) described in "Predictive Modeling of Medical Measures on HPP Using Representations," HbA1C a k-fold model on the AEGIS CGM data and aggregating the results into a predicted HbA1C%. Participants were then ranked by these predicted values, again splitting into top and bottom quartiles. We generated Kaplan–Meier curves for each group and performed a log-rank test to compare their time-to-diabetes outcomes, thus testing whether the stratification by GluFormer-derived versus blood-measured HbA1C% differed in separating those who would develop diabetes from those who remained diabetes-free.

**ROC AUC Analysis for Diabetes Prediction at Different Time Points**

To evaluate the predictive performance of GluFormer-derived representations in forecasting the onset of diabetes, we conducted Receiver Operating Characteristic (ROC) curve analyses at three distinct temporal horizons: at the time of Continuous Glucose Monitoring (CGM) collection (baseline), two years, and four years post-CGM monitoring. The study cohort comprised 337 prediabetic individuals, defined by baseline HbA1C% levels ranging from 5.7% to 6.4%, as per established clinical guidelines. Diabetes diagnosis was determined based on blood HbA1C% test, or blood fasting glucose during follow-up assessments conducted at the two-year and four-year marks. For each time point, we developed binary classification models using three different sets of predictors: (1) GluFormer-derived representations, generated by processing baseline CGM data through the pre-trained GluFormer model to obtain 1,024-dimensional feature vectors; (2) Glucose Management Indicator (GMI) scores, calculated using the iglu package [37,38]; and (3) baseline HbA1C% levels, serving as a traditional glycemic marker.

Logistic regression was employed to construct the classification models, with each set of predictors used independently to forecast diabetes onset. To assess model performance, ROC curves were plotted, and the Area Under the Curve (AUC) was computed for each predictor across all time points.

**Net Benefit Analysis for Diabetes Prediction at Different Time Horizons**

To evaluate the clinical utility of GluFormer-derived predictions in diabetes forecasting, we employed Decision Curve Analysis [44] [45] (DCA) to assess the net benefit of GluFormer compared to the Glucose Management Indicator (GMI) across varying threshold probabilities and temporal horizons. The analysis was conducted at three distinct time points: baseline (time of CGM collection), two years, and four years post-CGM monitoring. For each time horizon, we generated predicted probabilities of diabetes onset using both GluFormer and GMI models. Probabilities were calculated using logistic regression fitted on both the representations (as was done for the ROC AUC Analysis, described in the section above**)**, as well as GMI using the *predict_proba* method in sklearn. The true diabetes outcomes were determined based on blood HbA1C% test, or blood fasting glucose during follow-up assessments conducted at the two-year and four-year marks. . Utilizing the dcurves Python package, we calculated the net benefit for each model across a range of threshold probabilities (0 to 1 in increments of 0.01), which represent the clinician's willingness to treat based on predicted risk. This approach allows for the comparison of models by quantifying the trade-off between true positive and false positive classifications, thereby providing insight into the clinical relevance of each predictive model.

For the DCA, net benefit curves were plotted alongside baseline strategies of "treat all" and "treat none" to contextualize the performance of GluFormer and GMI. The net benefit was calculated using the formula:

Net Benefit = (True Positives / N) - (False Positives / N) × (Threshold Probability / (1 − Threshold Probability))
where N is the total number of participants.

### CGM-Derived Clinical Metrics Using iglu

To extract meaningful clinical insights from Continuous Glucose Monitoring (CGM) data, we utilized the R package iglu [37,38]) . Iglu is designed to derive a comprehensive set of metrics from CGM data, which are pivotal in assessing glucose control and variability across individuals.

### Functionality and Application of iglu

Iglu integrates advanced computational algorithms to process CGM data, providing more than just basic reading and organizing functionalities. Unlike other tools which may offer limited analysis capabilities, iglu supports a full spectrum of CGM-derived metrics. This tool facilitates an in-depth evaluation of glucose dynamics, enabling the categorization of glucose management into several clinically relevant aspects [43].

Mean-glucose measures: Metrics such as Mean Glucose Levels and estimated A1C (eA1C) are calculated to evaluate glycemic control.
Postprandial Glucose Adaptation: Measures like Standard Deviation (SD) and Mean Amplitude of Glycemic Excursions (MAGE) assess responses to meal intake.
Composite and Range Metrics: This includes Time in Range (TIR) and the J-Index, which provide insights into overall glucose exposure and variability.

All iglu measurements used in this study are detailed Table 2.

### Predictive Modeling of Medical Measures on HPP Using representations

To evaluate the predictive power of GluFormer representations for future health outcomes, we conducted a series of analyses targeting a broad range of clinical metrics. These metrics included blood tests, body measurements, anthropometric data, and body composition parameters such as muscle and fat mass. The primary goal was to assess the ability of GluFormer representations to forecast these health metrics over different time horizons, thereby validating their clinical applicability.

For each health metric, we trained a ridge regression model using either iglu measures and GluFormer output representations as input features. The representations, representing the CGM data for each participant, were used to predict various clinical metrics recorded for those individuals. Specifically, we used a KFold cross-validation approach, dividing the data into 5 folds and iteratively using each fold as the test set while aggregating results across all folds. We used this aggregated result to compute a Pearson correlation between observed and predicted values for all samples in the set. This process was repeated 10 times with different random seeds to ensure robustness, and the results were averaged, with standard deviations calculated. We evaluated the significance of our predictions using a permutation test across 100 different seeds, considering results as significant when random performance did not exceed the model performance more than 5 times (P-value < 0.05).

**Methodology for Out‑of‑Cohort Generalization**

Here, we specifically sought to determine how well GluFormer, trained **only** on the HPP training set, would generalize to these OOD datasets. First, each external dataset was fed into the pre‑trained GluFormer to obtain CGM representations. Next, using the same ridge regression approach described in the "Predictive Modeling of Medical Measures on HPP Using Representations" section, we evaluated how effectively these embeddings predict relevant clinical measures in the new populations. This procedure ensures consistency across all evaluations and provides a direct assessment of GluFormer's robust transferability to diverse real‑world settings.

**Out-of-Cohort Generalization**

To evaluate the generalizability of our model to out-of-distribution (OOD) data, we focused on assessing performance using all datasets that have CGM data, including the Human Phenotype Project (HPP), PNP1, PNP3, BREACPNT, PREDICT, T1DM, GDM, BREAD, IL Healthy, US Healthy, US-Obese, Colas 2019, JDRF CGM RCT, Shanghai T2DM, and Hypoglycemia in Older Adults. Using these datasets we wanted to evaluate the performance of GluFormer under different clinical conditions (T1D, T2D, GDM, etc..), different devices (iPro2, Dexcom, Medtronic etc..), and different geographies (Australia, US, Spain, China).

**Data Preparation and Embedding for Out-of-Cohort**

To be consistent with the HPP dataset preprocessing, glucose measurements from the external cohorts were discretized using the same bins that were established in the Tokenization section, based on the 80% training partition of the 10,812 participant dataset. If the data was from clinical trials, we further divided into two distinct phases: pre-intervention and post-intervention. We embedded the full sequence of glucose measurements directly into the 1024-dimensional space, applying the max-pooling to generate a single representative vector for each phase of the study.

**Predicting outcome of Randomized Clinical Trials**

For the Randomized Clinical Trials (RCTs) outcome prediction, we used the pre-intervention CGM GluFormer representations and added a binary variable to indicate the allocation of the participants. Each RCT had 2 intervention arms, so each representation vector was increased to be of size 1,025, where the last entry was either 0 or 1 depending on their allocation. We then used either iglu or the representations to predict the primary and secondary outcomes of the study, evaluating the results using Pearson correlation. The RCT cohorts in this category are PREDICT, BREACP, and PNP3.

For the open-source datasets, we focused on primary outcomes and, using the same scheme, attempted to predict the primary outcome for each dataset:

JDRF: We aimed to predict HbA1C% at 26 weeks using CGM data from 1-6 months, evaluating the results using Pearson correlation.

Shanghai T2D: We aimed to predict HbA1C% after 26 weeks using 2 weeks of baseline CGM data, evaluating the results using Pearson correlation.

Colas: We aimed to predict baseline HbA1C% using 2 weeks of CGM data, evaluating the results using Pearson correlation.

Hypoglycemia: We used 1 week of CGM data to predict whether there would be an event of hypoglycemia in the following weeks, evaluating the results using ROC AUC.

Aleppo 2017: We aimed to predict HbA1C% at baseline of study using 1 week of baseline CGM data, evaluating the results using Pearson correlation.

Chase 2005: We aimed to predict HbA1C% at baseline of study using 1 week of baseline CGM data, evaluating the results using Pearson correlation.

Wadwa 2023: We aimed to predict HbA1C% after 13 and 26 weeks using 1 week of baseline CGM data, evaluating the results using Pearson correlation.

### Predictive Modeling of Randomized Clinical Trials

The predictive analysis was again conducted using ridge regression, consistent with our previous methodological approach to ensure comparability. We used ridge regression with KFold optimized over 10 different seeds. The folds were randomly assigned based on participant ids. P-values were obtained over a permutation test over 100 different seeds.

### Comparative Analysis of SSL, Plain Transformer, and CNN Models

This section outlines the comparative performance of four distinct models—GluFormer, a Transformer with the same architecture as GluFormer, Frozen GluFormer representations, and Convolutional Neural Network (CNN) - in downstream task of prediction HbA1C% from CGM, as detailed in Supp Figure S2. Each model was trained using the same dataset, evaluated on the same validation and test sets to ensure comparability. The split was the same 80% train, 10% validation, 10% test, as was done for all other models.

### Fine Tuning the Pretrained CGM Transformer

To tailor the GluFormer architecture for the specific task of HbA1C% prediction, we modified its output mechanism. Initially trained to predict the next token, the model's output layer transformed representations into the token space. For fine-tuning, we replaced this output layer with a 1D adaptive pooling layer that aggregates the 1200 representations into a single vector. We then added a small, 3-layer MLP with GELU activation and a single neuron output to predict the measure from this vector. The fine-tuned model was optimized using Mean Squared Error (MSE) for regression tasks and employed the AdamW optimizer.

### Training CGM Transformer from Scratch (No SSL)

To evaluate the intrinsic benefit of SSL, a CGM Transformer model with the same architecture (including the transformer backbone, the 1d pooling, and the 3-layer MLP) was trained from scratch with randomly initialized weights. This approach aimed to isolate the effect of SSL from the architectural benefits. The model was optimized using MSE, with AdamW.

### Fine Tuning Frozen Transformer representations

In this variant, the Transformer's weights were frozen, and only the added 1D adaptive pooling layer and the 3-layer MLP were trained. This model setup tested the efficacy of the transformer representations when not allowed to adapt during the training of the downstream task. Optimization and hyperparameter tuning followed the same protocol as the other models, focusing on MSE and AdamW optimization.

**CNN Model**

A convolutional neural network (CNN) was also tested for comparison. The CNN model was extensively tuned across a range of hyperparameters, including different kernel sizes, numbers of layers, and MLP configurations. The best-performing setup in the validation set featured a five-layer CNN with a kernel size of 3, followed by a 3-layer MLP. This model was similarly optimized using MSE with AdamW.

**Adjustments for Long CGM Sequences**

Some external datasets featured very long CGM recordings that exceeded our system's memory constraints. Specifically: Wadwa: We split the CGM sequence into two intermediate chunks and embedded each chunk separately. We then computed the mean of these two embeddings to obtain one final representation per individual. Aleppo: The maximum sequence length was around 26k readings (after removing the final day and re‑sampling at 15‑minute intervals). To accommodate the model's capacity, we took the last 15k measurements of each sequence, embedded them, and used those embeddings for downstream analyses.

These adjustments ensure that our embedding process remains tractable for longer CGM data while preserving as much relevant information as possible, and adjusted for the different tasks for each dataset.

**Temporal Modeling**

To capture the temporal impacts of glucose fluctuations over time, we incorporated date & time information into our model. We explored two methods to directly integrate temporal information into the representations used by our transformer model.

<u>1. Temporal Positional Encoding</u>

Building on the positional encoding framework presented in "Attention is All You Need," we introduced additional sine and cosine functions to represent temporal dimensions. This method, we named "temporal positional encoding", extends traditional positional encoding by incorporating time-specific waveforms corresponding to minute, hour, day of the week, and month.

To tailor these encodings to the physiological context of glucose measurements, we adjusted the phase and wavelength of the sine and cosine functions to align with real-world time cycles.

2. Learned Temporal representations

Our second approach involved the use of learned representations for temporal values—minute, hour, day of the week, and month. These representations were added directly to the corresponding data representations within the transformer architecture. Unlike the fixed mathematical formulations of sinusoidal encodings, these learned representations were optimized during training, allowing the model to dynamically adjust and refine its understanding of time as it related to physiological changes and glucose measurements.

In practice, the learned temporal representations outperformed the sinusoidal temporal positional encoding is the generation task over the test set (see Fig.S5).

**GluFormer with Diet Tokens**

**Preprocessing of CGM and Diet Data**

To integrate continuous glucose monitoring (CGM) data with dietary logs, we established a preprocessing pipeline that combines these datasets, ensuring accurate alignment and comprehensive representation of nutritional intake alongside glucose measurements. Glucose measurements were recorded every 15 minutes, and dietary intake was aligned to the nearest 15 minute glucose measurement. Additionally, the diet was further broken into its macronutrient values.

Data Collection and Initial Processing: We collected CGM and dietary data from 10,844 participants. The CGM device recorded glucose levels subcutaneously at regular intervals, while participants logged dietary intake through a mobile application, detailing their consumption of calories, carbohydrates, proteins, lipids, and water.

Temporal Alignment and Data Cleaning: Dietary entries were aligned with CGM timestamps to ensure temporal correspondence between nutrient intake and glucose readings. We excluded any dietary data not temporally coinciding with CGM records. Additionally, glucose values were clipped to a physiological range of 40 to 500 mg/dL, and nutrient values exceeding the 99th percentile were trimmed to reduce the impact of outliers.

Caloric and Meal Filters: Entries from participants who logged less than 1,000 calories throughout the monitoring period were removed. Days with total caloric intake outside the range of 500 to 7,000 calories, or with fewer than three meal logs, were also excluded to maintain consistency in dietary patterns.

Data Imputation: In rare instances of device recording lapses, missing CGM measurements were linearly imputed to maintain a continuous glucose profile.

Meal Consolidation and Adjustment: Multiple dietary logs recorded within the same hour were consolidated into a single meal entry to simplify the dataset. Optionally, we adjusted meal timings to correlate better with observed glucose spikes, enhancing the model's ability to associate dietary intake with glycemic responses. This was done by observing that high sugar meals that did not have a subsequent spike in blood glucose, were probably mis-logged, and so we looked in a window of an hour before and after the meal to see it we can locate a glucose spike - if we found one, we moved the logging to be 15 minutes before the spike.

Data Binning and Tokenization: Nutrients were categorized into quantile-based bins, facilitating uniformity in representation. Similarly, minutes were binned into 15-minute intervals to synchronize with the frequency of CGM recordings. Glucose measurements and nutrient intakes were then discretized and tokenized into integers, providing a standardized scale for model input.

Statistics: The preprocessing steps reduced the initial dataset from 10,844 to 5,875 participants, with the number of analyzable days decreasing from 76,862 to 57,137. This refinement was crucial in ensuring that the dataset comprised only the most reliable and representative entries for model training.

Usage: The models that used this preprocessing were the models using temporal information, as well as the diet variant. Previous results, used the previous preprocessing that does not take into account temporal or nutrient based information.

**Tokenization and Temporal Encoding**

To effectively process and analyze the CGM and dietary data within our GluFormer + Diet model, a comprehensive tokenization strategy was adopted as detailed above. This approach not only standardizes the data but also integrates crucial temporal information, facilitating the model's ability to capture temporal dynamics associated with glucose fluctuations and dietary intakes.

One-Dimensional Sequence Formation: Both CGM and dietary data were tokenized into a unified one-dimensional sequence. This linear sequence format is crucial for the model, allowing it to process time-series data as a continuous stream of events, which is essential for capturing the dynamics of glucose responses and nutrient effects over time.

Incorporation of Temporal Tokens: Alongside each glucose and diet token, we included four additional temporal tokens representing the minute, hour, day of the week, and month. These tokens are vital for providing the model with context about the timing of glucose readings and dietary logs, enriching the input data with layers of temporal granularity. For diet-related tokens, time does not progress between entries on macronutrients of the same log.

Tokenization Details: All glucose measurements and nutrient values were discretized into predefined bins before tokenization, ensuring that each entry conforms to a standardized integer value. Glucose values were tokenized as before - into integer units, and diet nutrients were binned into bins based on quantiles.

The nutrients used for the diet are:

1. Calories (kcal)
2. Carbohydrates (grams)
3. Protein (grams)
4. Caffeine (mg)
5. Water (grams)
6. Total Lipid (grams)
7. Alcohol (grams)
8. Sugars Total (grams)

The bin values:

energy_kcal: [0.02, 31.68, 84.52, 146.04, 195.00, 262.12, 340.58, 423.10, 518.50, 623.53, 745.56, 891.08, 1069.25, 1317.02, 1726.57, 5832.24]

carbohydrate_g: [0.00, 3.47, 9.05, 15.00, 21.60, 28.87, 36.13, 43.08, 51.81, 61.97, 73.86, 88.02, 106.90, 132.38, 177.13, 608.09]

protein_g: [0.00, 0.66, 1.63, 2.81, 4.42, 6.80, 9.61, 13.64, 18.23, 23.76, 30.46, 38.61, 49.34, 65.22, 93.21, 354.53]

caffeine_mg: [0.30, 188.52, 377.04, 754.08, 2639.28]

water_g: [0.00, 4.56, 36.15, 78.24, 129.46, 182.26, 247.63, 311.44, 389.16, 468.20, 498.60, 575.58, 687.39, 855.37, 1109.35, 4507.10]

totallipid_g: [0.00, 0.41, 0.89, 2.59, 5.68, 8.74, 11.97, 15.84, 19.97, 24.95, 30.71, 37.91, 47.38, 60.38, 83.04, 306.73]

alcohol_g: [0.02, 0.88, 2.60, 6.60, 9.90, 13.20, 18.00, 20.00, 26.40, 28.80, 33.00, 49.50, 66.00, 172.84]
sugarstotal_g: [0.00, 1.25, 2.82, 4.67, 6.88, 9.27, 12.12, 14.70, 18.02, 22.04, 26.04, 31.28, 37.90, 48.22, 65.61, 266.78]

This method simplifies the model's processing by reducing the complexity of input data and ensuring consistency across the dataset.

Parallel Temporal Information: To ensure that the model recognizes the sequence of events accurately, temporal tokens accompany each glucose or diet token without advancing time unnecessarily during diet entries. This approach keeps the temporal data consistent and precise, reflecting the actual timing of meals and glucose measurements without artificial shifts.

**Transformer Configuration**

The GluFormer + Diet model employs a transformer architecture that has been specifically tailored to handle both CGM and dietary data effectively. This setup enables the model to learn from the complex relationships between nutrient intake and glucose levels while incorporating temporal dynamics. Here are the key components of the model's architecture:

Embedding Size: Each input token is represented in a 1024-dimensional space.

Heads and Layers: The model features 8 attention heads and 10 transformer layers.

Feedforward Dimension: A dimension of 2048 for the feedforward networks in each transformer block.

**Additional Training Parameters**

The model is trained on sequences that include both glucose and diet tokens. However, the diet tokens serve exclusively as contextual information. The predictive task is focused on forecasting glucose tokens only. This selective prediction approach helps the model specialize in glucose dynamics while still understanding the impact of dietary inputs as contextual modifiers.

**Positional and Modality Encodings**

Positional Encoding: Traditional positional encodings are utilized to preserve the order of tokens within the sequence, which is crucial for the model to understand sequence-dependent changes in glucose levels.

Modality Tokens: Each token type (glucose, calories, sugars, carbohydrates, proteins, lipids, alcohol, water) is associated with a learned modality embedding. These representations are added to the input representations and provide the model with information about the nature of each token, enhancing its ability to process mixed data types effectively. This is similar to segment representations in BERT [60], where they add learned representations to indicate sentence structure.

Temporal Positional Encoding: To integrate temporal information more directly, the model includes an additional positional encoding for time:

Non-Learned Temporal Encoding: The last eight dimensions of the embedding space are reserved for sinusoidal positional encodings that represent minute, hour, week, and month. Each unit of time is encoded using both sine and cosine functions, resulting in eight dimensions that help the model grasp the cyclic nature of time.

This encoding ensures that at any point in the sequence, the model has immediate access to precise temporal information, enhancing its ability to make time-sensitive predictions.

Temporal Learned Tokens

Alongside the static temporal encodings, the model also incorporates learned temporal tokens for minute, hour, day, and month. These tokens are pre-processed and added to each token's embedding before entering the transformer blocks. This setup allows the model to adjust its response based on learned patterns associated with specific times, providing a dynamic and responsive predictive mechanism.

**Statistical Analysis of Generation**

To evaluate the quality of generated samples from GluFormer, we used an autoregressive strategy to produce full days of CGM data on unseen test participants and then compared these generated time series to the true CGM data collected for those same days.

Glucose-Only GluFormer Variant

For the glucose-only version of GluFormer (i.e., the model without diet tokens), we provided one full day of an individual's CGM readings (4X24 = 96 measurements, sampled every 15 minutes) as "context." The model then inferred the subsequent day's glucose values in an autoregressive manner, predicting each next glucose token based only on previously generated or observed tokens (while masking future tokens).

To assess generation quality we computed a variety of standard CGM metrics (e.g., mean glucose, %time in range, coefficient of variation) using the iglu package in R on both the generated and actual CGM days. We then compared these metrics by calculating Pearson correlations between the observed and generated composite scores.

We also computed the Pearson correlation and MAE between the observed and generated glucose time series (i.e., data points at each 15-minute interval).
We report two-tailed p-values for each Pearson correlation using a t-test to determine statistical significance.

Multimodal GluFormer Variant (with Diet Tokens)

For the multimodal GluFormer model, which integrates both glucose and diet tokens, we provided one full day of CGM plus corresponding diet tokens as context. The model then autoregressively generated the subsequent day's CGM values. Whenever the participant ate, we inserted the relevant diet tokens (i.e., macronutrient bins) into the generated sequence at the appropriate time.

After generation, we removed the diet tokens in the output to focus on the generated glucose sequence. The same set of evaluation metrics was employed: Pearson correlation and MAE between the observed and generated glucose time series.

Context-Length Experiment

To investigate how the number of CGM days provided as context affects the quality of generated data, we conducted an experiment (Figures S3 & S4) where we varied the amount of CGM context (ranging from a single token up to multiple days). For each context length, the model generated a new day of CGM data, and we again calculated Pearson correlations between observed and generated CGM scores or the raw time series. We observed that more context about an individual's prior CGM data improves the generation fidelity (i.e., yields higher correlations and lower error between observed vs. generated curves).
All p-values for correlations were determined using a two-tailed t-test, and the MAE was similarly evaluated as an additional measure of discrepancy between observed and generated glucose signals.

**Optimization and Training Details**

For optimizing the GluFormer + Diet model, we utilized a carefully configured setup designed to promote stable learning while effectively minimizing the loss over time. Here's a detailed overview of the optimization parameters and training conditions:

Optimizer: AdamW, a variant of the Adam optimizer that incorporates weight decay, was selected for its robustness and effectiveness in managing sparse gradients and complex dependencies in high-dimensional data.

Learning Rate (lr): Set at 0.0001, this relatively low learning rate helps prevent the model from overshooting minima, allowing for finer adjustments in weight updates.

Beta Coefficients (b1 and b2): Values of 0.9 and 0.99 for the first and second moment estimates, respectively, balance the trade-off between momentum and stability.

Weight Decay: A value of 0.01 is used to regularize and prevent overfitting, particularly useful in complex models with a high capacity like this transformer.

Learning Rate Scheduler: Employing a gamma of 0.0001, the learning rate decays by this factor, gradually reducing the step size to fine-tune the model's weights as training progresses.

**Training Dynamics**

Epochs: The model is trained for 20 epochs, providing sufficient time for the complex patterns in the dataset to be learned without leading to excessive training times.

Dropout: Set at 0.2, dropout is used as a regularization method to prevent overfitting by randomly dropping units (along with their connections) during the training process.

Batch Size per GPU: 16 samples per batch,

**Hardware and Computational Details**

GPU: The training utilizes a single NVIDIA A100, known for its powerful computational capabilities and efficiency in handling large datasets and complex models.

Floating Point Precision: Training is conducted using 16-bit floating point precision (FP16), which reduces memory consumption and can speed up training without significantly impacting the accuracy of the results.

Gradient Clipping: A maximum gradient norm (max_grad_norm) of 1 is used to prevent the exploding gradient problem, which can lead to destabilized training dynamics.

Random Seed: A consistent numpy, torch, and random math seed of 42 is set to ensure reproducibility of the results, enabling consistent initialization and stochastic processes across different runs.

Warm-Up Phase: The model employs 100 warmup steps at the start of training. This gradual ramp-up in learning rate helps stabilize the model's parameters before entering the full training regime.

**Comparative Analysis Setup: GluFormer With and Without Diet**

To rigorously evaluate the impact of integrating dietary data into the GluFormer model, we conducted a comparative analysis between two versions of the model: one incorporating diet data (GluFormer + Diet) and the other excluding it (GluFormer). This comparison aims to assess the added value of dietary information in enhancing the model's predictive accuracy and understanding of glucose dynamics.

**Data Consistency Across Models**

Unified Preprocessing Pipeline: Both versions of the GluFormer model were trained using the same preprocessing pipeline detailed previously for the GluFormer and Diet model. This ensures that any observed differences in model performance can be directly attributed to the inclusion of dietary data, rather than discrepancies in data handling or preprocessing.

Identical Datasets: To ensure a fair comparison, the training, validation, and test sets were identical across both model variants. This approach eliminates variability in data distribution as a confounding factor, allowing for a clear assessment of the impact of diet data integration.

Hyper parameter search for these models was in this values:

Beta Coefficients (b1 and b2):

b1: {0.85, 0.88, 0.9}

b2: {0.9, 0.95, 0.99}

Diet Modeling (diet_modeling):

Values: {true, false} — Determines whether dietary data is incorporated into the model.

Dimension of Feedforward Network (dim_feedforward): {512, 2048}

Dropout Rate (dropout): {0, 0.1, 0.2} —

Number of Training Epochs (epochs):{10, 15, 20, 30}

Learning Rate Decay: {0.1, 0.01, 0.001} — Specifies the ration between starting lr (after warm up), and ending lr

Learning Rate (lr): {1e-06, 1e-05, 0.0001, 0.0005, 0.001}

Maximum Gradient Norm):{1, 10, 100}

Embedding Size (n_embd): {256, 512, 1024} — Size of each token embedding.

Number of Attention Heads (n_heads): {8, 16}

Number of Transformer Layers (n_layers): {8, 10, 12, 16}

Warmup Steps: {10, 100, 1000} — Number of steps to increase the learning rate at the beginning of training.

Weight Decay: {0, 0.001, 0.01, 0.1}

**Datasets used for study**

| Cohort Name | n | population | Country of origin | CGM | CGM Device |
|---|---|---|---|---|---|
| **HPP** [27] | 10,812 | Healthy | Israel | 2 weeks | FSLP, FSL IQ |
| **PNP1**[61], (NCT) | 926 | Healthy | Israel | 1 week | FSL |
| **PNP3**[62] (NCT) | 225 | Pre-diabetes | Israel | 26 weeks | FSL |
| **BREAD**[63] (NCT) | 20 | Healthy | Israel | 2 weeks | iPro2 |
| **BREACP**[64] (NCT) | 200 | Breast cancer survivor | Israel | 2 weeks | FSL |
| **PREDICT**[65] (NCT) | 264 | T2DM and prediabetes | Australia | 2 weeks | FSL |
| **GDM** | 549 | GDM | Israel | 2 weeks | FSL ,iPro |
| **IL Healthy** [66] | 1159 | Healthy | Israel | 2 weeks | iPro2; Medtronic |
| **T2D IL**[67], (NCT) | 23 | T2DM | Israel | 2 weeks | FSL |
| **T1DM**[68,69] (NCT) | 121 | T1DM | Israel | 2 weeks | Dexcom, Freestyle Navigator, FSL |
| **US Healthy** [67] | 327 | Healthy | US | 1 week | FSL |
| **US Obese**[70] (NCT) | 156 | Obese or T2DM or pre-diabetes | US | 2 weeks | FSL |
| **JDRF CGM RCT** [71] **(NCT)** | 451 | T1DM | US | 26 weeks | Freestyle Navigator, Dexcom SEVEN, Medtronic Minimed Paradigm |
| **Hypoglycemia** [70] | 200 | T1DM | US | 2 weeks | Dexcom SEVEN plus |
| **Colas 2019** [72] | 208 | Hypertension | Spain | 2 weeks | Minimed iPro |
| **Shanghai** [73] | 107 | T2DM | China | 2 weeks | Freestyle Libre H |
| **Wadwa 2023** [74] | 102 | T1DM | US | 13 weeks | Dexcom G6 |
| **Aleppo 2017** [75] | 226 | T1DM | US | 26 weeks | Dexcom G4 |
| **Chase 2005** [76] | 200 | T1DM | US | 26 weeks | GlucoWatch G2 Biographer |
| **AEGIS**[41] | 580 | Healthy& T2DM | Spain | 12 days | FreeStyle Libre Pro |

Table 1. Table of all cohorts used in this study. HPP is the main cohort, which we used for training the models. All others were used as validation for clinical measures prediction, and RCT results prediction.

| # | Parameter | Description |
|---|---|---|
| 1 | **ADRR** | Average Daily Risk Range; assesses daily glucose risk by evaluating the frequency and extent of hypo- and hyperglycemic episodes. |
| 2 | **COGI** | Continuous Overall Glycemic Index; evaluates overall glycemic control by considering both the mean glucose level and its variability. |
| 3 | **CV** | Coefficient of Variation; measures relative variability in glucose readings, calculated as the ratio of the standard deviation to the mean glucose level. |
| 4 | **eA1C** | Estimated A1C; provides an estimate of the average blood glucose level over the past 2-3 months, expressed as a percentage. |
| 5 | **GMI** | Glucose Management Indicator; estimates the laboratory A1C level based on average glucose measured by CGM. |
| 6 | **GRADE** | Glycemic Risk Assessment Diabetes Equation; quantifies the risk associated with hypo- and hyperglycemia by assigning weighted scores to glucose readings. |
| 7 | **GRADE_eugly** | GRADE for Euglycemia; evaluates the proportion of time glucose levels are within the normal range. |
| 8 | **GRADE_hyper** | GRADE for Hyperglycemia; assesses the severity and duration of high glucose episodes. |
| 9 | **GRADE_hypo** | GRADE for Hypoglycemia; assesses the severity and duration of low glucose episodes. |
| 10 | **HBGI** | High Blood Glucose Index; quantifies the risk of hyperglycemia by emphasizing higher glucose values. |
| 11 | **LBGI** | Low Blood Glucose Index; quantifies the risk of hypoglycemia by emphasizing lower glucose values. |
| 12 | **hyper_index** | Hyperglycemia Index; measures the extent and frequency of glucose readings above a specified upper limit. |
| 13 | **hypo_index** | Hypoglycemia Index; measures the extent and frequency of glucose readings below a specified lower limit. |
| 14 | **IGC** | Index of Glycemic Control; combines hyperglycemia and hypoglycemia indices to provide an overall assessment of glycemic control. |
| 15 | **IQR** | Interquartile Range; represents the middle 50% of glucose readings, indicating variability. |
| 16 | **J_index** | J-Index; combines mean glucose and standard deviation to assess overall glycemic control. |
| 17 | **M_value** | M-Value; evaluates glycemic control by considering deviations from an ideal glucose value, emphasizing both hypo- and hyperglycemia. |

| 18 | **MAD** | Mean Absolute Deviation; measures the average absolute difference between each glucose reading and the mean glucose level. |
|---|---|---|
| 19 | **MAGE** | Mean Amplitude of Glycemic Excursions; calculates the average of significant glucose swings, indicating glucose variability. |
| 20 | **above_140** | Percentage of readings above 140 mg/dL; indicates the proportion of time spent in hyperglycemia. |
| 21 | **above_180** | Percentage of readings above 180 mg/dL; indicates the proportion of time spent in significant hyperglycemia. |
| 22 | **above_250** | Percentage of readings above 250 mg/dL; indicates the proportion of time spent in severe hyperglycemia. |
| 23 | **below_54** | Percentage of readings below 54 mg/dL; indicates the proportion of time spent in severe hypoglycemia. |
| 24 | **below_70** | Percentage of readings below 70 mg/dL; indicates the proportion of time spent in hypoglycemia. |
| 25 | **in_range_63_140** | Percentage of readings between 63 and 140 mg/dL; indicates the proportion of time spent in the target glucose range. |
| 26 | **in_range_70_180** | Percentage of readings between 70 and 180 mg/dL; indicates the proportion of time spent in the standard target glucose range. |
| 27 | **range** | Difference between maximum and minimum glucose readings; indicates the overall spread of glucose values. |
| 28 | **SD** | Standard Deviation; measures the dispersion of glucose readings around the mean, indicating variability. |
| 29 | **Min.** | Minimum glucose reading recorded. |
| 30 | **1st Qu.** | First Quartile (25th percentile) of glucose readings. |
| 31 | **Median** | Median (50th percentile) of glucose readings. |
| 32 | **Mean** | Average glucose reading. |
| 33 | **3rd Qu.** | Third Quartile (75th percentile) of glucose readings. |
| 34 | **Max.** | Maximum glucose reading recorded. |
| 35 | **Conga** | Continuous Overall Net Glycemic Action; measures short-term glycemic variability over a specified period. |
| 36 | **GVP** | Glucose Variability Percentage; quantifies the percentage of glucose readings that deviate from the mean, indicating variability. |
| 37 | **MODD** | Mean of Daily Differences; assesses day-to-day variability in glucose readings by averaging absolute differences between consecutive days. |
| 38 | **SD.Roc** | Standard Deviation of Rate of Change; measures variability in the rate at which glucose levels change over time. |
| 39 | **CV_Measures_Mean** | Mean of Coefficient of Variation Measures; average of CV values calculated over specified intervals. |

| 40 | **CV_Measures_SD** | Standard Deviation of Coefficient of Variation Measures; variability of CV values calculated over specified intervals. |
|---|---|---|
| 41 | **AUC** | Area Under the Curve; quantifies the total exposure to glucose levels over a specified period. |
| 42 | **MAG** | Mean Absolute Glucose; average of absolute glucose values over a specified period. |
| 43 | **SDw** | Within-day Standard Deviation; measures variability in glucose readings within a single day. |
| 44 | **SDhhmm** | Standard Deviation at specific hours and minutes; assesses variability at particular times of day. |

Table 2. Table of all CGM derived composite scores [37,38].

## Declarations

### Acknowledgments

We thank all Segal lab members, Pheno.AI data science members, and NVIDIA Tel Aviv Research group members. Special Thank you to Dr. Elad Barkan and Dr. Assaf Shocher for useful discussions.

### Ethical approval

All participants signed an informed consent form upon arrival to the research site. All identifying details of the participants were removed prior to the computational analysis. The 10K cohort study is conducted according to the principles of the Declaration of Helsinki and was approved by the Institutional Review Board (IRB) of the Weizmann Institute of Science.

**Data availability**

The data used in this paper is part of the Human Phenotype Project (HPP) and is accessible to researchers from universities and other research institutions at: https://humanphenotypeproject.org/data-access. Interested bona fide researchers should contact info@pheno.ai to obtain instructions for accessing the data.

**Code availability**

Implementation of GluFormer is available at: https://github.com/Guylu/GluFormer

**Competing Interests**

G.S. and H.R. are employees in Pheno.AI Ltd. a biomedical data science company from Tel-Aviv, Israel. E.S. is a paid consultant of Pheno.AI Ltd. G.L's work was done during an internship at NVIDIA Research. Other authors declare no competing interests.

**Author contribution**

G.L conceived the project, designed and conducted all analyses, interpreted the results, and wrote the manuscript. G.S developed protocols, interpreted the results, and wrote the manuscript. A.G designed pipelines and created preprocessing scripts. S.S interpreted the results and wrote the manuscript. J.R.G & D.S.B acquired the PREDICT cohort. F.G wrote the manuscript J.M interpreted the results, and wrote the manuscript. S.M guided computational analyses. E.M guided computational analyses and managed code running infrastructure. G.C directed the project. H.R conceived and directed the project and analyses, designed the analyses, interpreted the results and wrote the manuscript. E.S. conceived and supervised the project and analyses, designed the analyses, interpreted the results and wrote the manuscript.

## Supplementary Appendix

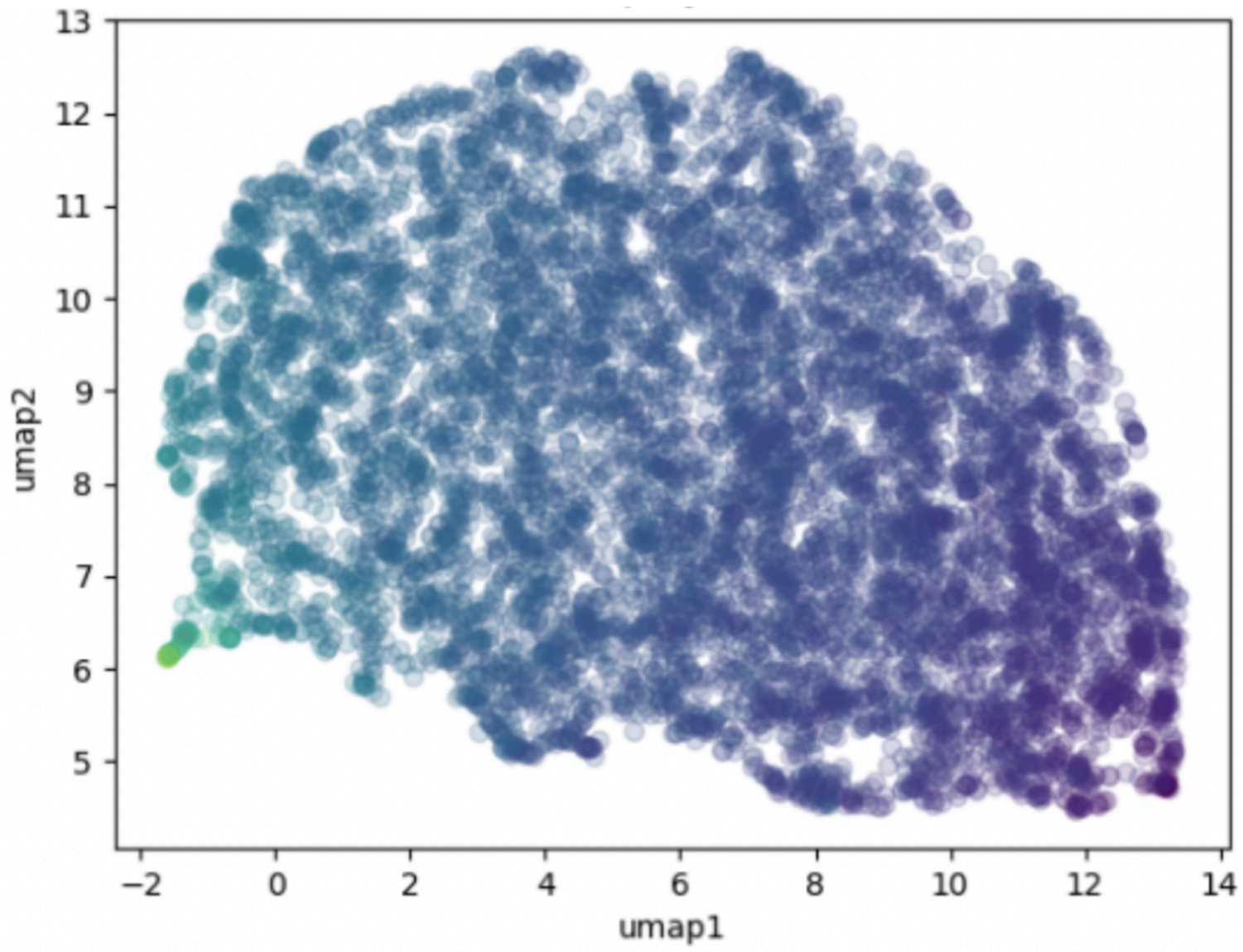

**Figure S1.A: UMAP of GluFormer representations of HPP cohort, colored by their independently taken blood HbA1C% results.**

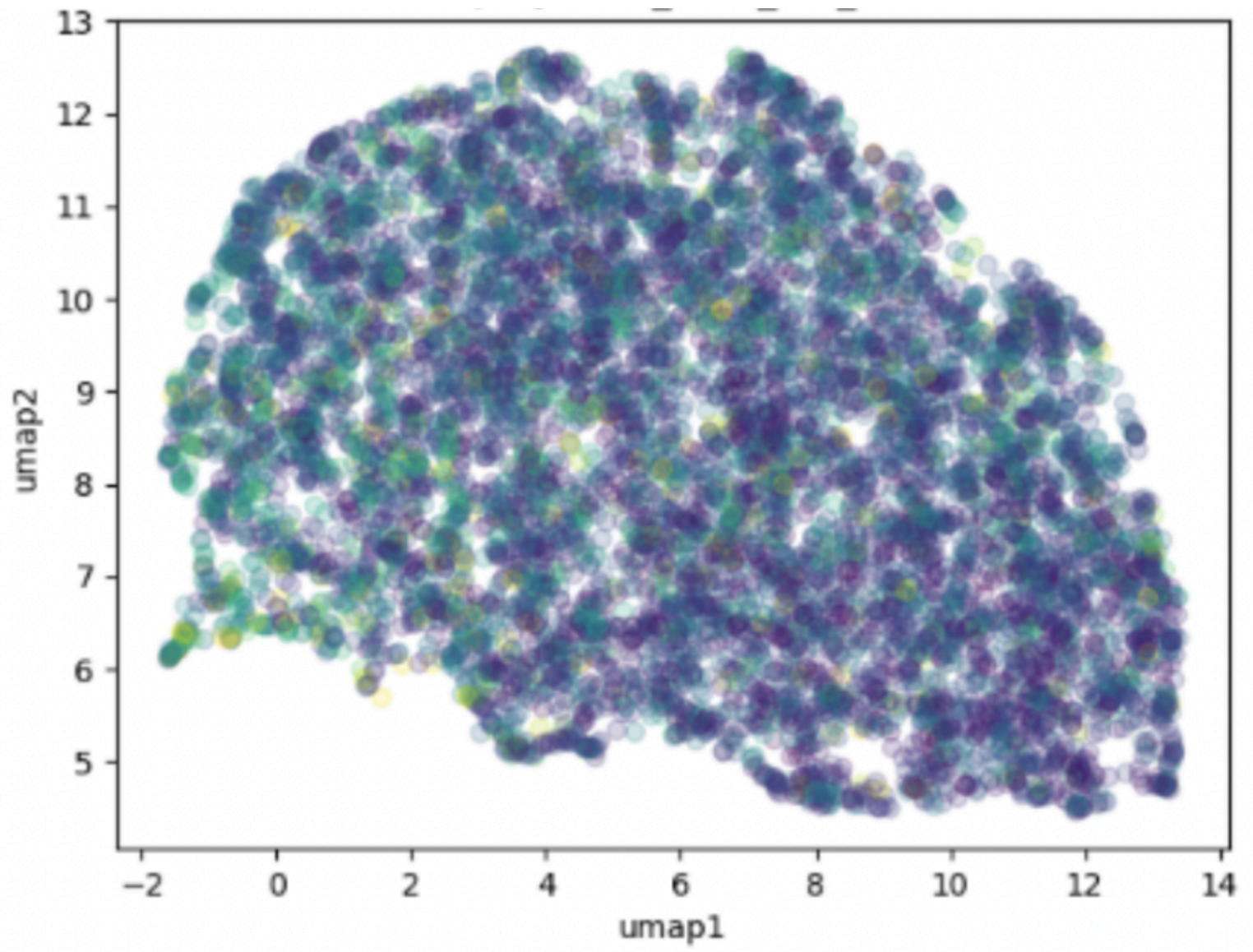

**Figure S1.B: UMAP of GluFormer representations of HPP cohort, colored by their independently taken DXA visceral fat mass results.**

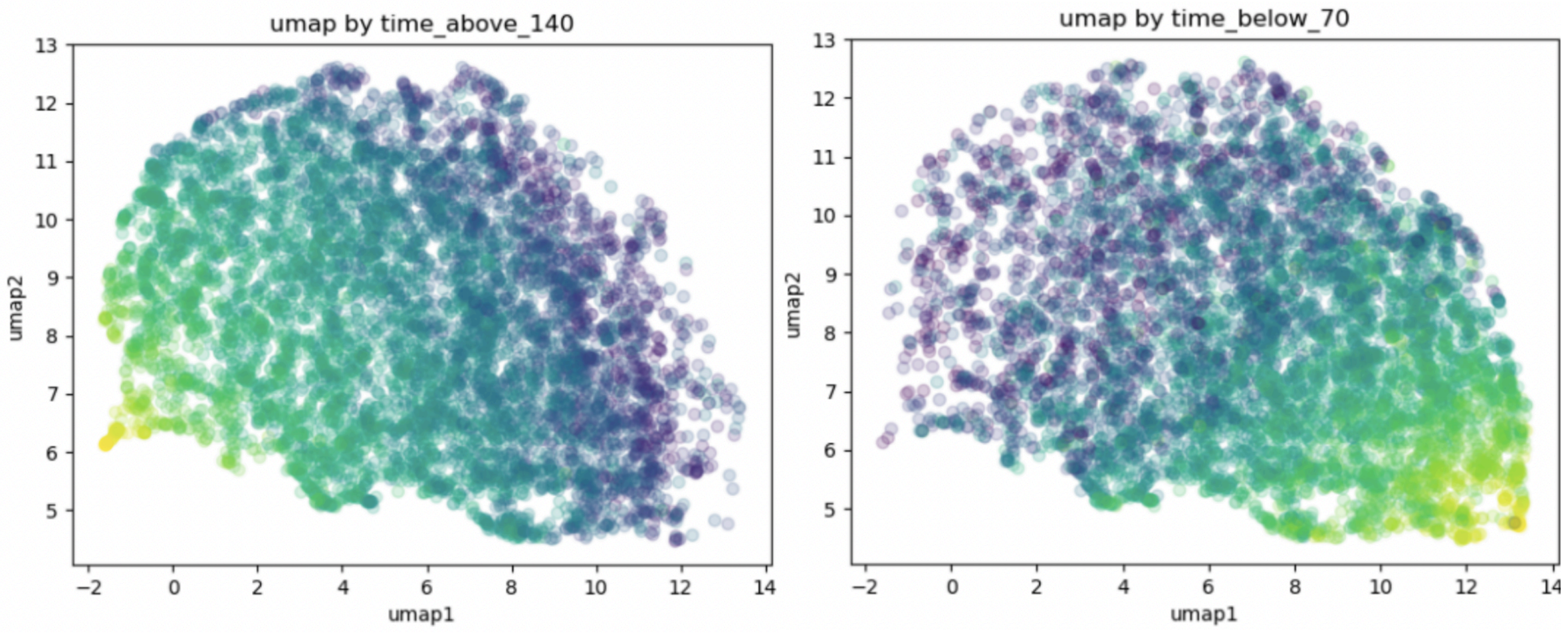

**Figure S1.C: UMAP of GluFormer representations of HPP cohort, colored by the percentage of time their CGM values were above 140 (left), and below 70 (right)**

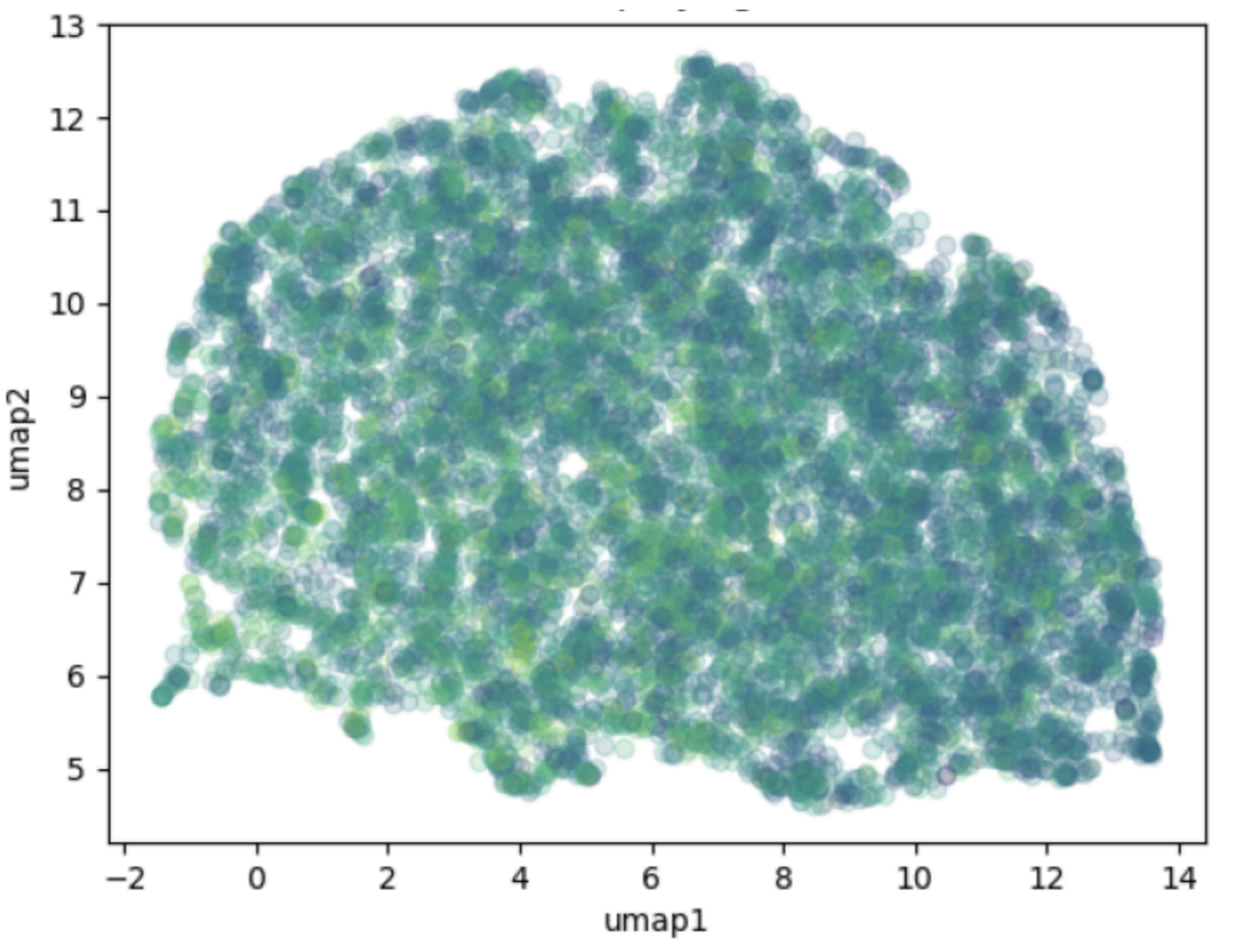

**Figure S1.D: UMAP of GluFormer representations of HPP cohort, colored by participants' age at CGM recording time.**

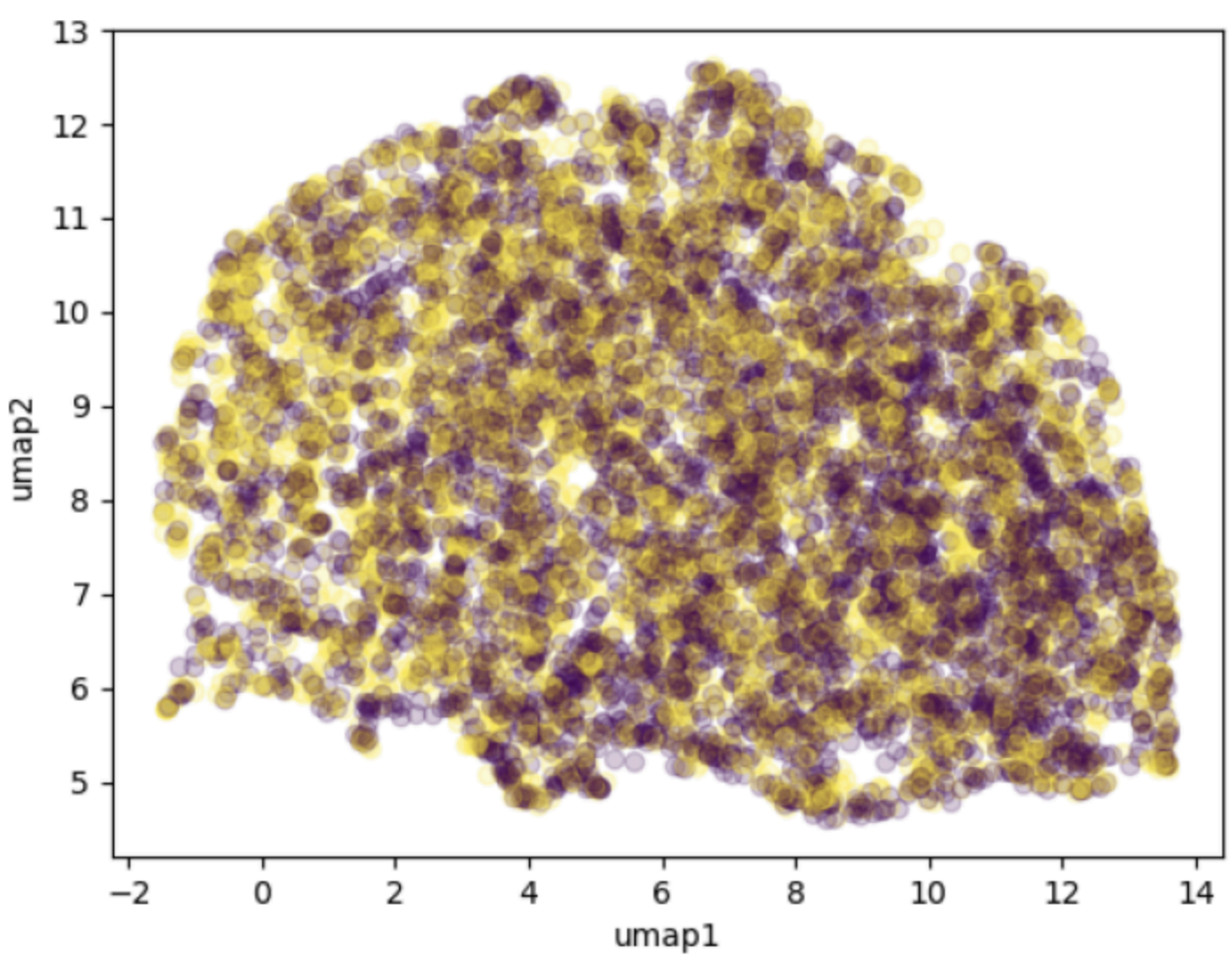

**Figure S1.E: UMAP of GluFormer representations of HPP cohort, colored by participants' biological sex.**

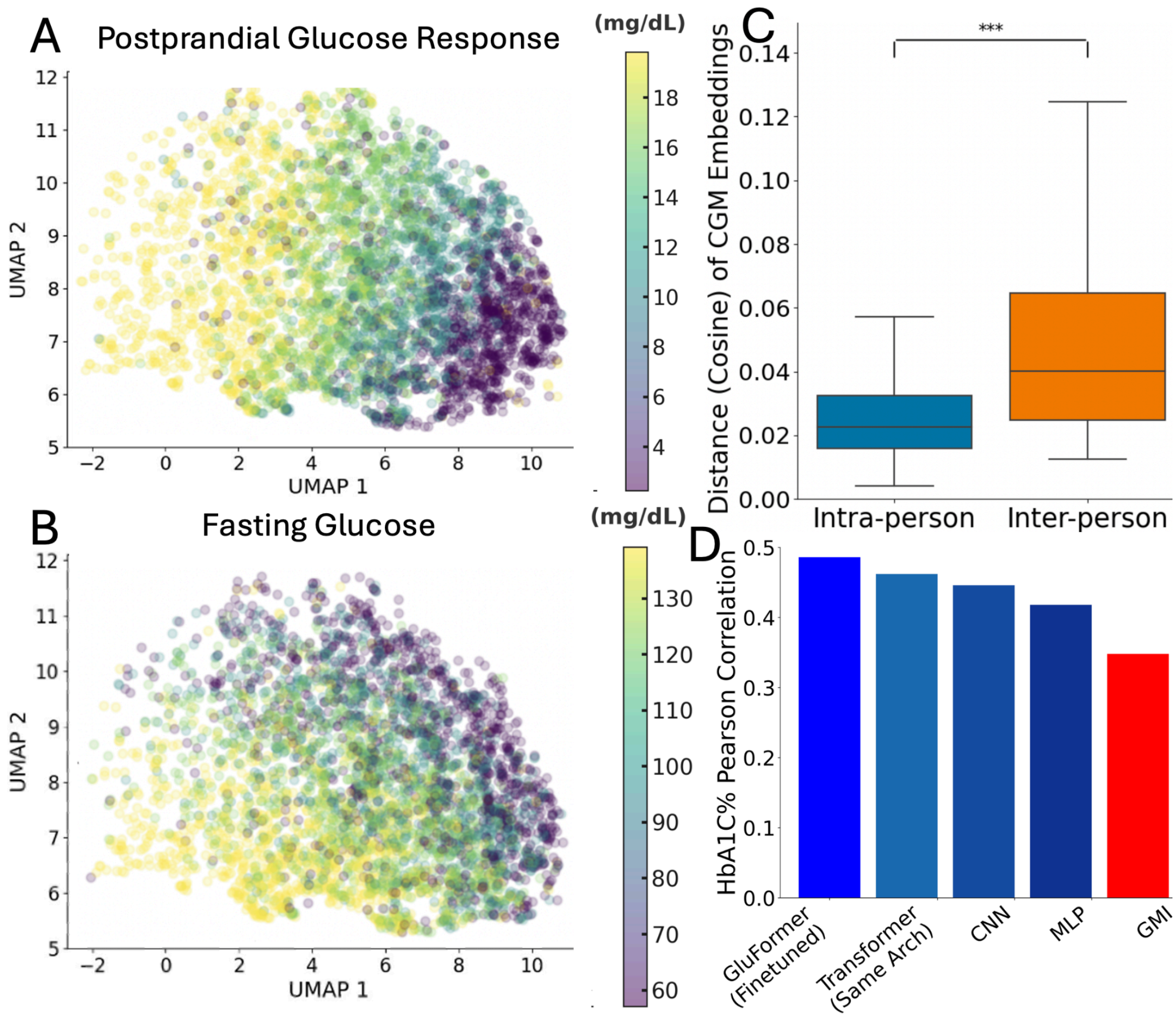

**Figure S2: Evaluation of GluFormer's Representations.**

**A. & B.** UMAP visualization of CGM representation from our model. Here we show how the representations relate to 2 clinical measurements not seen by the model during training. **A.** Plot shows a UMAP colored by Postprandial Glucose Response (PPGR), showing that UMAP dimension 1 captures the diversity in PPGR. Low PPGR values appear on the right, progressing to high PPGR on the left. **B.** Plot, colored by fasting glucose levels obtained from blood tests, shows that UMAP dimension 2 captures the range of fasting glucose levels, with lower levels on the right and higher levels on the left. These visualizations provide insights into how different clinical measures, crucial in endocrinology, could be associated with the learned CGM representations. **C.** A comparison of intra-participant and inter-participant cosine distances of CGM representations of the HPP. The "Intra Distances" (blue box plot) shows the distribution of cosine distances between representations of the same participant across different days (with no overlap), reflecting day-to-day variability in the individual's data. The "Inter

Distances" (orange box plot) shows the distribution of distances between representations from different participants, showing variation across individuals. There is an overlap in inter, intra-participant embedding distances meaning that there are some instances of participants who are very similar and some who are highly variable. The significant difference between the two, indicated by three asterisks, was tested using the Mann-Whitney test. **D.** A plot of the effectiveness of different models in predicting HbA1C% from CGM data, quantified by Pearson correlation coefficients. Blue hues indicate a deep learning model, red hues indicate extracted features followed by Ridge regression.

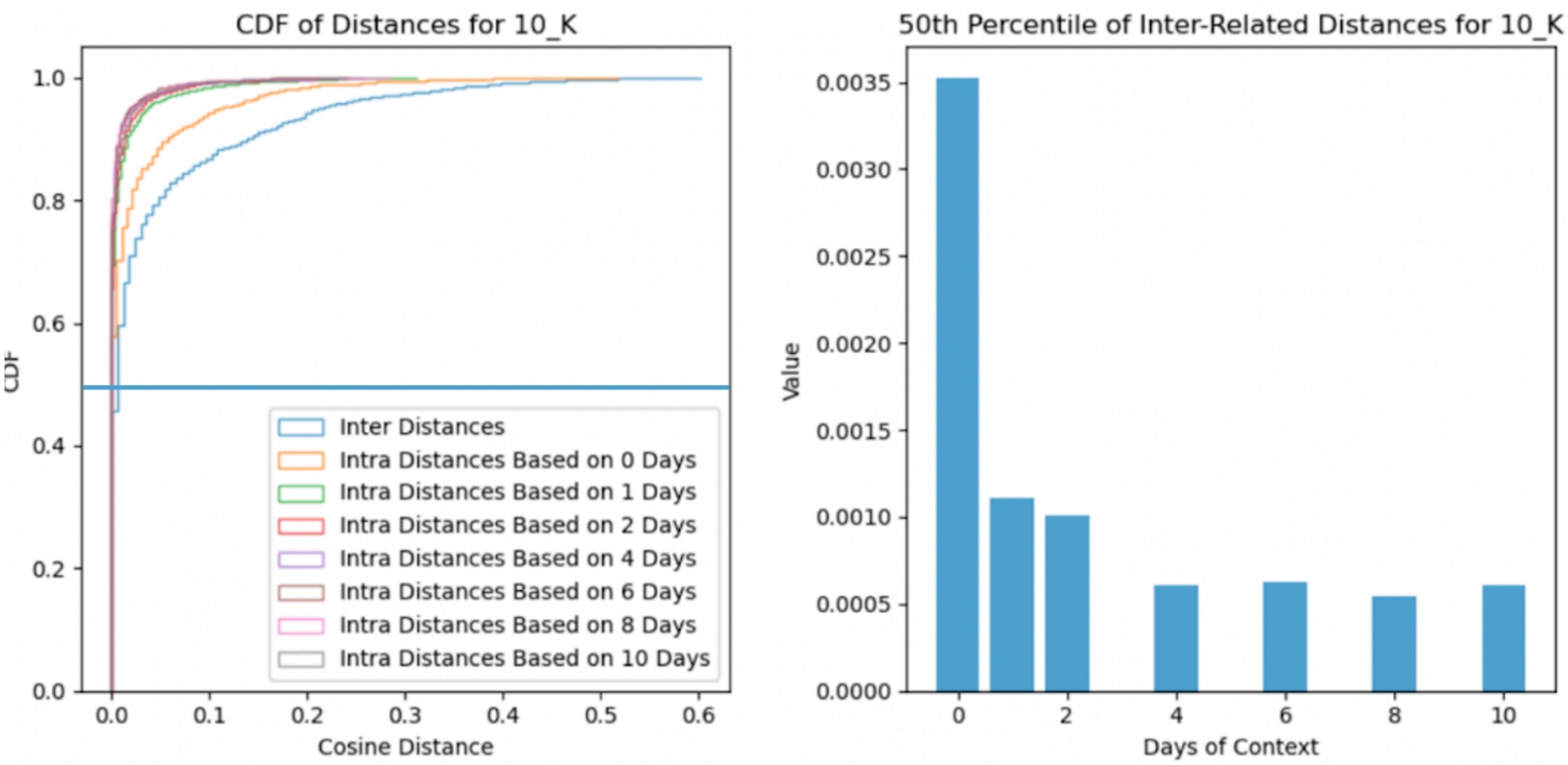

**Figure S3: CDF of distances of CGM-derived composite scores between observed HPP test set samples and generated samples given differing context length (left). Value of CDF at 50th percentile for differing context length. Distances become shorter as we add more context, suggesting GluFormer is able to make use of more context from participants to generate better CGMs.**

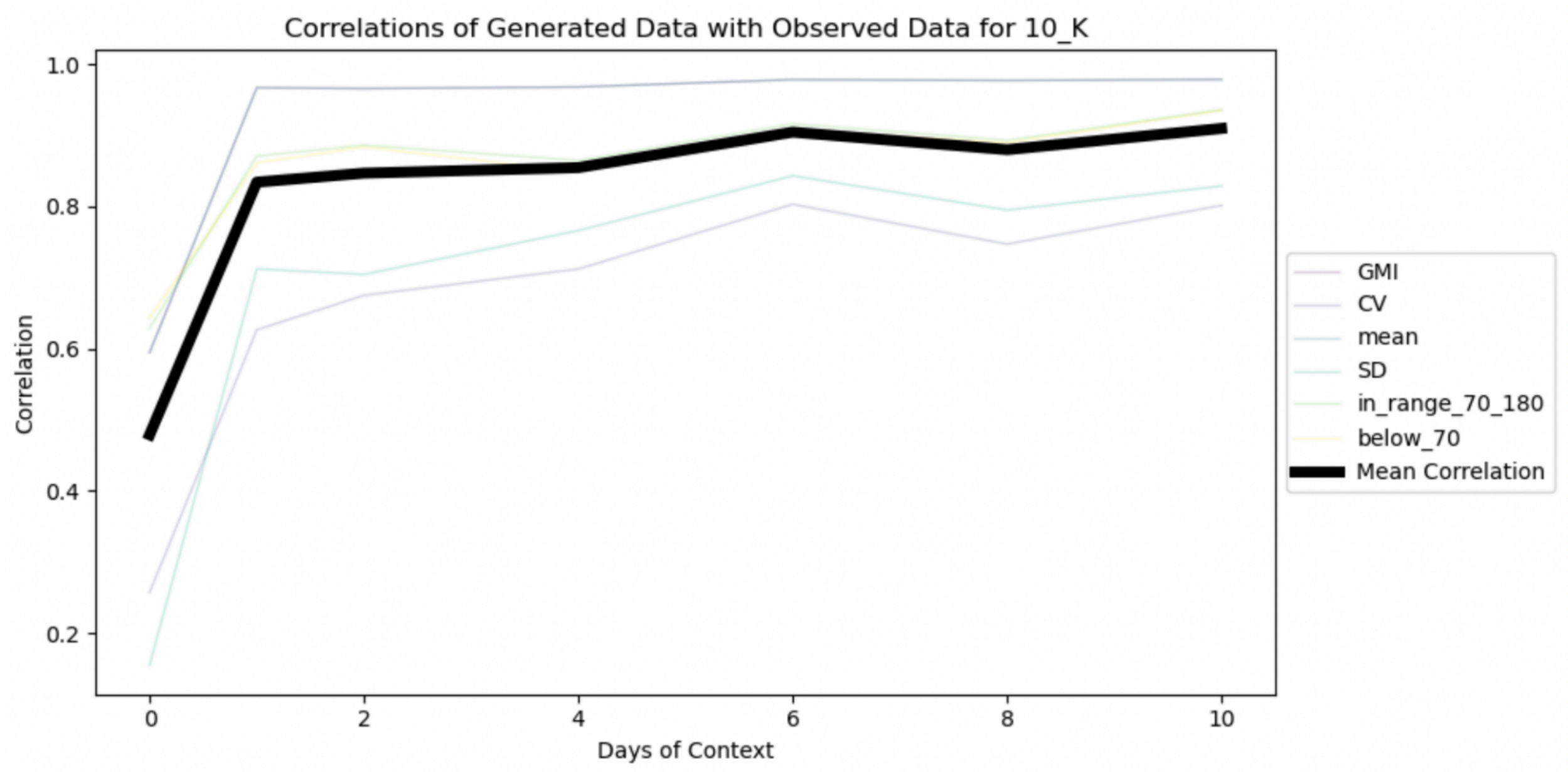

**Figure S4: Pearson correlations of CGM-derived composite scores between observed HPP test set samples and generated samples given differing context length. Correlations become higher for all CGM-derived composite scores as we add more context, suggesting GluFormer is able to make use of more context from participants to generate better CGMs.**

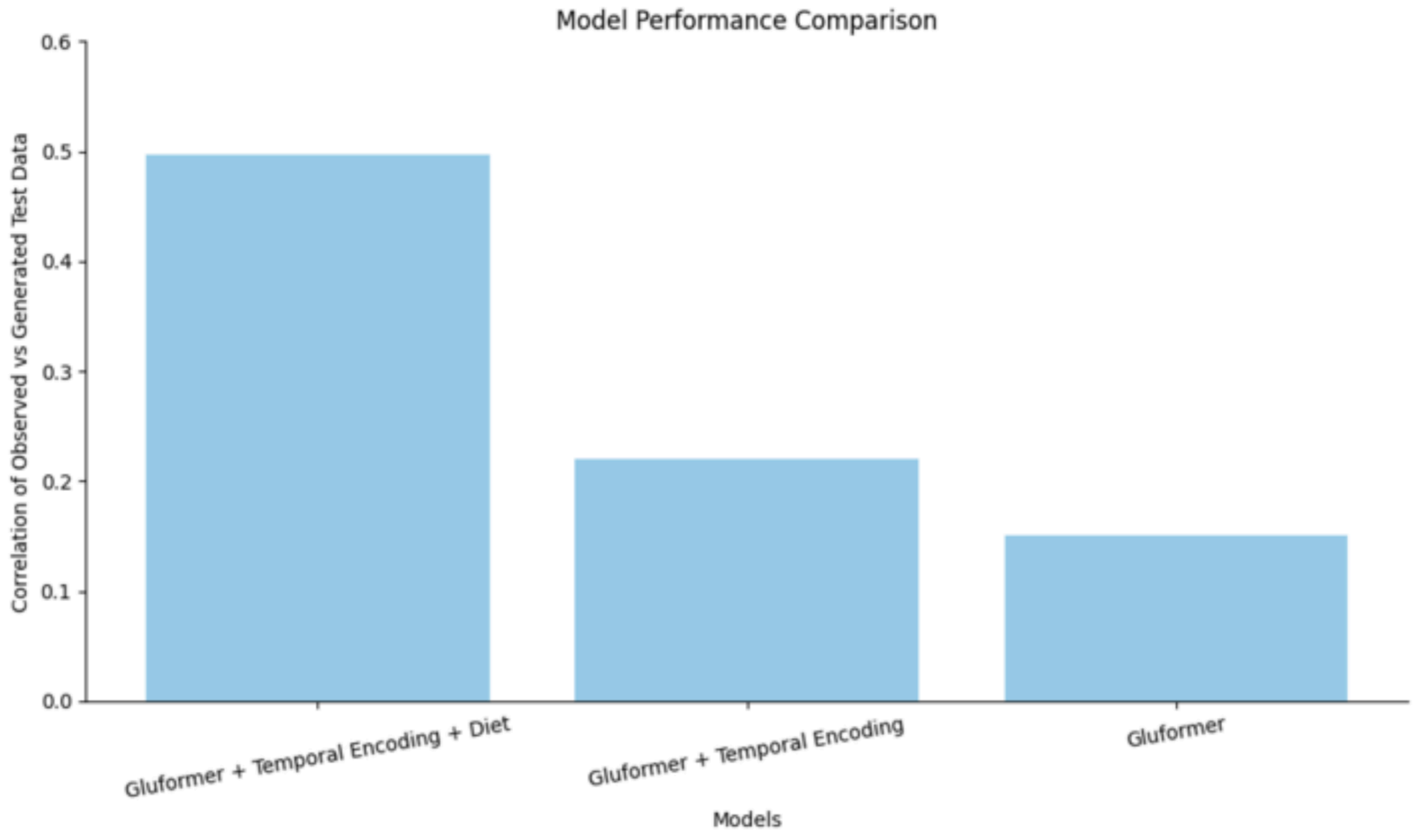

**Figure S5: Pearson correlations of generated CGM vs observed CGM on HPP test set. We can see that adding temporal date and time information increases performance by XXX, and adding diet tokens increases performance even more to 0.5.**

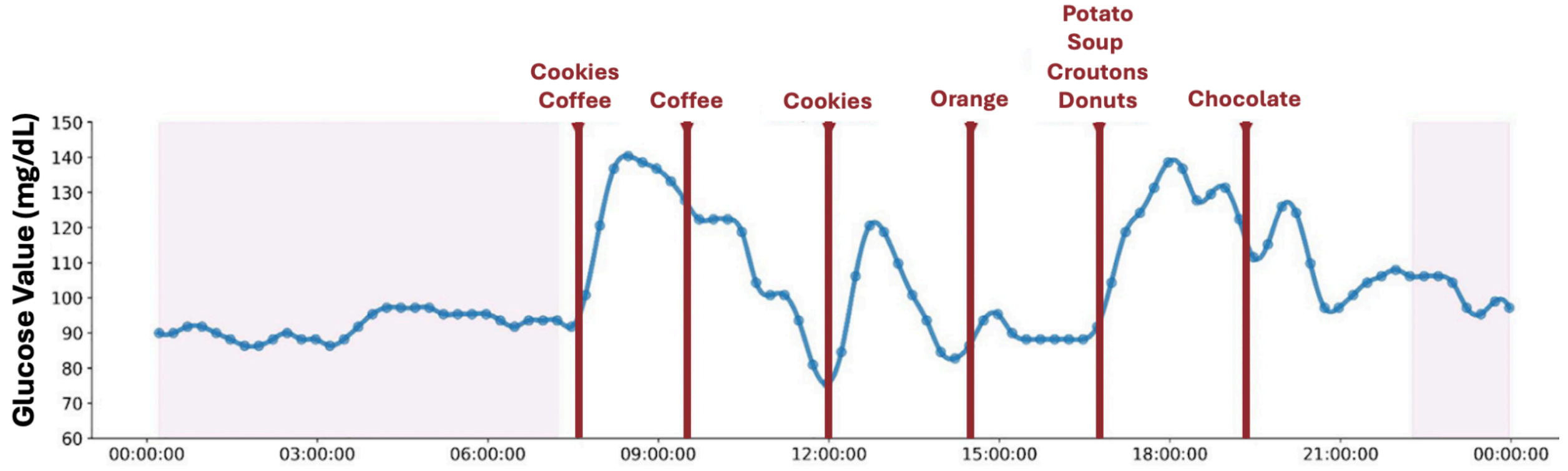

**Figure S6: Example of a typical CGM recording with diet information.**

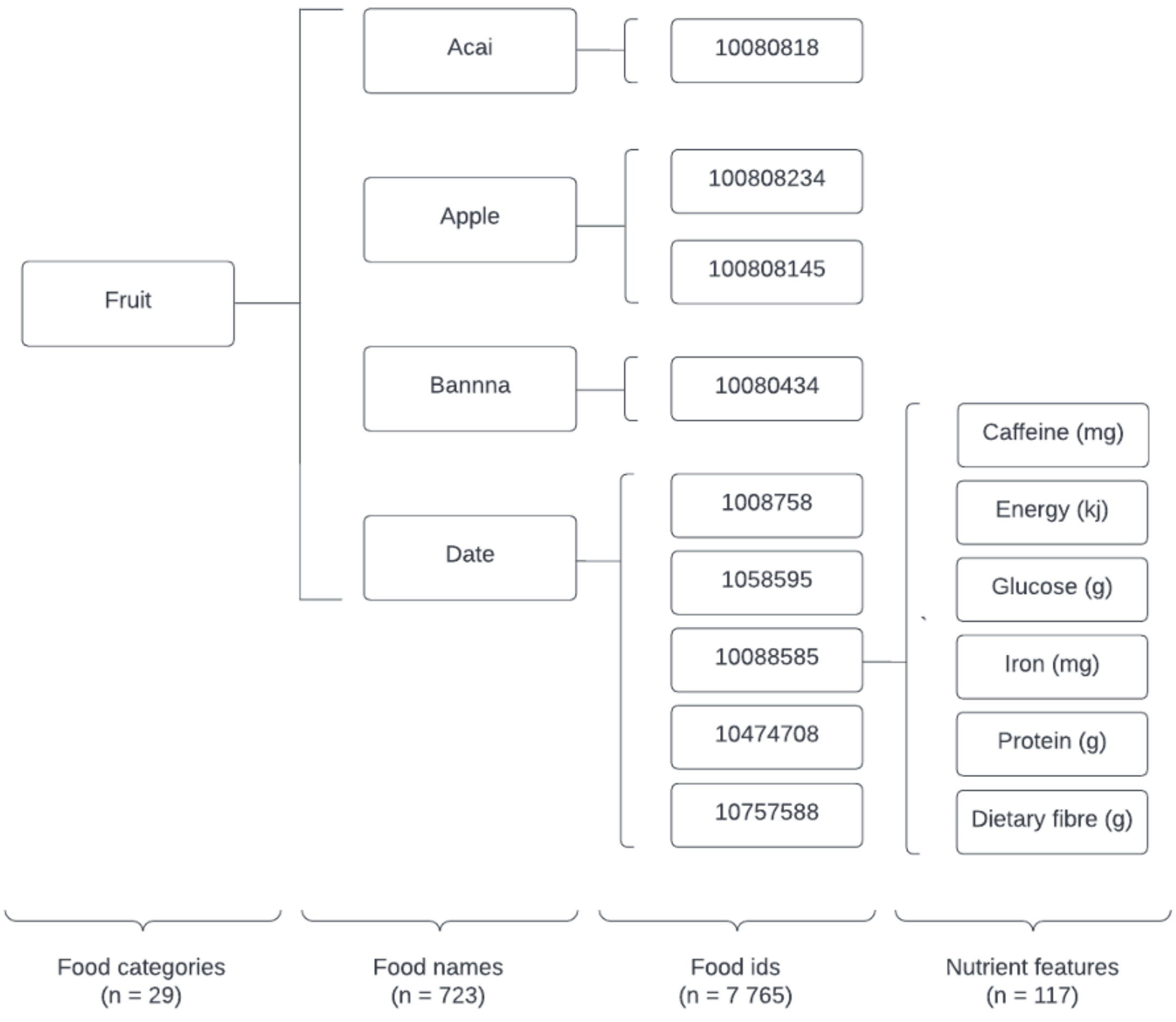

**Figure S7: Example of the hierarchy of food items from food categories, to names, and macro-nutrient features.**

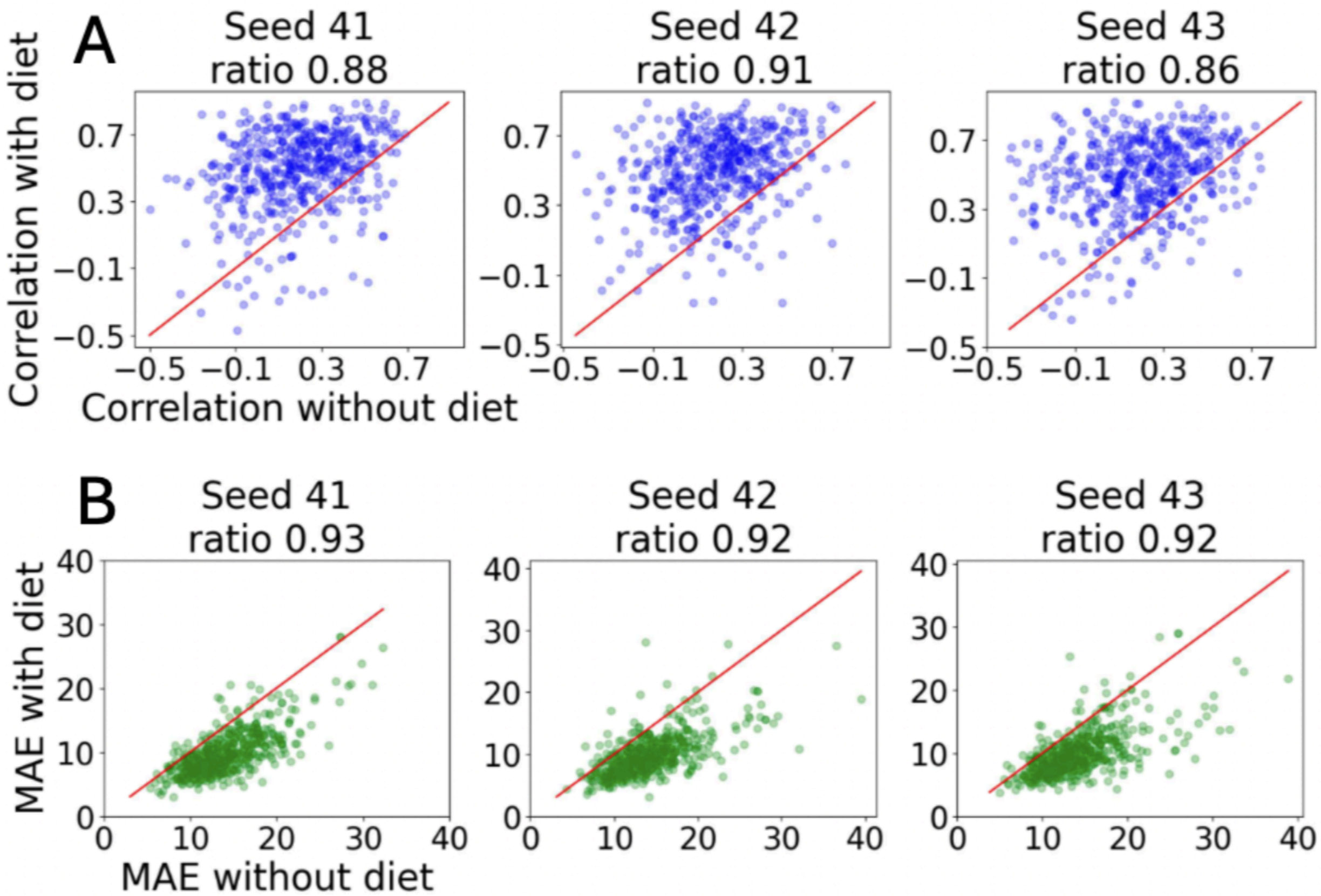

**Figure S8: Showing results of generation of CGM with and without diet for different random seeds, showing the generation is robust to seed configuration.**

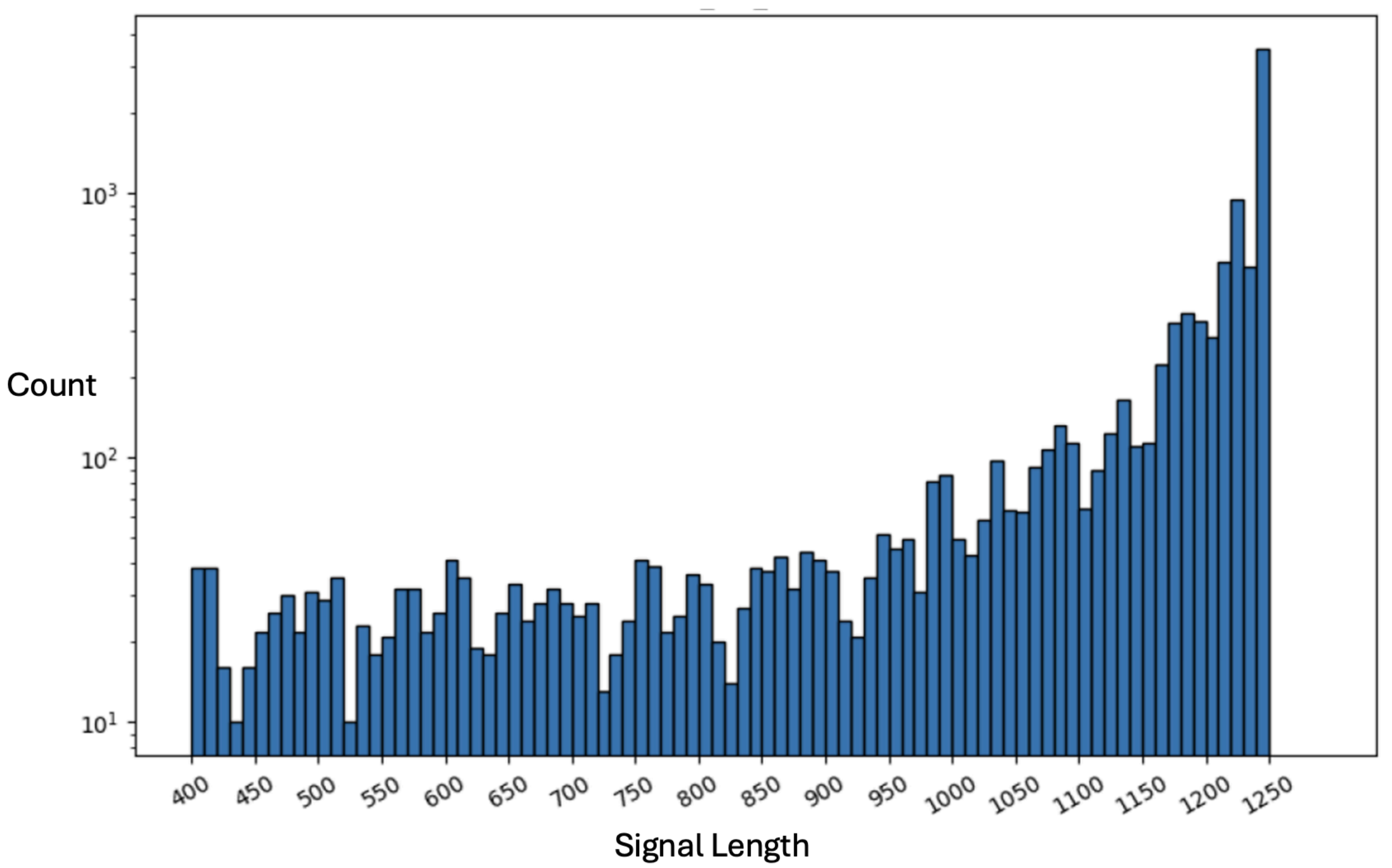

**Figure S9: Histogram of lengths of CGM recordings on HPP. This shows the reasoning behind choosing the cap to be at 1200 measures per sample.**

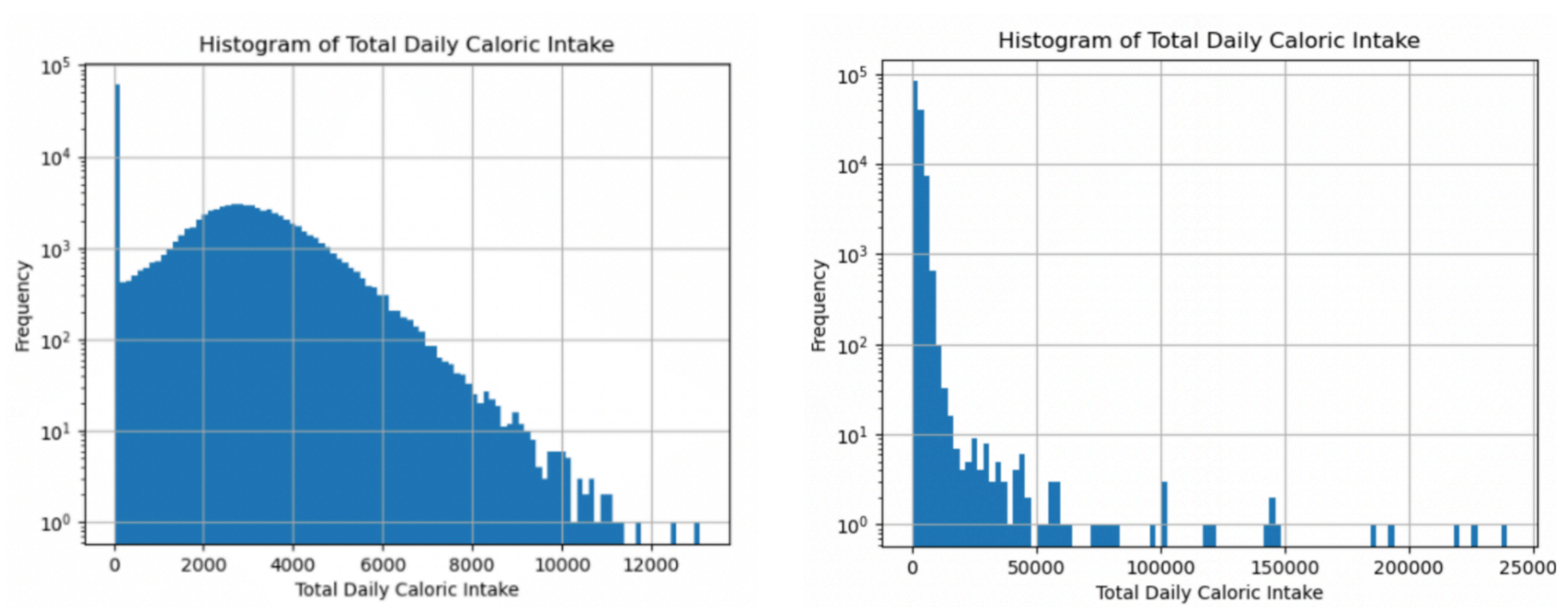

**Figure S10: Histogram of daily caloric intake before preprocessing (left), and after preprocessing (right).**

| Cohort | Below 70 | Mean | In Range 70-180 | Standard Deviation (SD) | Glucose Management Indicator (GMI) | Coefficient of Variation (CV) |
|---|---|---|---|---|---|---|
| HPP | 0.93 | 0.98 | 0.94 | 0.83 | 0.98 | 0.8 |
| T2D | 0.88 | 0.85 | 0.91 | 0.89 | 0.85 | 0.88 |
| PNP3 | 0.91 | 0.91 | 0.86 | 0.45 | 0.91 | 0.38 |
| T1DM | 0.57 | 0.85 | 0.67 | 0.5 | 0.85 | 0.57 |
| IL Healthy | 0.76 | 0.95 | 0.74 | 0.65 | 0.95 | 0.5 |
| PNP1 | 0.57 | 0.91 | 0.63 | 0.61 | 0.91 | 0.47 |
| Bread | 0.46 | 0.94 | 0.46 | 0.38 | 0.94 | 0.32 |
| GDM | 0.93 | 0.97 | 0.93 | 0.52 | 0.97 | 0.69 |
| BREACP | 0.73 | 0.99 | 0.72 | 0.91 | 0.99 | 0.78 |
| PREDICT | 0.73 | 0.96 | 0.8 | 0.64 | 0.96 | 0.42 |
| US Obese | 0.99 | 0.93 | 0.96 | 0.52 | 0.93 | 0.38 |
| US Healthy | 0.3 | 0.82 | 0.3 | 0.48 | 0.82 | 0.39 |

Table 3 - Results for figure 3C: This table compiles the correlations between real and generated CGM-based metrics across diverse cohorts not used for model training. These cohorts span healthy populations, individuals with type 1 and type 2 diabetes, gestational diabetes (GDM), and other metabolic conditions from multiple geographic regions. Correlations above 0.8 ($p < 0.001$) are observed in several subgroups, demonstrating GluFormer's robustness in generating clinically meaningful glucose profiles—even when applied to populations that differ substantially from the original training set.

| Measure | Male (45.7%) | Female (54.3%) | All |
|---|---|---|---|
| TIR | 94.64 (10.39) | 93.97 (10.31) | 94.28 (10.35) |
| GMI | 4.95 (0.41) | 4.87 (0.37) | 4.91 (0.39) |
| MAGE | 24.01 (6.5) | 24.03 (6.42) | 24.02 (6.45) |
| MODD | 13.0 (3.62) | 12.46 (3.18) | 12.7 (3.4) |
| CV | 15.54 (3.71) | 15.74 (3.5) | 15.65 (3.63) |
| HBGI | 0.15 (0.31) | 0.11 (0.22) | 0.13 (0.27) |
| LBGI | 2.34 (1.99) | 2.62 (1.87) | 2.49 (1.93) |

**Table 4. Statistics of key CGM-based composite scores.** Data from CGMap [43].: Characterizing continuous glucose monitor data in thousands of non-diabetic individuals - Ayya Keshet et.al.

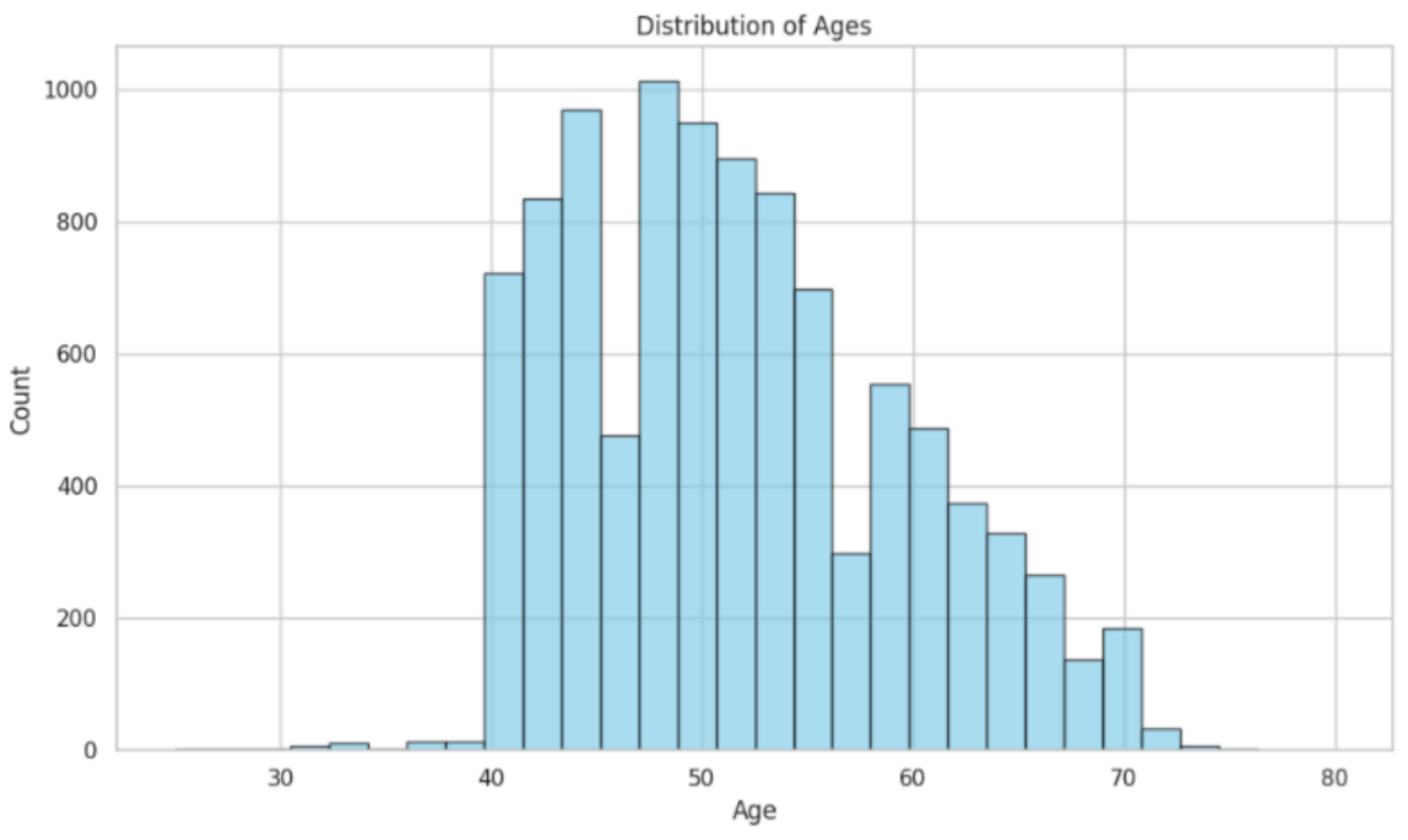

**Figure S11: HPP Distribution of Age**

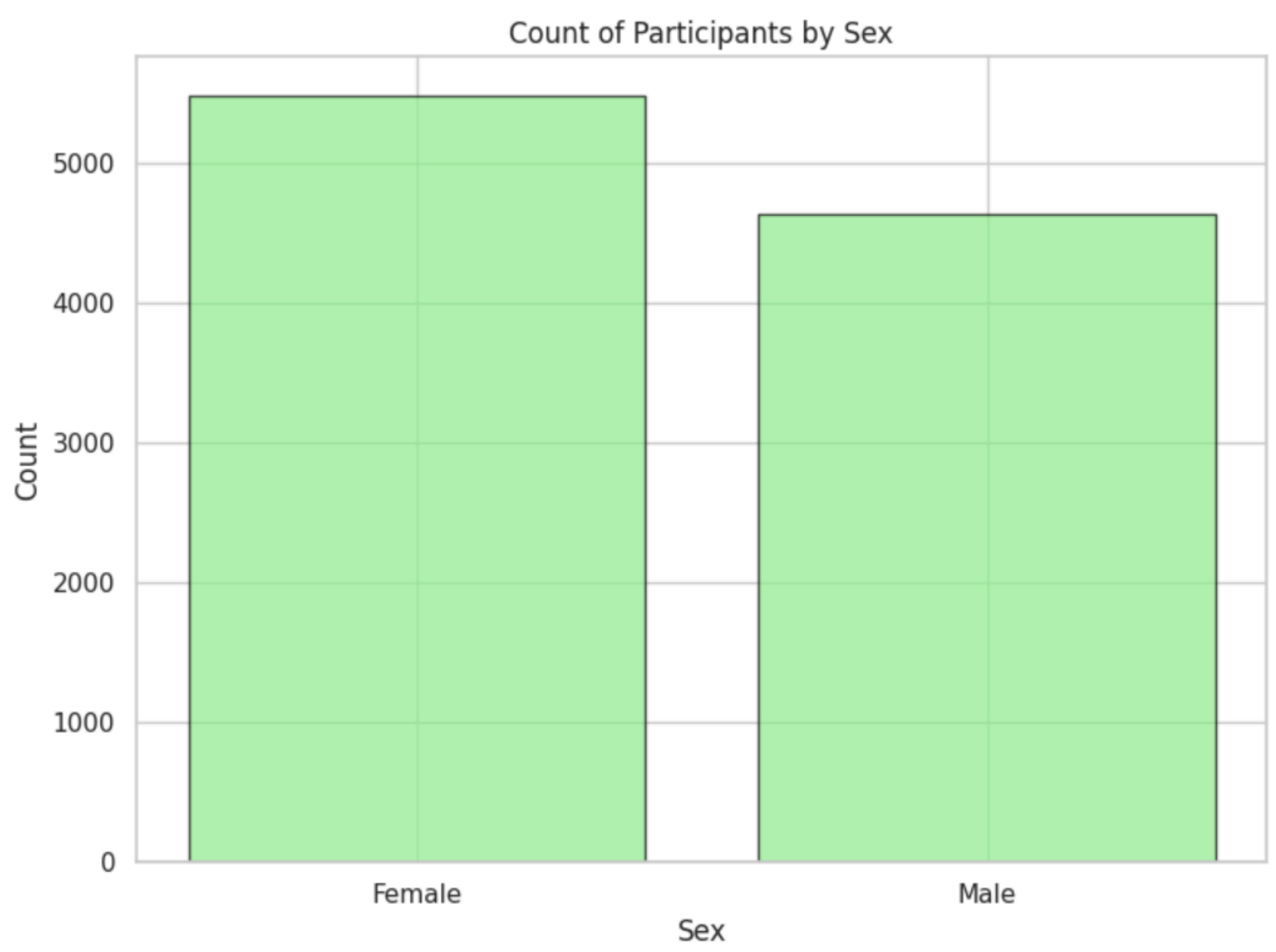

**Figure S12: HPP Sex Distribution**

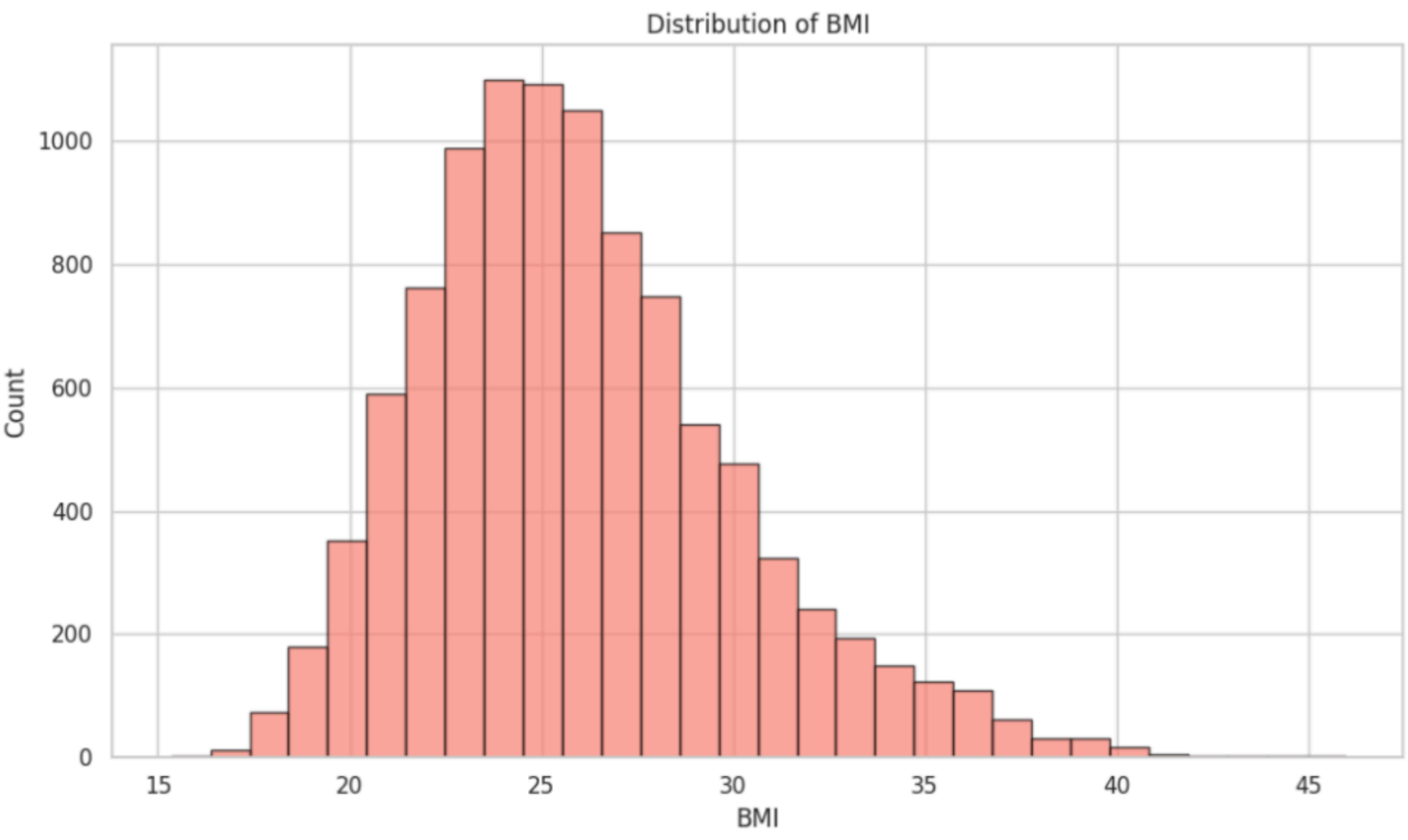

**Figure S13: HPP Distribution of BMI**

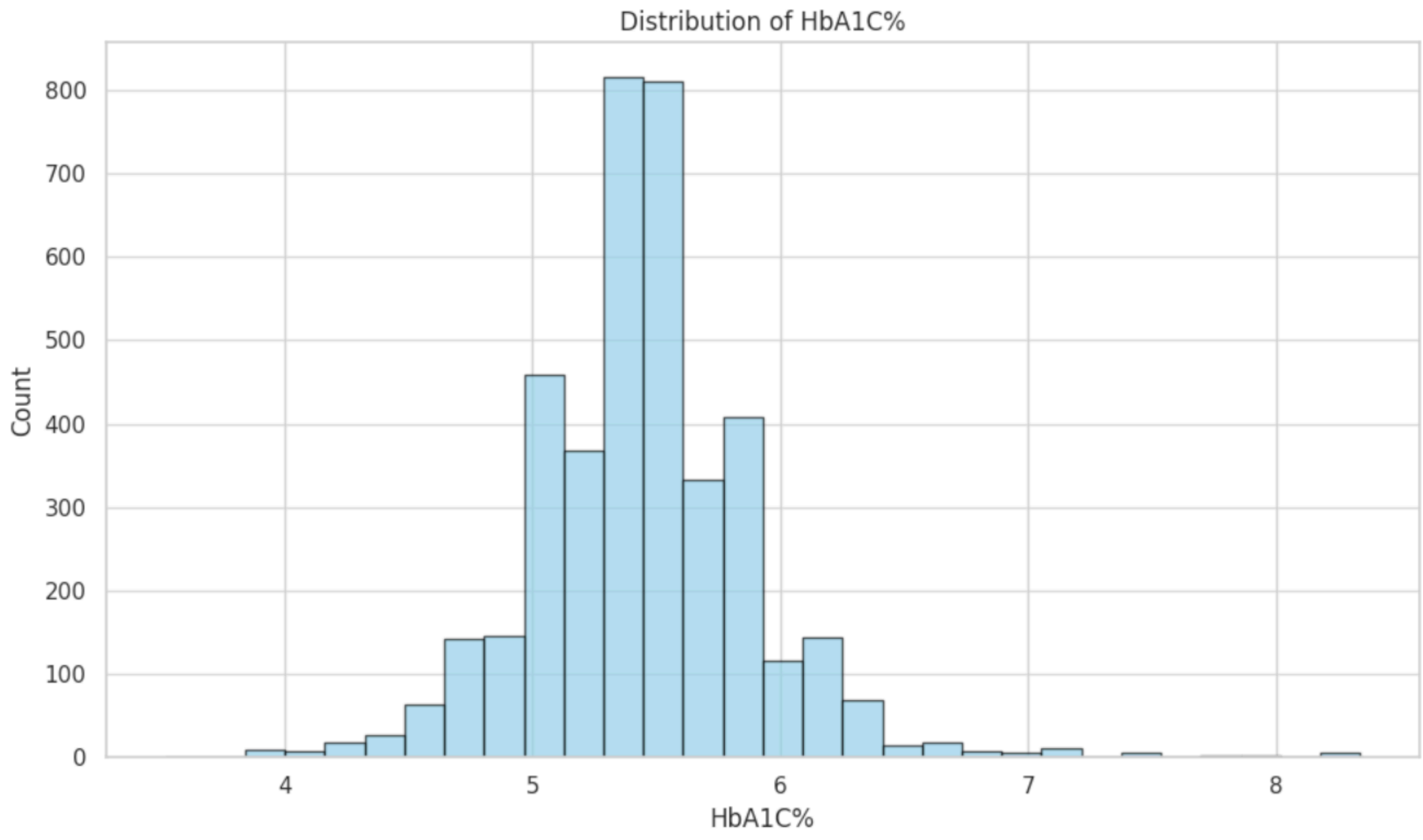

**Figure S14: HPP Distribution of HbA1C% (prediction covariant, post processing - clipping to 8 sigmas)**

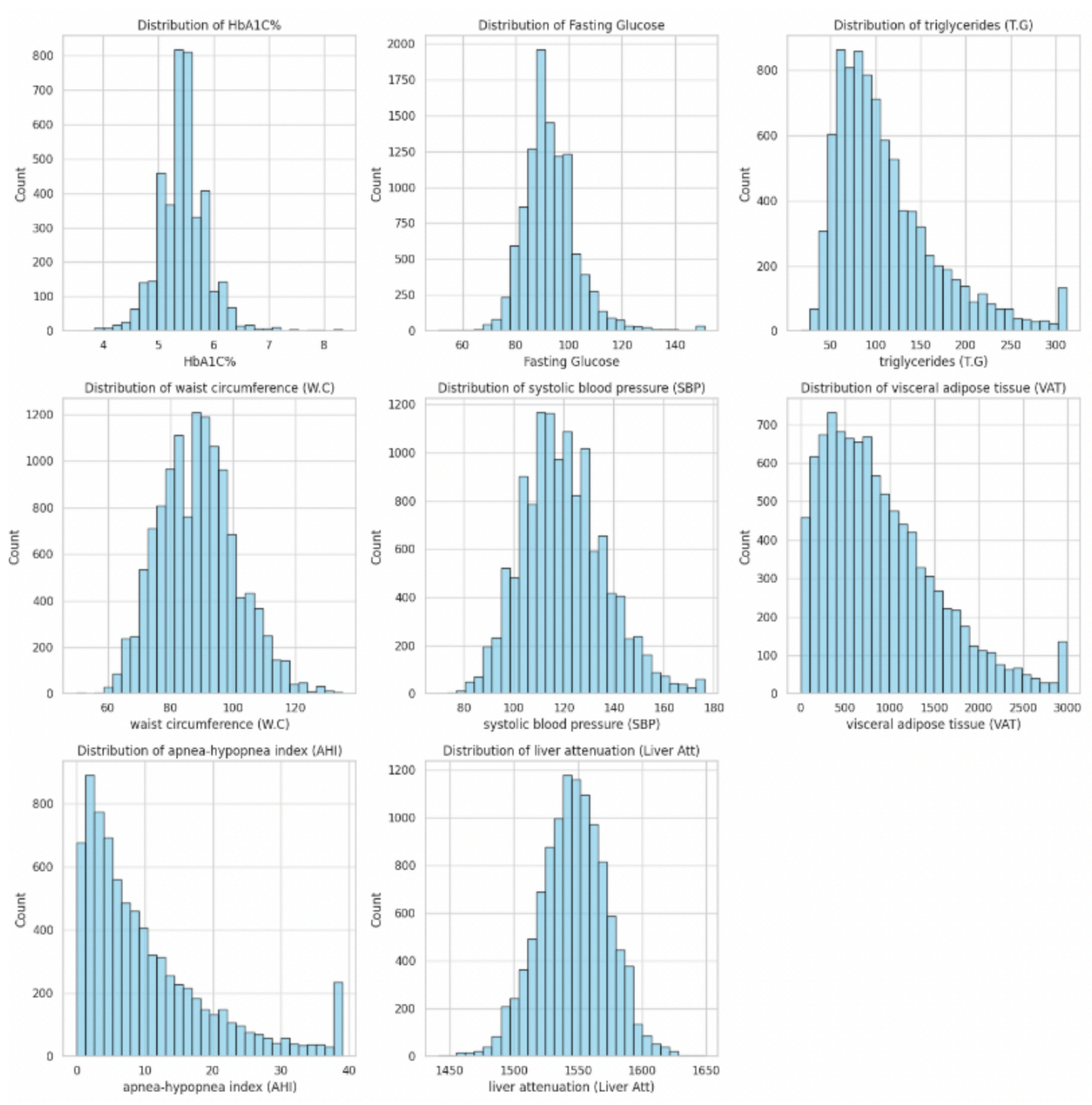

**Figure S15: HPP Distribution of Covariates used in study (prediction covariant - post processing - clipping to 8 sigmas)**

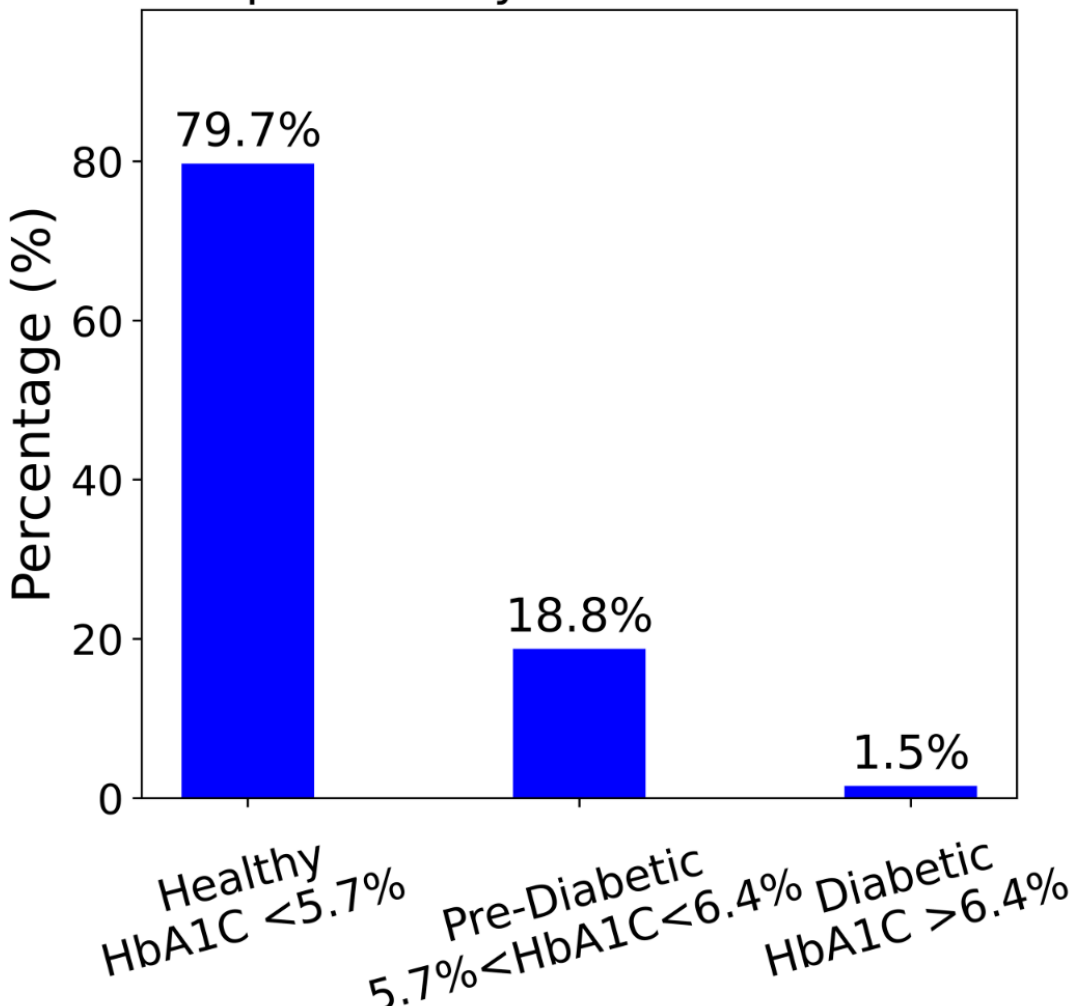

**Fig. S16. HPP Population by Diabetic Classification.** The bar chart illustrates the distribution of the HPP population according to diabetic classification based on HbA1C% levels. The majority (79.71%) of the population is classified as Healthy (HbA1C% < 5.7%), followed by 18.75% classified as Pre-Diabetic (5.7% ≤ HbA1C% < 6.4%), and a small fraction (1.54%) as Diabetic (HbA1C% ≥ 6.4%). Percentages are labeled above each bar for clarity.

**Demographics Information**

Data from individuals were included in the analysis. Of these, 53.8% were Ashkenaz, 8.1% were North African, 6.7% were Middle Eastern, 2.2% were Yemen , 2.3% were Sephardi, 2.5% were from other countries , and 24.4% reported mixed ethnic backgrounds.

**CGM Collection Time**

The continuous glucose monitoring (CGM) data were collected between December 2018 and May 2024.

<u>Additional Results:</u>

Aleppo 2017 HbA1C% Prediction:

GMI: mean: 0.67 std:  0.002

Gluforemer Representation: mean: 0.65, std  0.01

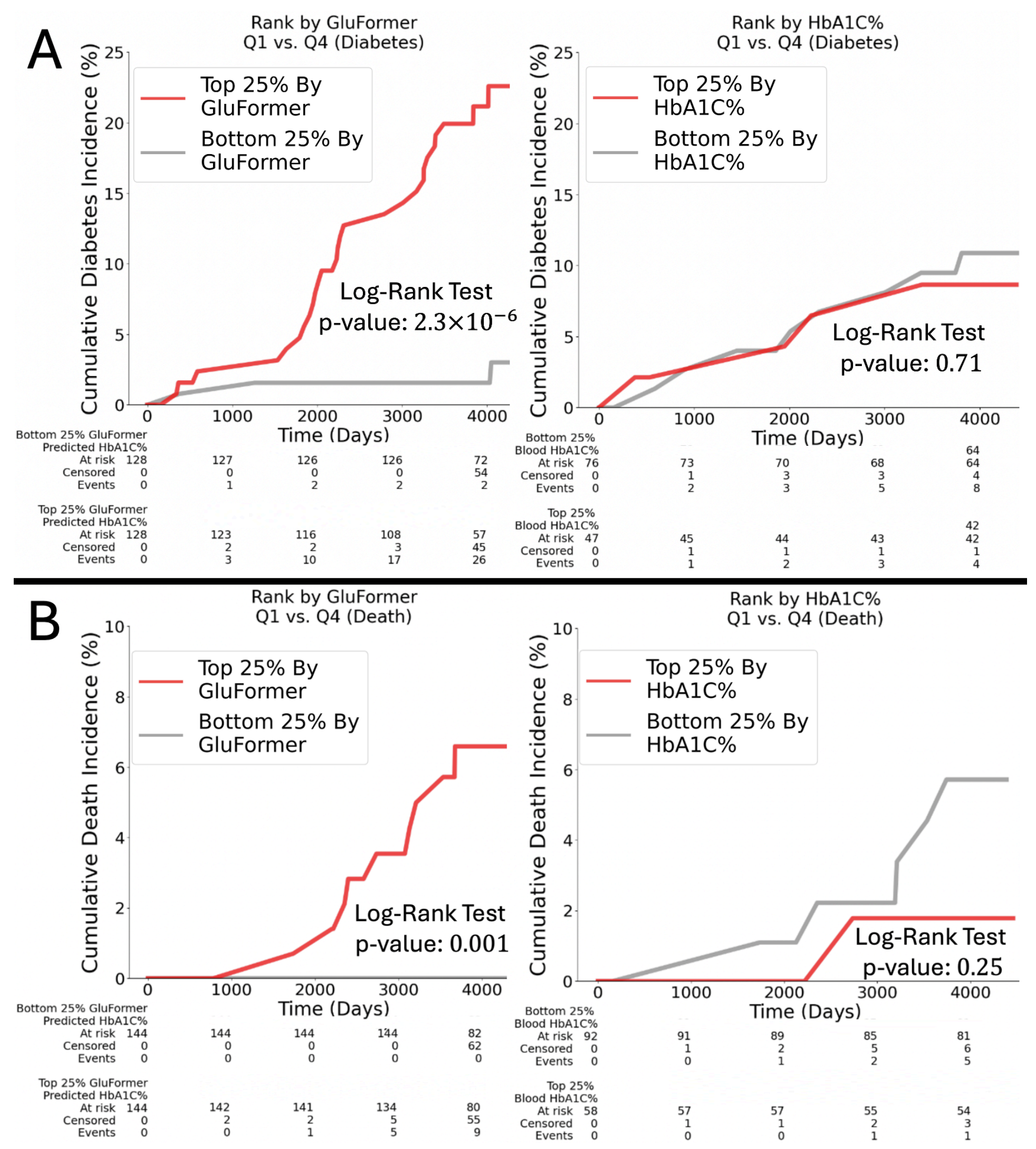

**Figure S17. Kaplan–Meier analysis of 580 adults in the AEGIS study [41], comparing diabetes-free survival over 12 years between quartiles of GluFormer - predicted versus measured (blood) HbA1C%.**

**A. (Left)** Participants were stratified by GluFormer-predicted HbA1C% into the top 25% (red) and bottom 25% (grey). Individuals in the top quartile developed diabetes at a higher rate, showing a significantly shorter diabetes-free survival (log-rank $p = 2.3\times10^{-6}$).

**(Right)** The same population was stratified by baseline blood HbA1C% into quartiles, resulting in nearly overlapping survival curves (log-rank $p = 0.71$ - not significant). This finding indicates that GluFormer captures additional glycemic-risk signals beyond those reflected by the standard HbA1C% blood test alone, enabling earlier and more accurate identification of high-risk individuals.

**B. (Left)** Participants were stratified by GluFormer-predicted HbA1C% into the top 25% (red) and bottom 25% (grey). Individuals in the top quartile had higher rates of cardiovascular death, showing a significantly shorter survival (log-rank $p = 0.001$).

**(Right)** The same population was stratified by baseline blood HbA1C% into quartiles, resulting in nearly overlapping survival curves (log-rank $p = 0.25$ - not significant).

**Additional information about the AEGIS dataset**

The AEGIS dataset comprises a longitudinal study of 580 adults from Spain, followed over a 12-year period to investigate long-term health outcomes, including diabetes incidence and cardiovascular mortality. During their enrollment, participants wore a Freestyle Libre Pro CGM device for roughly 12 days, providing detailed glucose profiles. In addition to healthy individuals, the AEGIS cohort includes participants with preexisting metabolic conditions such as type 2 diabetes. Accordingly, we note that some individuals may have been undergoing medication regimens for diabetes or other comorbidities, reflecting the real-world heterogeneity of clinical practices. This broad population distribution, coupled with long-term follow-up, makes the AEGIS dataset a particularly valuable resource for understanding how CGM-derived metrics can inform long-range cardiometabolic risk and disease progression.

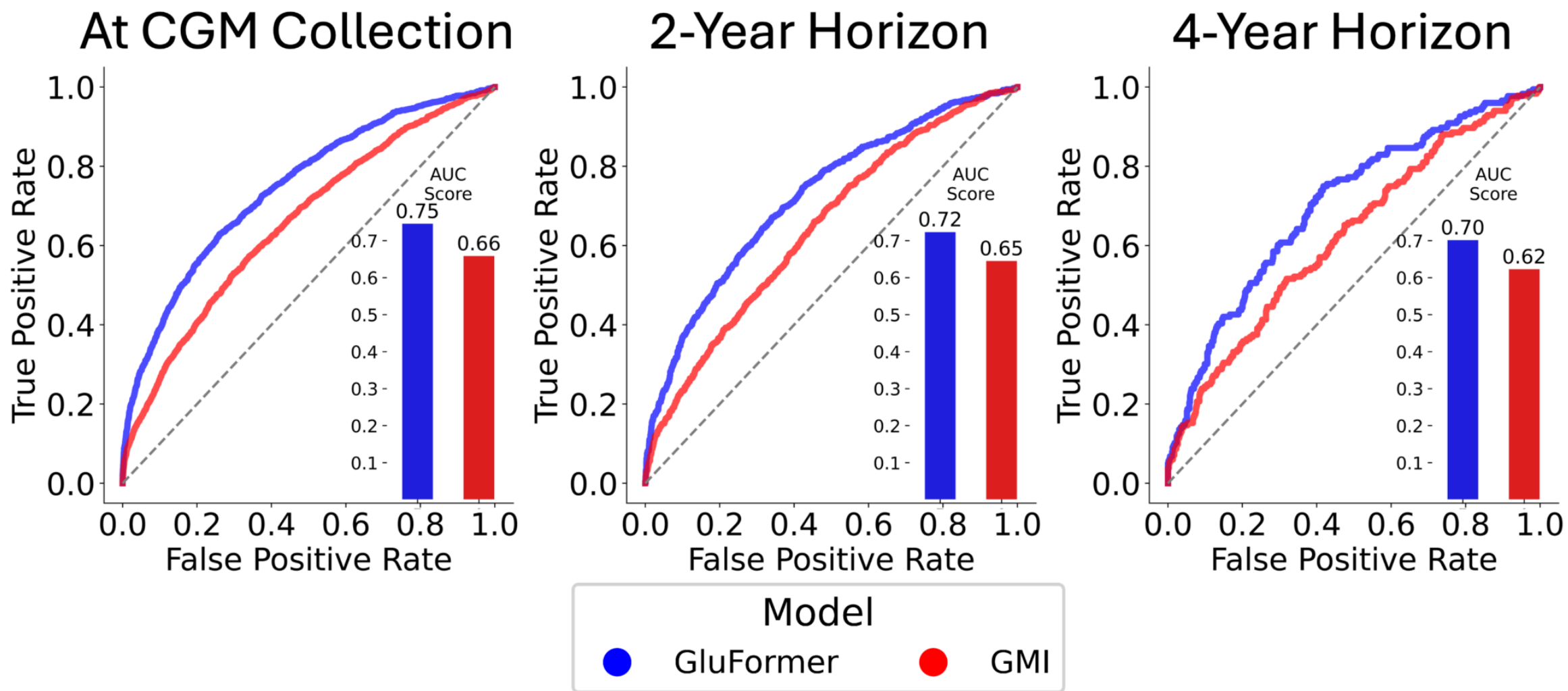

**Figure. S18. ROC AUC of Diabetes Prediction at Different Time Points**

Receiver Operating Characteristic (ROC) curves comparing the performance of GMI (red), and GluFormer (blue) for diabetes (as assigned by HbA1C% blood tests) prediction at three time points:

**At CGM Collection**: GluFormer with an AUC score of 0.75 and GMI of 0.66.

**2 Years in the Future**: GluFormer with an AUC score of 0.72 and GMI of 0.65.

**4 Years in the Future**: GluFormer with an AUC score of 0.70 and GMI of 0.62.

The results highlight the superior predictive power of GluFormer across all time points compared to GMI.

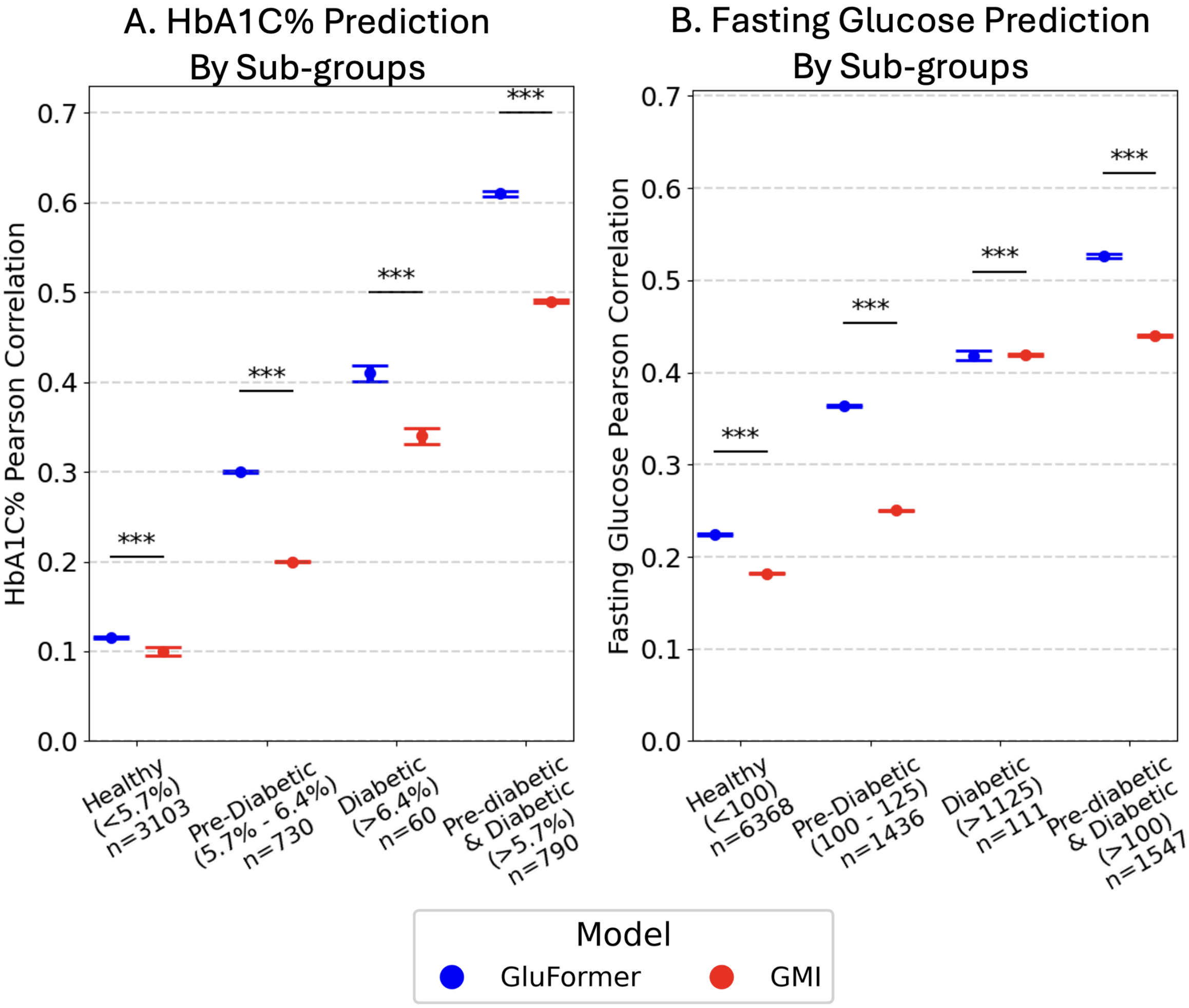

**Fig. S19. Prediction Performance of GluFormer vs. GMI Across Subgroups**

(A) HbA1C% Prediction by Subgroups: The plot shows Pearson correlation coefficients for HbA1C% predictions across subgroups (Healthy, Pre-Diabetic, Diabetic, and combined Pre-Diabetic & Diabetic) using GluFormer (blue), and GMI (red). GluFormer consistently outperforms other methods in all subgroups, with statistical significance (Mann-Whitney) indicated **(*p < 0.05, **p < 0.01, ***p < 0.001, n.s. = not significant**).

(B) Fasting Glucose Prediction by Subgroups: The plot displays Pearson correlation coefficients for fasting glucose predictions across subgroups. GluFormer demonstrates superior performance compared to alternative methods, with statistical significance across most subgroups. Sample sizes for each subgroup are provided below the x-axis.

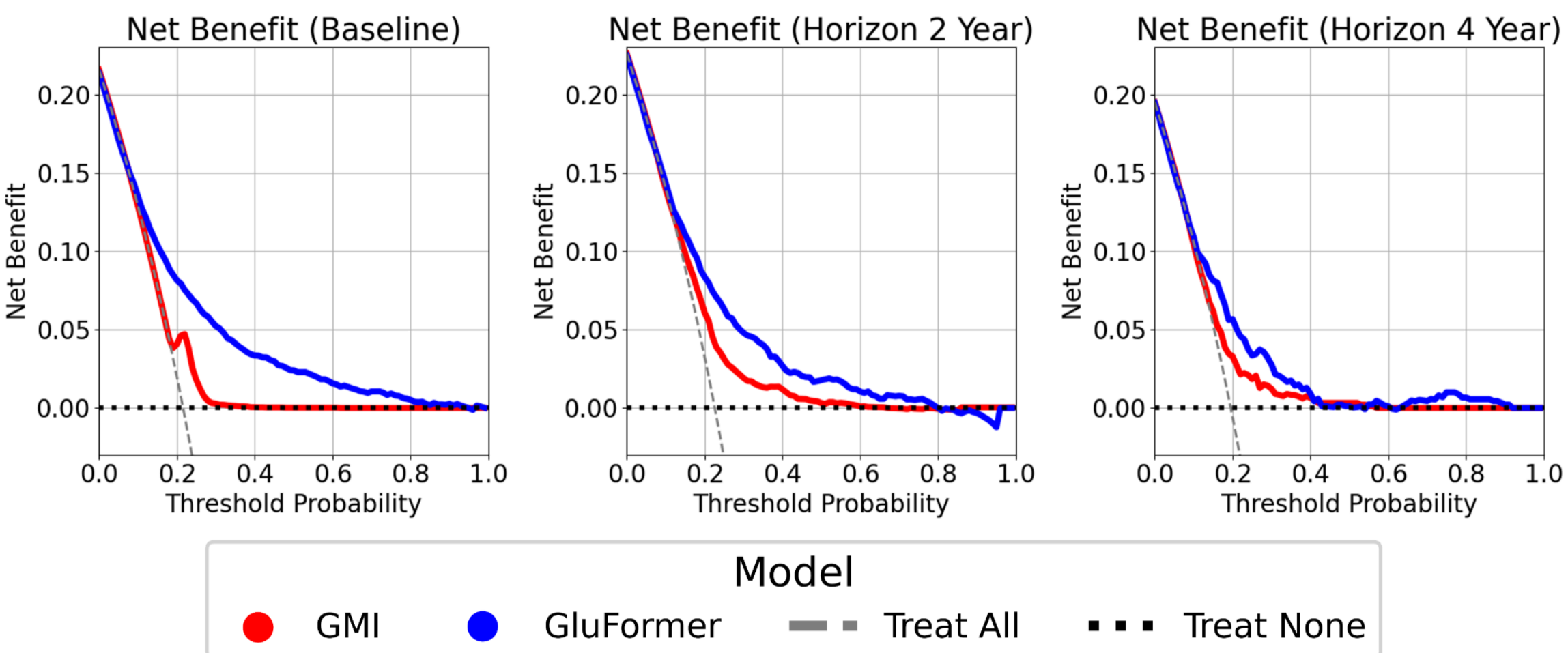

**Fig. S20. Net Benefit of Diabetes Prediction at Different Time Horizons**

Decision curve analysis evaluating the net benefit of diabetes prediction using GMI (red) and GluFormer (blue) across varying threshold probabilities at three time points:

1. **Baseline**: At the time of CGM collection, GluFormer demonstrates a consistently higher net benefit across a wide range of threshold probabilities compared to GMI.
2. **Horizon 2 Years**: GluFormer maintains superior net benefit at 2 years in the future across a wide range of threshold probabilities compared to GMI.
3. **Horizon 4 Years**: At 4 years in the future, GluFormer continues to show higher net benefit, further emphasizing its long-term predictive utility compared to GMI.

Bibliography

1. Saab, K. *et al.* Capabilities of Gemini Models in Medicine. *arXiv* (2024) doi:10.48550/arxiv.2404.18416.

2. [2406.06474] Towards a Personal Health Large Language Model. https://arxiv.org/abs/2406.06474.

3. Zhou, Y. *et al.* A foundation model for generalizable disease detection from retinal images. *Nature* **622**, 156–163 (2023).

4. Lutsker, G., Rossman, H., Godiva, N. & Segal, E. COMPRER: A Multimodal Multi-Objective Pretraining Framework for Enhanced Medical Image Representation. *arXiv* (2024) doi:10.48550/arxiv.2403.09672.

5. Zhang, J. *et al.* RETFound-enhanced community-based fundus disease screening: real-world evidence and decision curve analysis. *npj Digital Med.* **7**, 108 (2024).

6. Yuan, H. *et al.* Self-supervised learning for human activity recognition using 700,000 person-days of wearable data. *npj Digital Med.* **7**, 91 (2024).

7. Thapa, R. *et al.* SleepFM: Multi-modal Representation Learning for Sleep Across Brain Activity, ECG and Respiratory Signals. *arXiv* (2024) doi:10.48550/arxiv.2405.17766.

8. Chen, R. J. *et al.* Towards a general-purpose foundation model for computational pathology. *Nat. Med.* **30**, 850–862 (2024).

9. Krishnan, R., Rajpurkar, P. & Topol, E. J. Self-supervised learning in medicine and healthcare. *Nat. Biomed. Eng.* **6**, 1346–1352 (2022).

10. GBD 2021 Diabetes Collaborators. Global, regional, and national burden of diabetes from 1990 to 2021, with projections of prevalence to 2050: a systematic analysis for the Global Burden of Disease Study 2021. *Lancet* **402**, 203–234 (2023).

11. Khunti, K. *et al.* Diabetes and Multiple Long-term Conditions: A Review of Our Current Global Health Challenge. *Diabetes Care* **46**, 2092–2101 (2023).

12. Piché, M.-E., Tchernof, A. & Després, J.-P. Obesity phenotypes, diabetes, and cardiovascular

diseases. *Circ. Res.* **126**, 1477–1500 (2020).

13. Amiri Dash Atan, N. *et al.* Type 2 diabetes mellitus and non-alcoholic fatty liver disease: a systematic review and meta-analysis. *Gastroenterol. Hepatol. Bed Bench* **10**, S1–S7 (2017).
14. Ho, I. S. S. *et al.* Measuring multimorbidity in research: Delphi consensus study. *bmjmed* **1**, e000247 (2022).
15. Abudawood, M. Diabetes and cancer: A comprehensive review. *J. Res. Med. Sci.* **24**, 94 (2019).
16. Natesan, V. & Kim, S.-J. Diabetic Nephropathy - a Review of Risk Factors, Progression, Mechanism, and Dietary Management. *Biomol Ther (Seoul)* **29**, 365–372 (2021).
17. Andreoulakis, E., Hyphantis, T., Kandylis, D. & Iacovides, A. Depression in diabetes mellitus: a comprehensive review. *Hippokratia* **16**, 205–214 (2012).
18. Chaturvedi, S. K. *et al.* More anxious than depressed: prevalence and correlates in a 15-nation study of anxiety disorders in people with type 2 diabetes mellitus. *Gen. Psych.* **32**, e100076 (2019).
19. Kieu, A., King, J., Govender, R. D. & Östlundh, L. The benefits of utilizing continuous glucose monitoring of diabetes mellitus in primary care: A systematic review. *J. Diabetes Sci. Technol.* **17**, 762–774 (2023).
20. Moser, E. G., Crew, L. B. & Garg, S. K. Role of continuous glucose monitoring in diabetes management. *Avances en Diabetología* **26**, 73–78 (2010).
21. Battelino, T. *et al.* Continuous glucose monitoring and metrics for clinical trials: an international consensus statement. *Lancet Diabetes Endocrinol.* **11**, 42–57 (2023).
22. Holzer, R., Bloch, W. & Brinkmann, C. Continuous Glucose Monitoring in Healthy Adults-Possible Applications in Health Care, Wellness, and Sports. *Sensors* **22**, (2022).
23. Zahedani, A. D. *et al.* Digital health application integrating wearable data and behavioral patterns improves metabolic health. *npj Digital Med.* **6**, 216 (2023).
24. Shilo, S. *et al.* Continuous glucose monitoring and intrapersonal variability in fasting

glucose. *Nat. Med.* **30**, 1424–1431 (2024).

25. FDA Clears First Over-the-Counter Continuous Glucose Monitor | FDA. https://www.fda.gov/news-events/press-announcements/fda-clears-first-over-counter-continuous-glucose-monitor.
26. Vaswani, A. *et al.* Attention is all you need. *arXiv* (2017) doi:10.48550/arxiv.1706.03762.
27. Shilo, S. *et al.* 10 K: a large-scale prospective longitudinal study in Israel. *Eur. J. Epidemiol.* **36**, 1187–1194 (2021).
28. McInnes, L., Healy, J. & Melville, J. UMAP: Uniform Manifold Approximation and Projection for Dimension Reduction. *arXiv* (2018) doi:10.48550/arxiv.1802.03426.
29. Krizhevsky, A., Sutskever, I. & Hinton, G. E. ImageNet classification with deep convolutional neural networks. *Commun. ACM* **60**, 84–90 (2012).
30. Manning, C., Raghavan, P. & Schütze, H. *Introduction to information retrieval*. (Cambridge University Press, 2008).
31. Nathan, D. M. *et al.* The effect of intensive treatment of diabetes on the development and progression of long-term complications in insulin-dependent diabetes mellitus. *N. Engl. J. Med.* **329**, 977–986 (1993).
32. King, P., Peacock, I. & Donnelly, R. The UK prospective diabetes study (UKPDS): clinical and therapeutic implications for type 2 diabetes. *Br. J. Clin. Pharmacol.* **48**, 643–648 (1999).
33. LeCun, Y. & Yoshua, B. Convolutional networks for images, speech, and time series. *The handbook of brain theory and neural networks* **3361**, (1995).
34. LeCun, Y. *et al.* Handwritten Digit Recognition with a Back-Propagation Network. *Advances in Neural Information Processing Systems* (1989).
35. Bergenstal, R. M. *et al.* Glucose management indicator (GMI): A new term for estimating A1C from continuous glucose monitoring. *Diabetes Care* **41**, 2275–2280 (2018).
36. Balestriero, R. *et al.* A Cookbook of Self-Supervised Learning. *arXiv* (2023) doi:10.48550/arxiv.2304.12210.

37. Broll, S. *et al.* Interpreting blood GLUcose data with R package iglu. *PLoS ONE* **16**, e0248560 (2021).

38. Rodbard, D. New and improved methods to characterize glycemic variability using continuous glucose monitoring. *Diabetes Technol. Ther.* **11**, 551–565 (2009).

39. The Diabetes Prevention Program Research Group. The Diabetes Prevention Program. Design and methods for a clinical trial in the prevention of type 2 diabetes. *Diabetes Care* **22**, 623–634 (1999).

40. Knowler, W. C. *et al.* Reduction in the incidence of type 2 diabetes with lifestyle intervention or metformin. *N. Engl. J. Med.* **346**, 393–403 (2002).

41. Gude, F. *et al.* Glycemic variability and its association with demographics and lifestyles in a general adult population. *J. Diabetes Sci. Technol.* **11**, 780–790 (2017).

42. Wang, J. *et al.* Self-improving generative foundation model for synthetic medical image generation and clinical applications. *Nat. Med.* (2024) doi:10.1038/s41591-024-03359-y.

43. Keshet, A. *et al.* CGMap: Characterizing continuous glucose monitor data in thousands of non-diabetic individuals. *Cell Metab.* **35**, 758-769.e3 (2023).

44. Vickers, A. J., Van Calster, B. & Steyerberg, E. W. Net benefit approaches to the evaluation of prediction models, molecular markers, and diagnostic tests. *BMJ* **352**, i6 (2016).

45. [2412.10288] Performance evaluation of predictive AI models to support medical decisions: Overview and guidance. https://arxiv.org/abs/2412.10288.

46. Cersosimo, E., Solis-Herrera, C., Trautmann, M. E., Malloy, J. & Triplitt, C. L. Assessment of pancreatic β-cell function: review of methods and clinical applications. *Curr. Diabetes Rev.* **10**, 2–42 (2014).

47. Abdul-Ghani, M. A. *et al.* The relationship between fasting hyperglycemia and insulin secretion in subjects with normal or impaired glucose tolerance. *Am. J. Physiol. Endocrinol. Metab.* **295**, E401-6 (2008).

48. Brown, T. B. *et al.* Language models are few-shot learners. *arXiv* (2020)

doi:10.48550/arxiv.2005.14165.

49. Mwaniki, P. *et al.* Using self-supervised feature learning to improve the use of pulse oximeter signals to predict paediatric hospitalization. *Wellcome Open Res.* **6**, 248 (2021).

50. Danne, T. *et al.* International consensus on use of continuous glucose monitoring. *Diabetes Care* **40**, 1631–1640 (2017).

51. [2005.10111] The Effectiveness of Discretization in Forecasting: An Empirical Study on Neural Time Series Models. https://arxiv.org/abs/2005.10111.

52. FDA approves first continuous glucose monitoring system with a fully implantable glucose sensor and compatible mobile app for adults with diabetes | FDA. https://www.fda.gov/news-events/press-announcements/fda-approves-first-continuous-glucose-monitoring-system-fully-implantable-glucose-sensor-and.

53. Cefalu, W. T. *et al.* A global initiative to deliver precision health in diabetes. *Nat. Med.* **30**, 1819–1822 (2024).

54. Ahlqvist, E., Prasad, R. B. & Groop, L. Subtypes of type 2 diabetes determined from clinical parameters. *Diabetes* **69**, 2086–2093 (2020).

55. Xiong, Z. *et al.* How generalizable are foundation models when applied to different demographic groups and settings? *NEJM AI* (2024) doi:10.1056/AIcs2400497.

56. Ansari, A. F. *et al.* Chronos: Learning the Language of Time Series. *arXiv* (2024) doi:10.48550/arxiv.2403.07815.

57. Oord, A. van den *et al.* WaveNet: A Generative Model for Raw Audio. *arXiv* (2016) doi:10.48550/arxiv.1609.03499.

58. Oord, A. van den, Kalchbrenner, N. & Kavukcuoglu, K. Pixel Recurrent Neural Networks. *arXiv* (2016) doi:10.48550/arxiv.1601.06759.

59. Loshchilov, I. & Hutter, F. Decoupled Weight Decay Regularization. *arXiv* (2017) doi:10.48550/arxiv.1711.05101.

60. Devlin, J., Chang, M.-W., Lee, K. & Toutanova, K. BERT: Pre-training of deep bidirectional

transformers for language understanding. *arXiv* (2018) doi:10.48550/arxiv.1810.04805.

61. Zeevi, D. *et al.* Personalized nutrition by prediction of glycemic responses. *Cell* **163**, 1079–1094 (2015).
62. Ben-Yacov, O. *et al.* Personalized Postprandial Glucose Response-Targeting Diet Versus Mediterranean Diet for Glycemic Control in Prediabetes. *Diabetes Care* **44**, 1980–1991 (2021).
63. Korem, T. *et al.* Bread Affects Clinical Parameters and Induces Gut Microbiome-Associated Personal Glycemic Responses. *Cell Metab.* **25**, 1243-1253.e5 (2017).
64. Rein, M. S. *et al.* BREAst Cancer Personalised NuTrition (BREACPNT): dietary intervention in breast cancer survivors treated with endocrine therapy - a protocol for a randomised clinical trial. *BMJ Open* **12**, e062498 (2022).
65. Htet, T. D. *et al.* Rationale and design of a randomised controlled trial testing the effect of personalised diet in individuals with pre-diabetes or type 2 diabetes mellitus treated with metformin. *BMJ Open* **10**, e037859 (2020).
66. Mendes-Soares, H. *et al.* Assessment of a personalized approach to predicting postprandial glycemic responses to food among individuals without diabetes. *JAMA Netw. Open* **2**, e188102 (2019).
67. Rein, M. *et al.* Effects of personalized diets by prediction of glycemic responses on glycemic control and metabolic health in newly diagnosed T2DM: a randomized dietary intervention pilot trial. *BMC Med.* **20**, 56 (2022).
68. Shilo, S. *et al.* Prediction of personal glycemic responses to food for individuals with type 1 diabetes through integration of clinical and microbial data. *Diabetes Care* **45**, 502–511 (2022).
69. Shilo, S. *et al.* The gut microbiome of adults with type 1 diabetes and its association with the host glycemic control. *Diabetes Care* **45**, 555–563 (2022).
70. Weinstock, R. S. *et al.* Risk factors associated with severe hypoglycemia in older adults with

type 1 diabetes. *Diabetes Care* **39**, 603–610 (2016).

71. Juvenile Diabetes Research Foundation Continuous Glucose Monitoring Study Group *et al.* Continuous glucose monitoring and intensive treatment of type 1 diabetes. *N. Engl. J. Med.* **359**, 1464–1476 (2008).

72. Colás, A., Vigil, L., Vargas, B., Cuesta-Frau, D. & Varela, M. Detrended Fluctuation Analysis in the prediction of type 2 diabetes mellitus in patients at risk: Model optimization and comparison with other metrics. *PLoS ONE* **14**, e0225817 (2019).

73. Zhao, Q. *et al.* Chinese diabetes datasets for data-driven machine learning. *Sci. Data* **10**, 35 (2023).

74. Wadwa, R. P. *et al.* Trial of Hybrid Closed-Loop Control in Young Children with Type 1 Diabetes. *N. Engl. J. Med.* **388**, 991–1001 (2023).

75. Aleppo, G. *et al.* REPLACE-BG: A Randomized Trial Comparing Continuous Glucose Monitoring With and Without Routine Blood Glucose Monitoring in Adults With Well-Controlled Type 1 Diabetes. *Diabetes Care* **40**, 538–545 (2017).

76. Chase, H. P. *et al.* A randomized multicenter trial comparing the GlucoWatch Biographer with standard glucose monitoring in children with type 1 diabetes. *Diabetes Care* **28**, 1101–1106 (2005).